\documentclass[fleqn,usenatbib]{mnras}

\usepackage{newtxtext,newtxmath}

\usepackage[T1]{fontenc}
\usepackage{ae,aecompl}

\DeclareRobustCommand{\VAN}[3]{#2}
\let\VANthebibliography\thebibliography
\def\thebibliography{\DeclareRobustCommand{\VAN}[3]{##3}\VANthebibliography}

\usepackage{graphics}
\usepackage{pgfplotstable}
\usepackage{natbib}
\usepackage{float}
\usepackage{placeins}
\usepackage{textcomp}
\usepackage[figuresright]{rotating}
\usepackage{graphicx}
\usepackage{caption}
\usepackage{subcaption}
\usepackage{array}
\usepackage{url}
\usepackage{hyperref}
\usepackage{geometry} 
\usepackage{endnotes}
\usepackage{adjustbox}
\usepackage{longtable} 
\usepackage{pdflscape} 
\usepackage{mathtools} 
\usepackage{amsmath}
\usepackage{amssymb}
\usepackage{makecell}
\newcolumntype{P}[1]{>{\raggedleft\arraybackslash}m{#1}}
\newcolumntype{Q}[1]{>{\raggedright\arraybackslash}m{#1}}

\usepackage{dcolumn}
\usepackage{bm}
\usepackage{xtab}
\usepackage{booktabs}
\usepackage[T1]{fontenc}
\usepackage[utf8]{inputenc}
\usepackage{lineno}
\usepackage[graphicx]{realboxes}

\usepackage{siunitx}
\pgfplotsset{compat=1.16}

\usepackage{graphicx}	
\usepackage{amsmath}	

\newcolumntype{R}[1]{>{\raggedleft\arraybackslash}m{#1}}
\newcolumntype{L}[1]{>{\raggedright\arraybackslash}m{#1}}

\usepackage{endnotes}

\let\footnote=\endnote
\usepackage{xpatch}
\makeatletter
\xpretocmd{\theendnotes}{%
  \xpatchcmd{\@makeenmark}{\hbox{\@textsuperscript{\normalfont\@theenmark}}}{%
   \hbox{\normalfont[\theenmark].\space}%
 }{}{}%
}{}{}

\title[The catalogue of optical GRBs with redshift (Part 1)]{An Optical Gamma-Ray Burst Catalogue with Measured Redshift PART I: Data Release of 535 Gamma-Ray Bursts and Colour Evolution}

\author[M. G. Dainotti et al.]{M. G. Dainotti$^{1,2,3,4,5}$\thanks{E-mail: maria.dainotti@nao.ac.jp (NAOJ,SOKENDAI)},
B. De Simone$^{6,7}$,
R. F. Mohideen Malik$^{8,9}$\thanks{The second and third author have contributed equally},
V. Pasumarti$^{10}$,
D. Levine$^{11}$,
\newauthor
N. Saha$^{12,13}$\thanks{The fifth and sixth author have contributed equally},
B. Gendre$^{14}$,
D. Kido$^{15}$,
A. M. Watson$^{16}$,
R. L. Becerra$^{8,17}$,
S. Belkin$^{18,19,20}$,
S. Desai$^{10}$,
\newauthor
A. C. C. do E. S. Pedreira$^{16}$,
U. Das$^{21}$,
L. Li$^{22}$,
S. R. Oates$^{23}$,
S. B. Cenko$^{24,25}$,
A. Pozanenko$^{18,26}$,
\newauthor
A. Volnova$^{18}$,
Y. -D. Hu$^{27,28}$,
A. J. Castro-Tirado$^{27,29}$,
N. B. Orange$^{30,31}$,
T. J. Moriya$^{1,2,20}$,
N. Fraija$^{16}$,
Y. Niino$^{32}$,
\newauthor
E. Rinaldi$^{33}$,
N. R. Butler$^{34}$,
J. d. J. G. González$^{16}$,
A. S. Kutyrev$^{24,25}$,
W. H. Lee$^{16}$,
X. Prochaska$^{35}$,
\newauthor
E. Ramirez-Ruiz$^{35}$,
M. Richer$^{16}$,
M. H. Siegel$^{36}$,
K. Misra$^{37}$,
A. Rossi$^{38}$,
C. Lopresti$^{39}$,
U. Quadri$^{40}$,
\newauthor
L. Strabla$^{40}$,
N. Ruocco$^{41}$,
S. Leonini$^{42}$,
M. Conti$^{42}$,
P. Rosi$^{42}$,
L. M. T. Ramirez$^{42}$,
S. Zola$^{43}$,
\newauthor
I. Jindal$^{44}$,
R. Kumar$^{45}$,
L. Chan$^{46}$,
M. Fuentes$^{47}$,
G. Lambiase$^{6,7}$,
K. K. Kalinowski$^{43,48}$,
W. Jamal$^{49}$
\\
$^{1}$Division of Science, National Astronomical Observatory of Japan, 2-21-1 Osawa, Mitaka, Tokyo 181-8588, Japan\\
$^{2}$The Graduate University for Advanced Studies (SOKENDAI), Shonankokusaimura, Hayama, Miura District, Kanagawa 240-0115\\
$^{3}$Space Science Institute, 4765 Walnut St Ste B, Boulder, CO 80301, USA\\
$^{4}$Nevada Center for Astrophysics, University of Nevada, 4505 Maryland Parkway, Las Vegas, NV 89154, USA\\
$^{5}$Bay Environmental Institute, P.O. Box 25 Moffett Field, CA, California\\
$^{6}$Dipartimento di Fisica, Universit\'a di Salerno, Via Giovanni Paolo II, 132 I-84084 Fisciano (SA), Italy\\
$^{7}$INFN, Sezione di Napoli, Gruppo collegato di Salerno, Italy\\
$^{8}$Department of Physics, University of Rome "Tor Vergata", Italy\\
$^{9}$Department of Astronomy, University of Belgrade, Serbia\\
$^{10}$Department of Physics, Indian Institute of Technology Hyderabad, Kandi, Telangana-502284, India\\
$^{11}$Department of Astronomy, California Institute of Technology, Pasadena, CA, 91125, USA\\
$^{12}$Department of Physical Science, Indian Institute of Science Education and Research Kolkata, Mohanpur-741246, India\\
$^{13}$Astronomisches Rechen-Institut, Zentrum für Astronomie der Universität Heidelberg (ZAH), Mönchhofstr. 12-14, 69120 Heidelberg, Germany\\
$^{14}$The University of Western Australia - OzGrav, School of Physics (M013), 35 Stirling Highway, 6009 Perth, Australia\\
$^{15}$Department of Physics, Faculty of Science, the University of Tokyo, Japan\\
$^{16}$Instituto de Astronomia, Universidad Nacional Autónoma de México, Circuito Exterior, C.U., A. Postal 70-264, 04510, CDMX, Mexico\\
$^{17}$Instituto de Ciencias Nucleares, Universidad  Nacional Aut\'onoma de M\'exico, Apartado Postal 70-264,04510 M\'exico, CDMX, Mexico\\
$^{18}$Space Research Institute of the Russian Academy of Sciences (IKI), 84/32 Profsoyuznaya Street, Moscow, 117997, Russia\\
$^{19}$National Research University “Higher School of Economics”, Myasnitskaya ul. 20, Moscow, 101000 Russia\\
$^{20}$School of Physics and Astronomy, Monash University, Clayton, Victoria 3800, Australia\\
$^{21}$School of Physical Sciences, National Institute of Science Education and Research, Bhubaneswar, India \\
$^{22}$ICRA Net, Piazza della Repubblica 10, I-65122 Pescara, Italy\\
$^{23}$Physics Department, Lancaster University, Bailrigg, Lancaster LA1 4YB, UK \\
$^{24}$Astrophysics Science Division, NASA Goddard Space Flight Center, MC 661, Greenbelt, MD 20771, USA\\
$^{25}$Joint Space-Science Institute, University of Maryland, College Park, MD 20742, USA\\
$^{26}$Institute of Physics and Technology, Institutskiy Pereulok, 9, Dolgoprudny, 141701, Russia\\
$^{27}$Instituto de Astrof\'isica de Andaluc\'ia (IAA-CSIC), Glorieta de la Astronom\'ia s/n, E-18008, Granada, Spain\\
$^{28}$INAF - Osservatorio Astronomico di Brera, Via E. Bianchi 46, I-23807, Merate (LC), Italy \\
$^{29}$Departamento de Ingenier\'ia de Sistemas y Autom\'atica, Escuela de Ingenier\'ias, Universidad de M\'alaga, C. Dr. Ortiz Ramos sn, 29071, M\'alaga, Spain\\
$^{30}$OrangeWave Innovative Science, LLC, Moncks Corner, SC 29461, USA\\
$^{31}$Etelman Observatory Research Center, University of the Virgin Islands, St. Thomas 00802, USVI, USA\\
$^{32}$Institute of Astronomy, The University of Tokyo, 2-21-1 Osawa, Mitaka, Tokyo 181-0015, Japan\\
$^{33}$Interdisciplinary Theoretical and Mathematical Sciences (iTHEMS) Program, RIKEN, Wako, Saitama 351-0198, Japan\\
$^{34}$School of Earth and Space Exploration, Arizona State University, Tempe, AZ85287, USA\\
$^{35}$Department of Astronomy and Astrophysics, UCO/Lick Observatory, University of California, Santa Cruz, 1156 High Street, Santa Cruz, CA 95064, USA\\
$^{36}$Department of Astronomy \& Astrophysics, The Pennsylvania State University, University Park, PA 16802, USA\\
$^{37}$ARIES, Manora Peak, Nainital 263001 India\\
$^{38}$INAF, Osservatorio di Astrofisica e Scienza dello Spazio, via Piero Gobetti 93/3, 40024, Bologna, Italy\\
$^{39}$GAD-Gruppo Astronomia Digitale, Italy, GAD -Observatory - UAI-SSV GRB Section\\
$^{40}$Osservatorio Bassano Bresciano (565), Bassano Bresciano, Italy - UAI-SSV GRB Section\\
$^{41}$Osservatorio Astronomico “Nastro Verde” (C82), Sorrento, Italy - UAI-SSV GRB Section\\
$^{42}$Montarrenti Observatory (C88), Siena, Italy - UAI-SSV GRB Section\\
$^{43}$Astronomical Observatory, Jagiellonian University, ul. Orla 171, 30-244 Kraków, Poland\\
$^{44}$Indian Institute of Technology Kharagpur, Kharagpur, West Bengal 721302, India\\
$^{45}$Department of Physics, Indian Institute of Technology, Kanpur 208016, India\\
$^{46}$Assets High School, Honolulu, HI 96817, USA\\
$^{47}$Department of Aerospace Engineering, University of Michigan, Ann Arbor, MI 48109, USA\\
$^{48}$Department of Physics and Astronomy, Aarhus University, Ny Munkegade 120, DK-8000 Aarhus C, Denmark\\
$^{49}$Lebanese International University, 9RWV+M37, 26, Leban\\
}
\date{Accepted XXX. Received YYY; in original form ZZZ}

\pubyear{2024}

\begin{document}
\label{firstpage}
\pagerange{\pageref{firstpage}--\pageref{lastpage}}
\maketitle

\begin{abstract}
We present the largest optical photometry compilation of Gamma-Ray Bursts (GRBs) with redshifts ($z$). We include 64813 observations of 535 events (including upper limits) from 28 February 1997 to 18 August 2023. We also present a user-friendly web tool
\href{https://github.com/SLAC-Gamma-Rays/grbLC}{\textit{grbLC}} which allows users to visualise photometry, coordinates, redshift, host galaxy extinction, and spectral indices for each event in our database. Furthermore, we have added a Gamma-ray Coordinate Network (GCN) scraper that can be used to collect data by gathering magnitudes from the GCNs.
The web tool also includes a package for uniformly investigating colour evolution. We compute the optical spectral indices for 138 GRBs, for which we have at least 4 filters at the same epoch in our sample, and craft a procedure to distinguish between GRBs with and without colour evolution.
By providing a uniform format and repository for the optical catalogue, this web-based archive is the first step towards unifying several community efforts to gather the photometric information for all GRBs with known redshifts.
This catalogue will enable population studies by providing light curves (LCs) with better coverage since we have gathered data from different ground-based locations. Consequently, these LCs can be used to train future LC reconstructions for an extended inference of the redshift. The data gathering also allows us to fill some of the orbital gaps from Swift in crucial points of the LCs, e.g., at the end of the plateau emission or where a jet break is identified.
\end{abstract}

\begin{keywords}
catalogue -- Gamma-Ray Burst
\end{keywords}



\section{Introduction}\label{sec:intro}
Gamma-ray bursts (GRBs) are among the most intense explosions in the universe, releasing from $10^{46}$ to $10^{54}$ ergs in $\gamma$-rays in time scales ranging from a fraction of seconds up to a few hours. GRBs have two main phases: the prompt emission, which is the initial explosion typically observed from hard X-rays to $\ge100$ MeV $\gamma$-rays, though sometimes in optical as well \citep{Vestrand2005Natur,Blake2005Natur,Beskin2010ApJ}; the afterglow \citep[e.g.,][]{1997Natur.387..783C, 1997Natur.386..686V,1998A&A...331L..41P,Gehrels2009ARA&A,Wang2015}, which is the long-lasting emission in X-ray, optical, and occasionally radio wavelengths following the prompt phase.
GRBs are typically categorized as either Short GRBs (SGRBs) or Long GRBs (LGRBs), based on their prompt duration: SGRBs have $T_{90}\leq 2\mathrm{\,s}$ and LGRBs have $T_{90} \ge 2\mathrm{\,s}$, where $T_{90}$ represents the duration over which a burst emits between $5\%$ and $95\%$ of its prompt emission total measured counts \citep{mazets1981catalog, kouveliotou1993identification}. 
LGRBs originate from the deaths of massive stars \citep{1993ApJ...405..273W,1998ApJ...494L..45P, Woosley2006ARA&A,Cano2017}, while SGRBs are produced by the combination of two compact objects, such as two neutron stars (NSs) \citep[NSs,][]{Eichler1989,Kochanek93,2011MNRAS.413.2031M}, or a NS and a black hole (BH)~\citep[NS-BH,][]{1992ApJ...395L..83N}.
The central engines of these models are thought to be either hyper-accreting BHs or fast-spinning newborn NSs, called magnetars.

We provide details of the current challenges in GRB studies and the importance of creating a uniform optical catalogue for a large sample of LCs with good data coverage:

\textbf{(1)} The transient nature of GRBs requires decisions about follow-up observations to be made swiftly, especially with unusual bursts such as high-redshift candidates, under-luminous GRBs, or SGRBs that could show an associated kilonova \citep[KN, see e.g.][]{rossi2020comparison,2022Natur.612..223R,Becerra2023}. 

In addition, GRB classification is challenged by the emerging presence of many sub-classes besides the LGRBs and SGRBs, which account for the diversity of GRB features.
Because of this complexity in GRB classification, a comprehensive collection of their complete optical LCs would aid in their categorization.

\textbf{(2)} Many GRB LCs show interesting features, such as flares, bumps, or plateaus, the understanding of which may give insight into the behaviours and progenitors of these GRBs. It is, therefore, crucial to have as complete coverage of the LCs as possible to better identify and characterize these features. In addition, complete optical LCs also allow for the estimation of the optical jet break time and comparison with jet breaks in other electromagnetic bands ($\gamma$-rays, X-rays, and radio), thus enhancing the investigation of chromatic or achromatic breaks in multiwavelengths \citep{Becerra2023b}.

Particularly the ``plateau" phase, found in X-ray, optical, and radio LCs, following the prompt emission and preceding the steep decay phase \citep{2006ApJ...647.1213O, 2007ApJ...669.1115S, Dainotti2013, 2020ApJ...904...97D, 2020ApJ...905L..26D,  2020ApJ...905..112F, 2022ApJ...925...15L, 2022ApJ...934..188F, 2023MNRAS.tmp.2216F,2022ApJS..261...25D,Becerra2023b} has been observed using data from the Neil Gehrels Swift Observatory (Swift, \citealt{2004ApJ...611.1005G}), which has allowed for earlier observations of GRB afterglows.
Many explanations have been proposed for the plateau, 
however, temporal gaps due to satellites orbital period, the lack of quick follow-up observations, and the process of stitching together multiple epochs make it difficult to draw a definitive conclusion. A catalogue that compiles data from different instruments and locations may allow some of these issues to be resolved.

Indeed, a large sample of GRB optical LCs will also allow testing of the standard fireball model \citep{1999PhR...314..575P}, which is currently the most accredited description of the GRB emission. Moreover, a collection of limiting magnitudes would help the community put new constraints on the GRB progenitor physics.

\textbf{(3)} GRB samples are often incomplete in redshift data, which hinders GRB population studies and the use of GRBs in cosmological studies to understand the early universe. These studies involve correlations among intrinsic GRB parameters, such as
the Dainotti 2D and Dainotti 3D fundamental plane relations \citep{Dainotti2008,Dainotti2010,Dainotti2011,Dainotti2013,Dainotti2016,Dainotti2020a,dainotti2015}, the Amati relation between $E_{\rm iso}$ and $E_{\rm peak}$ \citep{2009A&A...508..173A, 1999ApJ...511..550L}, the $E_{\gamma}-E_{\rm peak}$ relation \citep{2004ApJ...616..331G, 2006NJPh....8..123G}, the $L-V$ luminosity-variability relation \citep{2000ApJ...534..248N, 2000astro.ph..4176F},  the $L-E_{\rm peak}$ relation \citep{2004ApJ...609..935Y}, and the Energy-Hardness-Duration (EHD) relation, i.e. the combination of $E_{\rm iso}$ and $E_{\rm peak}$ versus $T_{90}$ \citep{Minaev2020}. An extensive catalogue of well-sampled GRB LCs with redshift is therefore vital to develop these correlations.

One key issue in addressing the challenges listed above is retrieving many optical data points, along with the colour evolution analysis of GRBs. Although several catalogues of optical observations 
exist, data from these and other databases, such as the GCN or the Swift catalogue, do not report data in a uniform format and do not always include host galaxy and spectral information. Having essential properties such as the coordinates, redshift, host extinction, the GRB classification reported in the literature, and the spectral index reported alongside the magnitudes in a uniform format would be helpful for the aforementioned future GRB studies. In this sense, the main advantage of this GRB catalogue is the uniformity in the data presentation. For each event the following are included: the time in seconds after the trigger, magnitude, magnitude error (1 $\sigma$), filter, magnitude system (AB, Vega), observatory+telescope+instrument, a flag for specifying if the Galactic extinction correction has been performed, the source of data (GCN, paper, etc.), and a flag for the quality of the photometry reported. Since this catalogue will be released to the community as an open project, we encourage the astronomical community to contribute to the catalogue by submitting their data to our system following this convention.

Indeed, a critical advantage of this catalog is its ability to report data points from several telescopes that, when combined, can highlight features that otherwise would remain hidden. This allows important information to be revealed even if one telescope carries only a few data points. A remarkable example of this is the Subaru data for the GRB 010222A, which are critically 
positioned at the end time of the plateau, providing a more explicit determination of the plateau's end time, \citep[see][]{2022arXiv220312908D}. A similar situation also applies to other LC segments.

The data from this catalogue will be available for download, reducing the time-consuming data collection and analysis of the LCs for the community. This work follows what has been initiated by several authors in the past, like the catalogues of \citet{Oates2009} and \citet{Roming2009,roming2017large} who focus on a specific instrument, i.e. the Swift Ultraviolet and Optical Telescope (UVOT). Here, we have expanded the analysis to a more extensive collection of different telescopes as done by \citet{kann2006signatures,2010ApJ...720.1513K,2011ApJ...734...96K,2024arXiv240300101K}. 

Having clarified the urgent need for this catalogue, we present the compilation of 535 optical LCs included in a new web-based repository. In this paper, we also discuss the analysis of the LC colour behaviour and present two new pieces of software that aid in developing this catalogue and will be available to users - a new {\sc python} package for processing GRB LCs and colour evolution and a {\sc Julia}-based web scraper for the data gathering.

The paper is organized as follows: In Section \ref{sec:sample}, we describe the selection process for obtaining the LCs. In Section \ref{methods}, we discuss our methodology for converting the data to a uniform format and determining the optical spectral index together with the colour behaviour for each GRB.
In Section \ref{sec:colorevolution}, we discuss our spectral analysis, our colour evolution analysis, and the division of GRBs into groups depending on the presence or lack of colour evolution based on our analysis. We also compare our results to those presented in the literature. In Section \ref{sec:website}, we outline the functionality of the new web-based optical LC repository. This web-based repository is created with a newly developed, open-access {\sc python} package designed for displaying the LCs, collecting information from NASA ADS sources (\url{https://ui.adsabs.harvard.edu/}) and GCNs (\url{https://gcn.gsfc.nasa.gov/}), converting the magnitudes into AB system, and correcting them for Galactic extinction, $k$-correction effect, and investigating the color evolution following the methodology of Section \ref{sec:colorevolution}. We present our conclusions in Section \ref{sec:conclusions}.
In Appendix \ref{sec:instruments}, we describe the instruments from which data were detected. In Appendix \ref{sec:kcorrectcases}, we discuss 5 peculiar GRBs for which a different spectral analysis is required. In Appendix \ref{sec:ZaninoniKann}, we discuss the spectral fitting approaches of \citet{kann2006signatures} and \citet{2013PhDT.......144Z}. In Appendix \ref{sec:disagreementexamples}, we discuss some examples of cases of disagreement between our analysis and the literature.

\vspace{-0.8cm}
\section{Data Sample}\label{sec:sample}
We compile a sample of 535 optical LCs from those published in the literature between February 28, 1997, and August 18, 2023, taking a part of the GRBs from \citet{2022ApJS..261...25D}.
We have included all the GRBs with measured and confirmed redshifts. The redshift is either photometric or spectroscopic or from the host galaxy identification and it is taken either from (\url{https://www.mpe.mpg.de/~jcg/grbgen.html}) or GRBOX (\url{https://sites.astro.caltech.edu/grbox/grbox.php}) with a confirmed optical counterpart according to Greiner's table. Generally, we do not report optically dark GRBs, although 28 cases considered dark bursts in the literature have been included in this catalogue.
This sample includes data privately shared by co-authors or by private communication. In particular, our sample includes 76 GRBs with data points from RATIR, 43 from \cite{Oates2009,Oates2012}, 25 from BOOTES (Alberto Castro-Tirado, 2023, private comm.), 34 from \citet[][2022, private comm.]{Li2012,Li2015,Li2018a}, 19 from \cite{Zaninoni2013}, 16 from Michael Hiram Siegel (2023, private comm.), 16 from \cite{Si2018} and references within, 15 from Alexei Pozanenko (2023, private comm.), 12 from the Kiso telescope (one of which elaborated by us), 11 from \cite{2009ApJ...693.1484C}, six from VIRT (Brice Orange, 2023, private comm.), and one from \cite{rossi2020comparison}. The rest of the information is taken from the literature such as GCNs, The Astronomer's Telegram (ATel), International Astronomical Union Circulars (IAUC), arXiv, and papers.
The repository contains 64813 data points in total. Of these, $54300$ are observed magnitudes, and $10513$ are limiting magnitudes. 
We formatted the collected data as shown in Table~\ref{tab:magfile} and stored them in space-separated \textit{.txt} files. For each GRB, we collect the following information: (1) midtime of the observation (in seconds) after the satellite trigger; (2) magnitude; (3) magnitude error (1 $\sigma$); (4) filter; (5) magnitude system; (6) observatory, telescope, and instrument; (7) a flag for Galactic extinction correction, where we assign 'y' if this correction has been applied in the literature, and 'n' when not; (8) source of the data (GCN, paper, arXiv, ATel, IAUC, or private communication); (9) a flag marking the points that are considered outliers (the discussion will be presented in Section \ref{sec:outliers3sigma}), (10) the photometric/spectroscopic/host galaxy redshift, and (11) the GRB classification. While the information from (1) to (9) are stored only in the single GRB files, the redshift (10) and the GRB classification (11) are collected also in Table 1 of the Online Materials. 
All the magnitudes presented in this paper are reported in the AB system and corrected for the Galactic extinction. Magnitudes without errors and limiting magnitudes have been reported with zero magnitude error.

\begin{table*}
    \centering
    \begin{tabular}{ccccccccc}
        \hline
        Time(s) & Mag & MagErr & Filter & System & Observatory/Telescope/Instrument & GalExtCorr & Source & Flag \\
        \hline
        19353.6	&	20.76	&	0.30	&	$U$	&	AB	&	CAHA/2.2m/CAFOS	&	y	&	\citet{1997IAUC.6657....1C}	& no \\
        19612	&	21.11	&	0.80	&	$I_C$	&	AB	&	Hubble/HST/STIS	&	y	&	\citet{1997ApJ...489L.127S} 	& no	\\
        20390	&	21.33	&	0.21	&	$V$	&	AB	&	Hubble/HST/STIS	&	y	&	\citet{1997ApJ...489L.127S}		& no \\
        20822	&	20.55	&	0.21	&	$I_C$	&	AB	&	Hubble/HST/STIS	&	y	&	\citet{1997ApJ...489L.127S} 	& no	\\
        21704.4	&	20.31	&	0.10	&	$I$	&	AB	&	KO/Keck-1/NIRC	&	y	&	\citet{1997IAUC.6666....1M} 	& no	\\
        \hline
    \end{tabular}
    \caption{In this table, we report an excerpt of GRB 970508A from the collected catalogue of 535 GRBs. The columns contain (1) midtime of the observation (in seconds) after the satellite trigger; (2) magnitude; (3) magnitude error (1 $\sigma$); (4) filter; (5) magnitude system; (6) observatory, telescope, and instrument; (7) flag for Galactic extinction correction, if already applied (y) or not applied yet (n); (8) source of the data; (9) flag for outlier (if the point is an outlier, \textit{no} otherwise).}
    \label{tab:magfile}
\end{table*}

The Greiner and GRBOX web pages have been used to search for the references, the respective literature, and the GCNs reported in the Table 1 of the Online Materials. There, along with the redshift, we collect and present the following information in Table 4: the coordinates (right ascension = RA, declination = Dec), from the XRT and UVOT; the optical spectral index, $\beta_\mathrm{opt}$ from the literature; the host extinction from the literature, including the magnitude extinction in the $V$-band, and the best fit dust model for the host galaxy. In addition, we record the GRB classes. The convention adopted for the spectral index in this catalogue is given by the following flux expression: $F(\nu,t)\sim t^{-\alpha} \nu^{-\beta_\mathrm{opt}}$, $\alpha$ being the temporal decay index \citep{Sari1998,Jakobsson2004}.

\vspace{-0.5cm}
\section{Methodology}\label{methods}

\subsection{Sample homogenization}\label{sample homogeneization}
The data in magnitudes are collected from GCNs, literature (refereed papers, arXiv, ATel, IAUC), and private communications. As a first step to homogenize the sample, the magnitudes are converted to the AB system and corrected for the Galactic extinction, as follows:

\vspace{-0.5cm}
\begin{equation}
    \text{mag}_{AB,\text{gal}} = \text{mag}-A_{\lambda,\text{gal}}+\mathrm{shift}_{\text{Vega}\rightarrow \text{AB}},
    \label{eq:magABextcorr}
\end{equation}

where, $\mathrm{mag}$ is the apparent magnitude value collected from the literature (in the observer frame), $A_\mathrm{\lambda,gal}=R_\lambda*E_\mathrm{B-V}$ is the Galactic extinction, $E_\mathrm{B-V}$ is the colour excess, and $R_\lambda$ is the extinction coefficient related to a given $\lambda$ band. To correct the magnitudes for the Galactic extinction, the dust maps of \cite{Schlafly2011} or the Asiago Database on Photometric Systems (ADPS, \citealt{AsiagoADPS}) are used.
The $\mathrm{shift}_{\text{Vega}\rightarrow \text{AB}}$ is the shift from Vega to AB system; it is given for a generic filter in \cite{Blanton2007}. 

\vspace{-0.5cm}
\subsection{Spectral data analysis, $k$-correction, and host extinction}\label{sec:spectral}
After converting the magnitudes to AB system and correcting for the Galactic extinction, the next step is to compute the optical spectral index ($\beta_\mathrm{opt}$) and the host galaxy extinction in the $V$-band ($A_{V}$) together with the dust model. 

To compute $\beta_\mathrm{opt}$ and $A_V$, we need at least four magnitudes, all in different filters, observed at coincident epochs (times). Given two filters, $f$ and $g$ observed at different epochs, the observations are considered to be coincident if the difference in time of the two observations is less than $0.025$ of the first epoch, as expressed in Equation \ref{eq:coincidenttimes}. This percentage has been estimated by taking the peak of the distribution of the magnitude errors. The typical GRB magnitudes are reported with errors of $\pm 0.04-0.05$ mag, or equivalently $\sim 4-5 \%$. This suggests that the simultaneity requirement can be relaxed up to half of the magnitude error ($\sim 0.025$) without significantly impacting the colour evolution. This criterion reads as follows:

\vspace{-0.4cm}
\begin{equation}
    \frac{|t_{f}-t_{g}|}{t_{f}}\leq 0.025\ \rightarrow t_{f}\,\text{and}\,t_{g}\,\text{are}\,\text{coincident},
    \label{eq:coincidenttimes}
\end{equation}

where $t_{f}$ and $t_{g}$ are the midtimes of the observations in seconds (after the satellite trigger) in the $f$ and $g$ bands, respectively. To estimate $\beta_\mathrm{opt}$, the spectral energy distribution (SED) of each GRB is fitted with a simple power law model according to the formulation $F_{\nu} = F_{0} \nu^{-\beta_\mathrm{opt}}$ shown in \citet{Sari1998,Jakobsson2004}, where $F_{0}$ is the normalization of the spectrum and $\nu$ is the frequency. We remark that, in the current analysis, the host galaxy extinction is included following the approach reported in \citet{2013PhDT.......144Z}. 
The formulation of \citet{kann2006signatures} can be proven to be equivalent to the one in \citet{2013PhDT.......144Z}. This calculation is reported in Appendix \ref{sec:ZaninoniKann}. 
The SED model $F_{\nu}$ can be rewritten in terms of the wavelength $\lambda (\propto 1/\nu)$:

\vspace{-0.5cm}
\begin{multline}
    \mathrm{mag}_\mathrm{AB,gal}(\lambda) = -2.5\beta_\mathrm{opt}\log_{10}(\lambda) \\ -2.5 A_V \biggr(\biggr[\frac{A_\lambda}{A_{V}}\biggr]_\mathrm{obs} - \biggr[\frac{A_\lambda}{A_{V}}\biggr]_{1+z}\biggr) + \mathrm{mag}_{0},
    \label{eq:magbeta}
\end{multline}

where $\mathrm{mag}_{0}$ is the magnitude value correspondent to $F_{0}$, $A_V$ is the host galaxy extinction in $V$-band, $A_\lambda/A_V$ is taken from the extinction map reported in \citet{Pei1992} for the three dust models: Milky Way (MW), Large Magellanic Cloud (LMC), and Small Magellanic Cloud (SMC), with $R_V = 3.08, 3.16, 2.93$ respectively, being $R_V=\frac{A_V}{E_{B-V}}$ a fixed number that corresponds to the ratio between the total extinction and the selective extinction in the $V$-band. 
The subscript $obs$ refers to the wavelength $\lambda$ in the observer frame. The subscript $1+z$ refers to the wavelength $\lambda$ in the host galaxy frame.

Using the Python package \textit{lmfit}, the SEDs are fitted through the weighted Levemberg-Marquardt method, where the inverse square of the magnitude errors are taken as weights. 
We often encounter multiple spectral indices from the different epochs of a particular GRB, but not all provide a reliable fitting. Thus, we exclude the cases where $|\sigma_{\beta_\mathrm{opt}}|>|\beta_\mathrm{opt}|$. The prior values for $A_V$ are constrained in the interval $(0,10)$, thus the SED fitting does not explore the parameter space region for which $A_V<0$, which is indeed non physical. We do not remove any outliers from the SED fitting. However, we have flagged 601 outliers across 45 GRBs with the notation \textit{SED} in the 9th flag column to indicate that the corresponding data points are outliers for the spectral fitting. Some GRBs are very well sampled in the optical wavelengths. Thus, the $0.025$ simultaneity criterion of Equation \ref{eq:coincidenttimes} allows multiple $\beta_\mathrm{opt}$ values at very close epochs.

Then, we choose the fitting with a probability value of $p > 0.05$, of this fit being computed not by chance, considering as the null hypothesis that the fit is not drawn by chance, which is accepted for $p > 0.05$. To estimate this probability, we integrate the inverse Gamma distribution from $\mu$ to $\infty$, where $\mu$ is the product between the reduced $\chi^2$ value and the number of data points involved in the fitting, see Equation \ref{eq:probability}.

\vspace{-0.4cm}
\begin{equation}
    P(X>\mu)=\frac{2^{-\mu/2}}{\Gamma(\mu/2)} \int^{+\infty}_{\mu}e^{-x/2}x^{-1+(\mu / 2)}dx
    \label{eq:probability}
\end{equation}

After estimating the $\beta_\mathrm{opt}>0$ values, we also explore the possible $\beta_\mathrm{opt}<0$ cases. The $\beta_\mathrm{opt}<0$ condition is rarely expected if compared with the rest of the cases: indeed, we find such a condition only for 298 spectral indices across 68 GRBs. One remarkable example is GRB 090510A for which we find $\beta_\mathrm{opt}=-0.853 \pm 0.535$, compatibly with the negative spectral index observed in $\gamma$-rays \citep{2010ApJ...716.1178A}. The selected $\beta_\mathrm{opt}$ values are reported in the Table 2 of the Online Materials. The $k$-correction and the host correction are applied to the AB magnitude corrected for Galactic extinction as below:

\vspace{-0.4cm}
\begin{equation}
    \mathrm{mag}_\mathrm{corr}(t) = \mathrm{mag}_{AB,gal}-k(t)-A_\lambda,
    \label{eq:maghostkcorr}
\end{equation}

where $k(t)=-2.5(\beta_\mathrm{opt}(t)-1)\log_{10}(1+z)$ is the $k$-correction, according to the formulation in \citet{Oke1968,Peterson1997,Greiner2015,Li2018a}. This treatment considers that the spectrum could show signs of evolution and the $\beta_\mathrm{opt}$ may depend on the epoch. To highlight which value of $\beta_\mathrm{opt}$ must be taken at each time, we have considered the following procedure. We first fit the spectral indices in our $\delta_t/t=0.025$ condition with the three extinction models (SMC, LMC, and MW). As dust model and $A_V$ value, we consider the ones given by the best fit identified through the maximum probability $P$. Then, concerning the spectral indices, we check which values among all the obtained $\beta_\mathrm{opt}$ agree with each other within 2 $\sigma$ with the iterative procedure discussed below. We first order the $\beta^{i}_\mathrm{opt}$ values of each $i$-th interval of a given GRB in the epoch $T$ and then we compare each spectral index with all the previous ones with a 2 $\sigma$ agreement criterion. If there is compatibility, the intervals are merged for that epoch and the procedure continues. If not, then a new epoch $T'$ is defined, and the procedure continues with that new epoch. All the successive epochs for which the $\beta_\mathrm{opt}$ value agrees within 2 $\sigma$ are joined in groups to form one or multiple larger epochs. For each enlarged epoch, the $\beta_\mathrm{opt}$ are then averaged to obtain only one value of the spectral index to be applied in the $k$-correction in the enlarged epoch using the prescriptions of the Central Limit Theorem (CLT), without refitting the SED. Instead, we have used several constraints from the $\delta_t/t$ condition, and we have used several measures of the spectral indices at several different epochs, and we have constructed a smaller error bar based on the CLT.

To double-check that our procedure is reliable, we have considered several GRB cases randomly selected in which the literature agrees with our results in terms of the absence or presence of the colour evolution. We have checked with these GRBs that while performing compatibility of the spectral index in 1 $\sigma$ leads to some discrepancy among our results and the literature, with the 2 $\sigma$ we almost have full compatibility.
The GRB cases on which we have tested our method are 071010A, 071025A, and 090313A for no colour evolution, and 060218A, 071031A, 100418A, and 130702A for colour evolution.  
For 5 GRBs (the 3$\%$ of the total 143), the 2 $\sigma$ compatibility approach described above does not allow a reliable $k$-correction process and these cases have been discussed as separate cases in the Appendix \ref{sec:kcorrectcases} and have not been included in our analysis. These GRBs are: 021004A, 110205A, 161219B, 171010A, and 171205A. For this reason, the number of GRBs with a reliable optical SED analysis reduces from 143 to 138.
The error on the $k$-correction is given by $\Delta k(t)=|-2.5\log_{10}(1+z) \times \sigma_{\beta_\mathrm{opt}(t)}|$. The $k$-correction is applied to all the GRBs that have the $\beta_\mathrm{opt} \pm \sigma \beta_\mathrm{opt}$ for which we have at least 4 data points available in different filters. For some GRBs, we only have limiting magnitudes in our catalogue, since the authors' data in the published literature are private, and thus for those, we are not able to compute $\beta$ and perform the $k$-correction. We calculate the error on the final magnitude value, $\mathrm{mag}_\mathrm{corr}$, by using the error on the $k$-correction and propagating them with their uncertainties from Equation \ref{eq:magABextcorr} and the error given by the host galaxy extinction, namely, $\Delta A_\lambda = \bigr|\frac{A_\lambda}{A_{V}}\bigr| \times \sigma_{A_V}$, $\sigma_{A_V}$ being the 1 $\sigma$ uncertainty on $A_V$. 

\vspace{-0.5cm}
\subsubsection{The optical and X-ray spectral indices comparison}\label{sec:spectraloptbeta}
\begin{figure}
    \centering
    \includegraphics[scale=0.28]{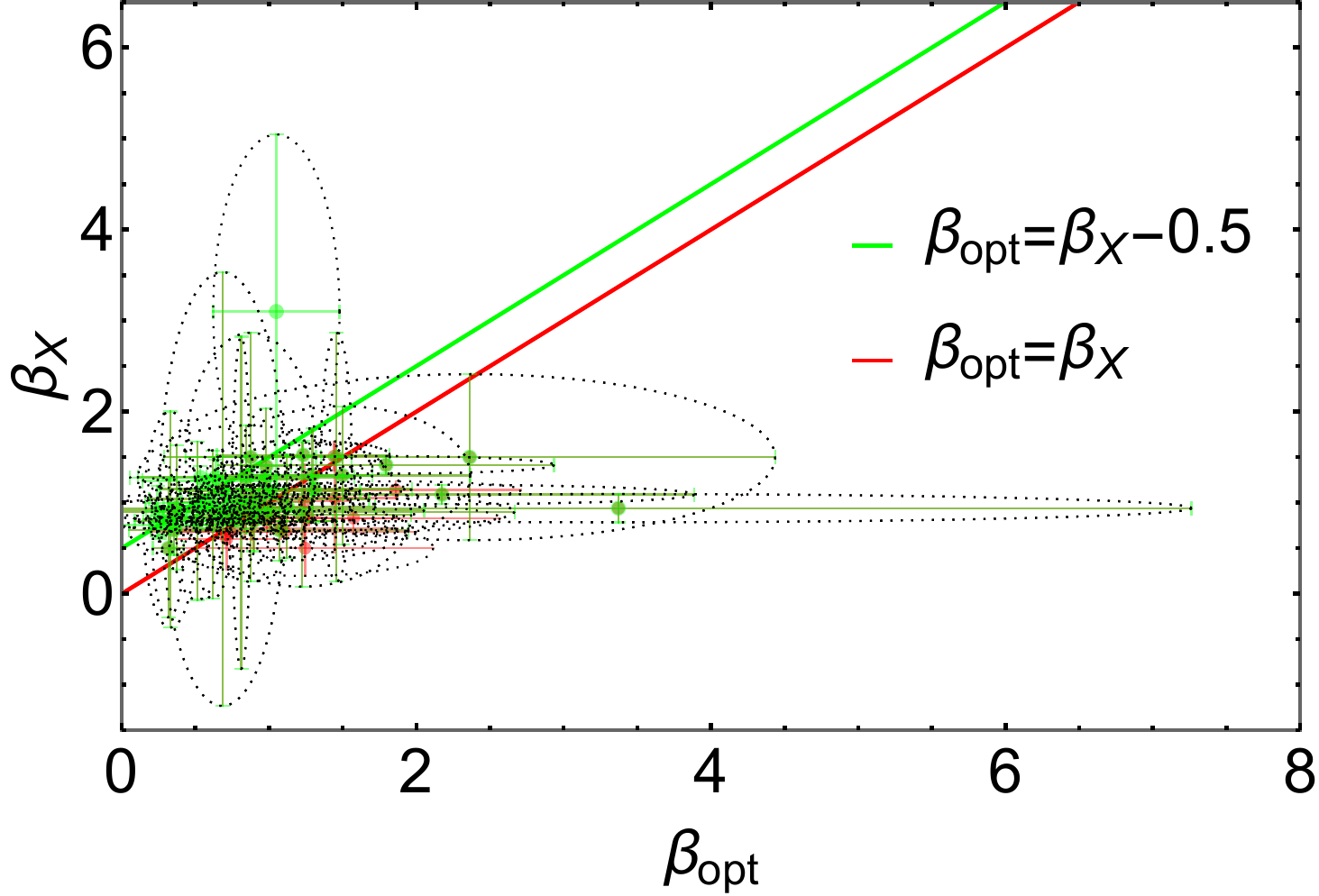}
    \includegraphics[scale=0.28]{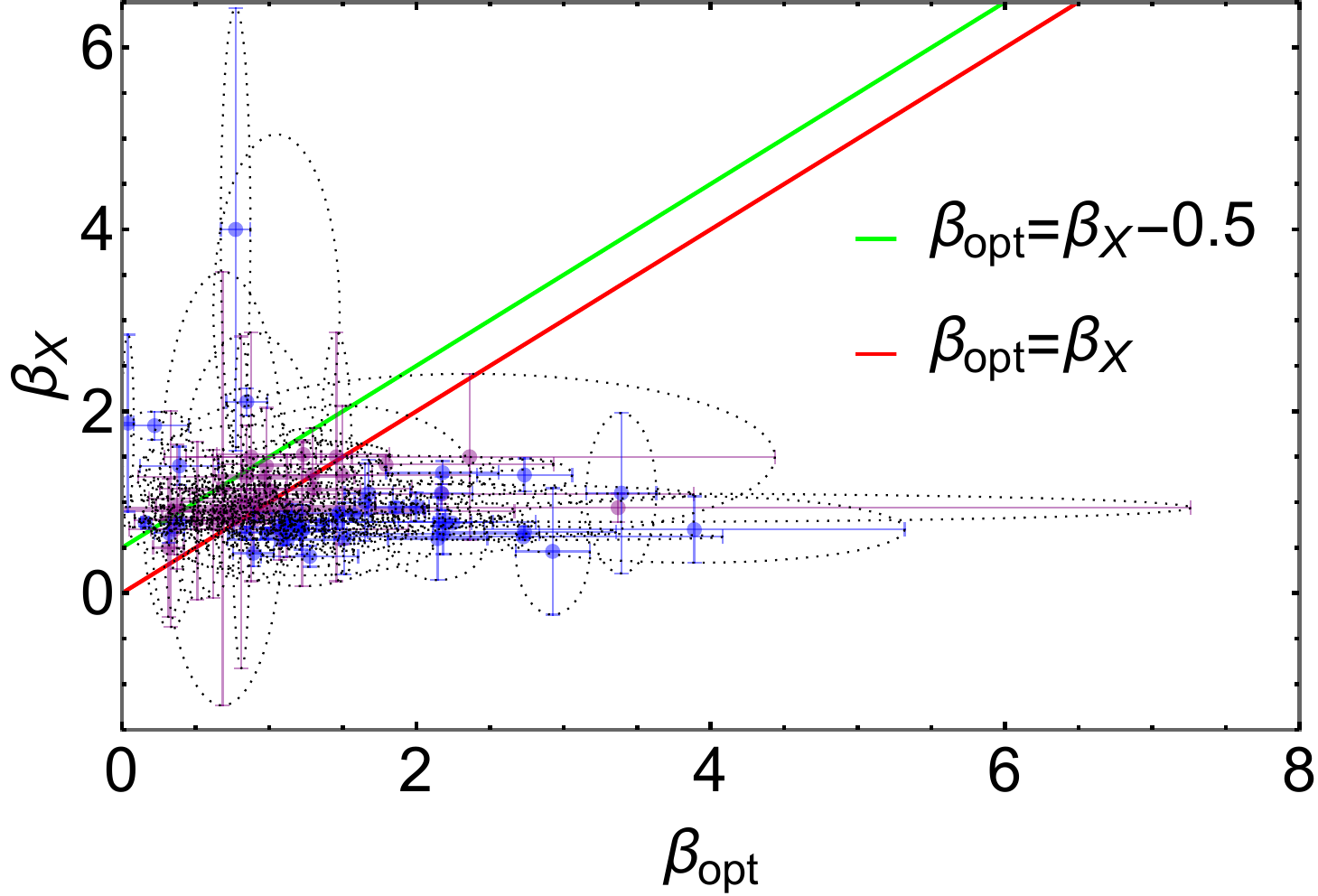}
    \caption{The comparison between the optical spectral indices (in the x-axis) and the X-ray ones (in the y-axis). \textbf{Upper panel.} The green data points are the $\beta_\mathrm{opt}$s that follow the $\beta_\mathrm{opt}=\beta_{X}-0.5$ relation (48 cases), while the red ones respect the $\beta_\mathrm{opt}=\beta_{X}$ relation (56). The red line is the relation $\beta_\mathrm{opt}=\beta_{X}$, while the green line represents $\beta_\mathrm{opt}=\beta_{X}-0.5$. \textbf{Lower panel.} The purple data are the GRBs that respect both the relations $\beta_\mathrm{opt}=\beta_{X}-0.5$ and $\beta_\mathrm{opt}=\beta_{X}$ (34), while the blue ones do not respect any of the two (43). The color-coding is the same in both panels.}
    \label{fig:betaoptbetaX}
\end{figure}

When performing the comparison between optical and the XRT spectral indices ($\beta_{X}$), we apply the same procedure described above to obtain the epochs in which the $\beta_{X}$ values are computed. Indeed, the larger epoch with the 2 $\sigma$ condition will allow the XRT spectral index epoch to be computed. The comparison of the $\beta_{X}$ values extracted from the Swift-XRT Spectrum repository with our $\beta_\mathrm{opt}$ is reported in Figure \ref{fig:betaoptbetaX}: there are 56 pairs compatible in 1 $\sigma$ with the $\beta_{X}=\beta_\mathrm{opt}$ relation, 48 pairs of $(\beta_\mathrm{opt},\beta_{X})$ that are compatible in 1 $\sigma$ with the $\beta_{X}=\beta_\mathrm{opt}-0.5$ relation, 34 that are compatible with both in 1 $\sigma$, 79 are the cases in which either of the two, but not both, are incompatible with this scenario. Here, we discuss the 43 cases where both the relations are not compatible with the optical and X-ray spectral indices. For these 43 cases, we estimate the difference $\Delta \beta_{opt,X}=|\beta_{opt}-\beta_{X}|$: the values of $\Delta \beta_{opt,X}$ are normally distributed with a mean value $\mu_{\Delta \beta_{opt,X}}=1.26$ and a standard deviation $\sigma_{\mu_{\Delta \beta_{opt,X}}}=0.78$. The z-scores, instead, computed as $Z_{score}=|\beta_{opt}-\beta_{X}|/\sqrt{\sigma^{2}_{\beta_{opt}}+\sigma^{2}_{\beta_{X}}}$, follow a uniform distribution with mean $\mu_{Z_{score}}=1.60$ and standard deviation $\sigma_{Z_{score}}=0.93$. Thus, overall they differ in less than 2 $\sigma$. The choice of the 2 $\sigma$ ensures a good compromise on the final selected epoch: it avoids spurious and artificial evolution in the spectral index due to the tiny epoch. On the other hand, this epoch will still be short enough to appreciate if there is a small undergoing evolution.

\vspace{-0.5cm}
\subsection{Description of the rescaling factor}
\label{sec:scaling}
To study how the colour of GRBs changes over time in our dataset, we consider the measurements taken with different filters rather than considering the results from the spectral index evolution. Such a choice is justified by the fact that the colour evolution in the optical filters could be observed even with only two filters with at least 3 data points at a coincident epoch, while our SED analysis requires at least four different bands at coincident epochs, given that we need the slope, the intercept, and the $A_V$. Nevertheless, in the discussion of the colour evolution results we have provided a comparison between the SED evolution and the colour evolution results in our analysis. If a GRB does not display colour evolution, magnitudes in different filters can be rescaled to the most numerous filter.
Performing this rescaling allows us to form the most complete picture of the GRB's behaviour, as some bands may contain points at critical times (i.e. at very early or late times, or even during the appearance of interesting features such as the plateau or a flare) that are not present in other bands. These points may be crucial to subsequent GRB studies, especially involving LC morphology - therefore, when possible, we aim to perform this rescaling. However, when a GRB does display colour evolution, rescaling colours might obscure unique characteristics of the surrounding environment that drive this colour change. The origins of such features include the common causes of late-time chromaticities, like in the case of dust destruction that leads to a decrease in the optical-NIR extinction \citep{2020ApJ...895...16H}, or even an associated Supernova (SN) that, when present at the end of the LC, reddens the colour of the GRB afterglow \citep{Cano2017}. Other late-time colour evolution cases are caused solely by the fading of the GRB afterglow with the subsequent emerging contribution of the host galaxy to the luminosity of the transient.
Conversely, early-time colour variations typically result from specific phenomena, such as the cooling break energy frequency crossing the observed band or the shift from a reverse-shock regime to a forward-shock regime \citep{Li2018a}. Host galaxy extinction presents another potential reason for colour changes: this occurs when dust reappears early on, following its initial destruction by intense GRB radiation, causing a shift to redder colours. However, the time scale for such a process to be set is $\tau \geq 6*10^{8}/n\,yr$, where $n$ is the density of the dust region \citep{2003ApJ...585..775P}. This is a process much slower than the time scale for the GRB afterglows to become no longer visible at optical wavelengths. To determine whether GRBs exhibit colour evolution, we start by calculating rescaling factors for each data point and then assess the behaviour of each filter. After determining the colour behaviour for each GRB, we perform the rescaling for the GRBs that do not undergo colour evolution.
We developed a customized {\sc python} script to calculate these rescaling factors, which will be included in the package.Here, we stress that the colour evolution analysis is propaedeutic and necessary for rescaling. The colour evolution analysis is described below.
Then, the rescaling factors are calculated as follows: given a GRB with magnitudes in the filters $f$ and $g$, the \textit{rescaling factor} is defined as the difference between the magnitude of the $f$ filter and $g$ filter at coincident epochs.  
Thus, the rescaling factor definition is:

\vspace{-0.4cm}
\begin{equation}
    rf_{f,g}=(\mathrm{mag}_{f}-\mathrm{mag}_{g})_{coincident},
    \label{eq:rescalingfactor}
\end{equation}

where $t_{f}$ and $t_{g}$ follow the condition expressed in Equation \ref{eq:coincidenttimes}.

\begin{figure*}
    \centering
    \includegraphics[width=0.47\textwidth]{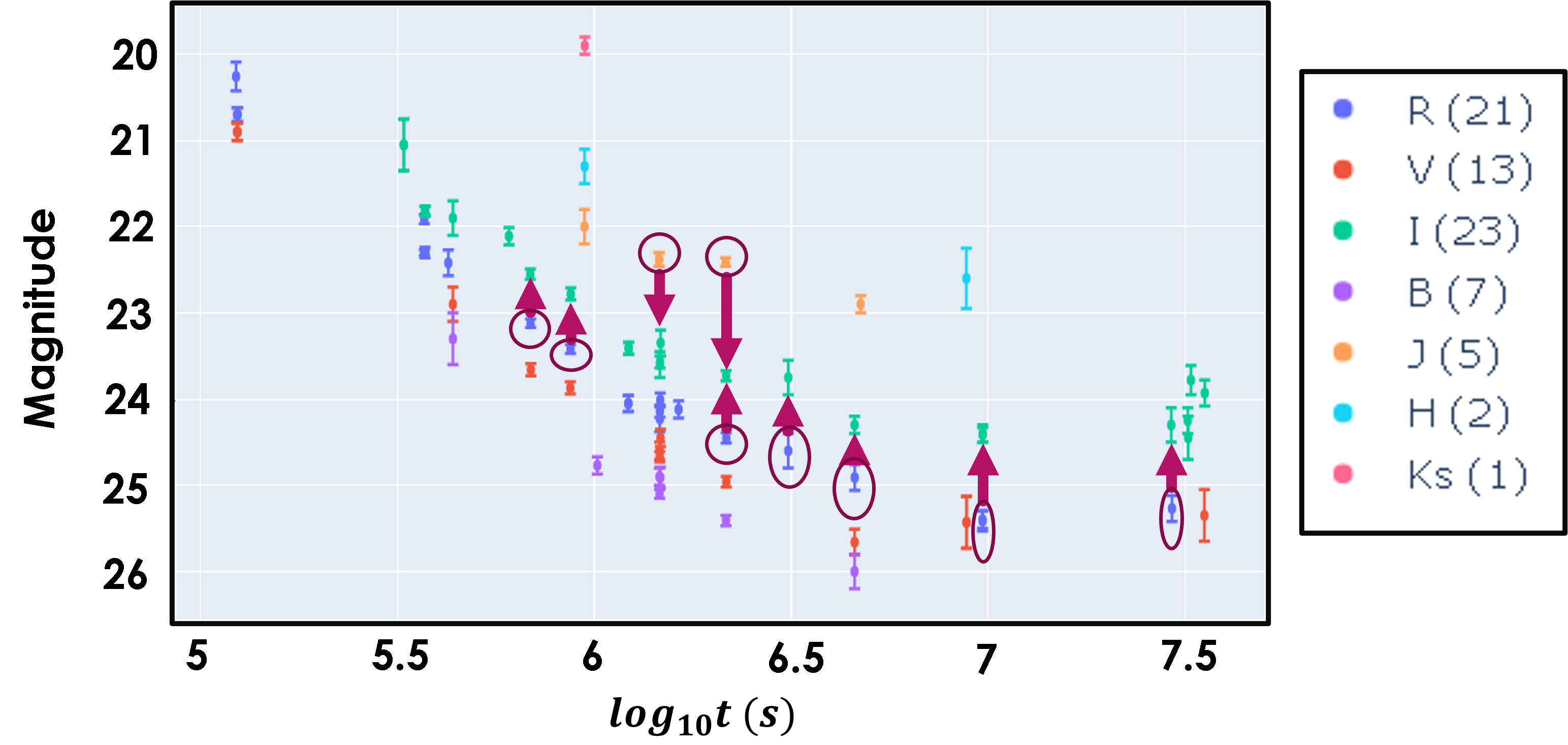}
    \includegraphics[width=0.47\textwidth]{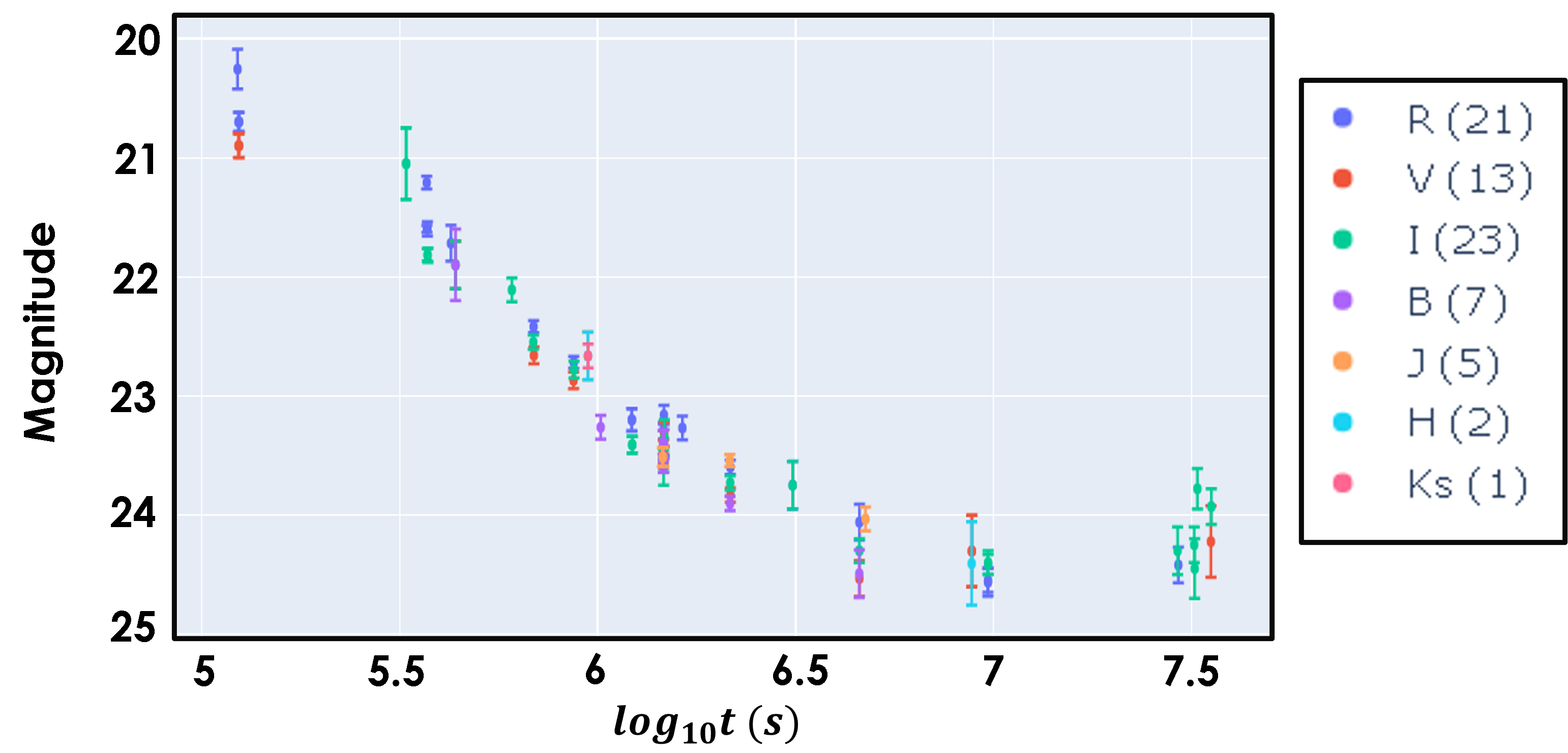}
    \includegraphics[width=0.47\textwidth]{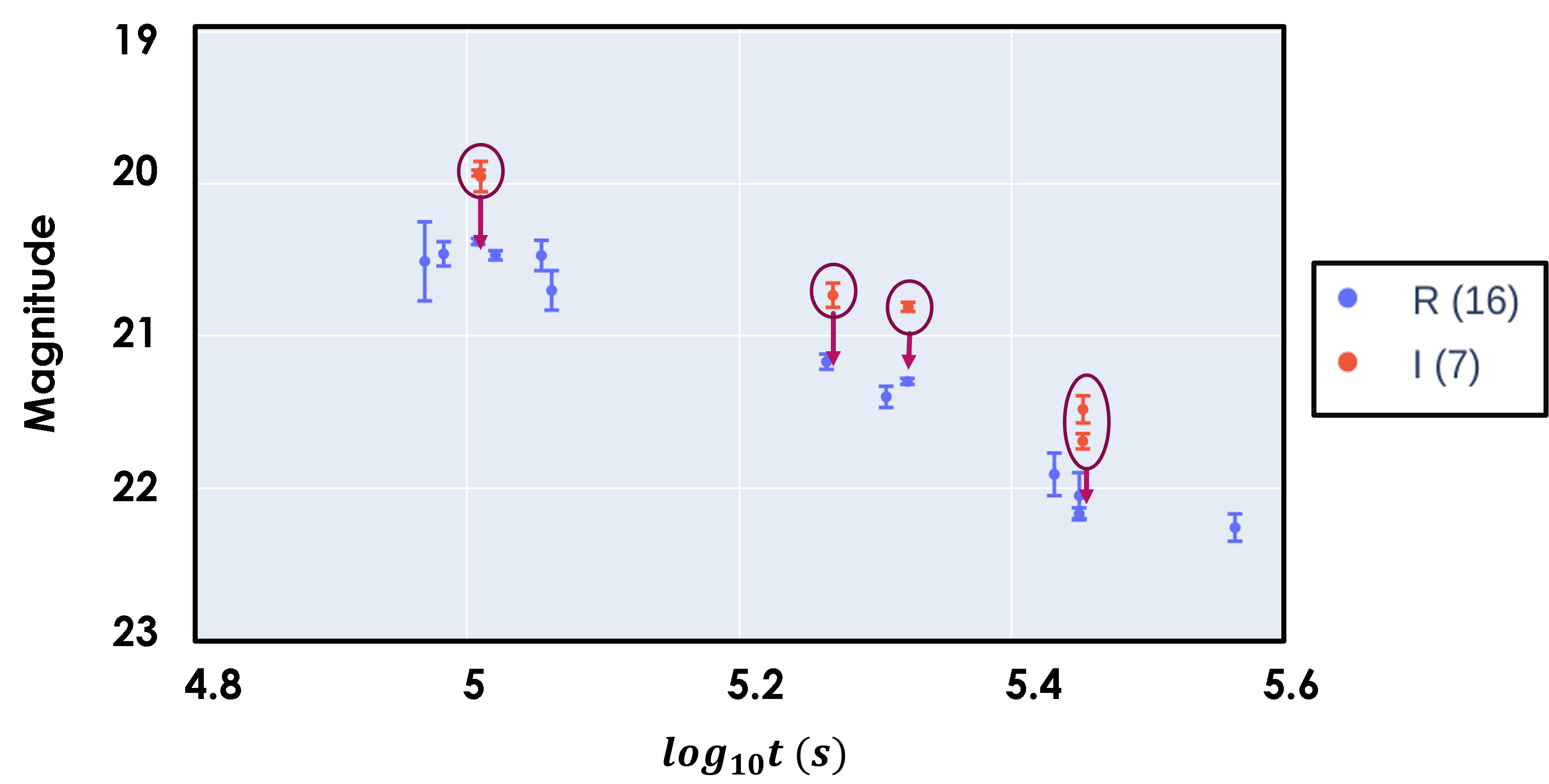}
    \includegraphics[width=0.47\textwidth]{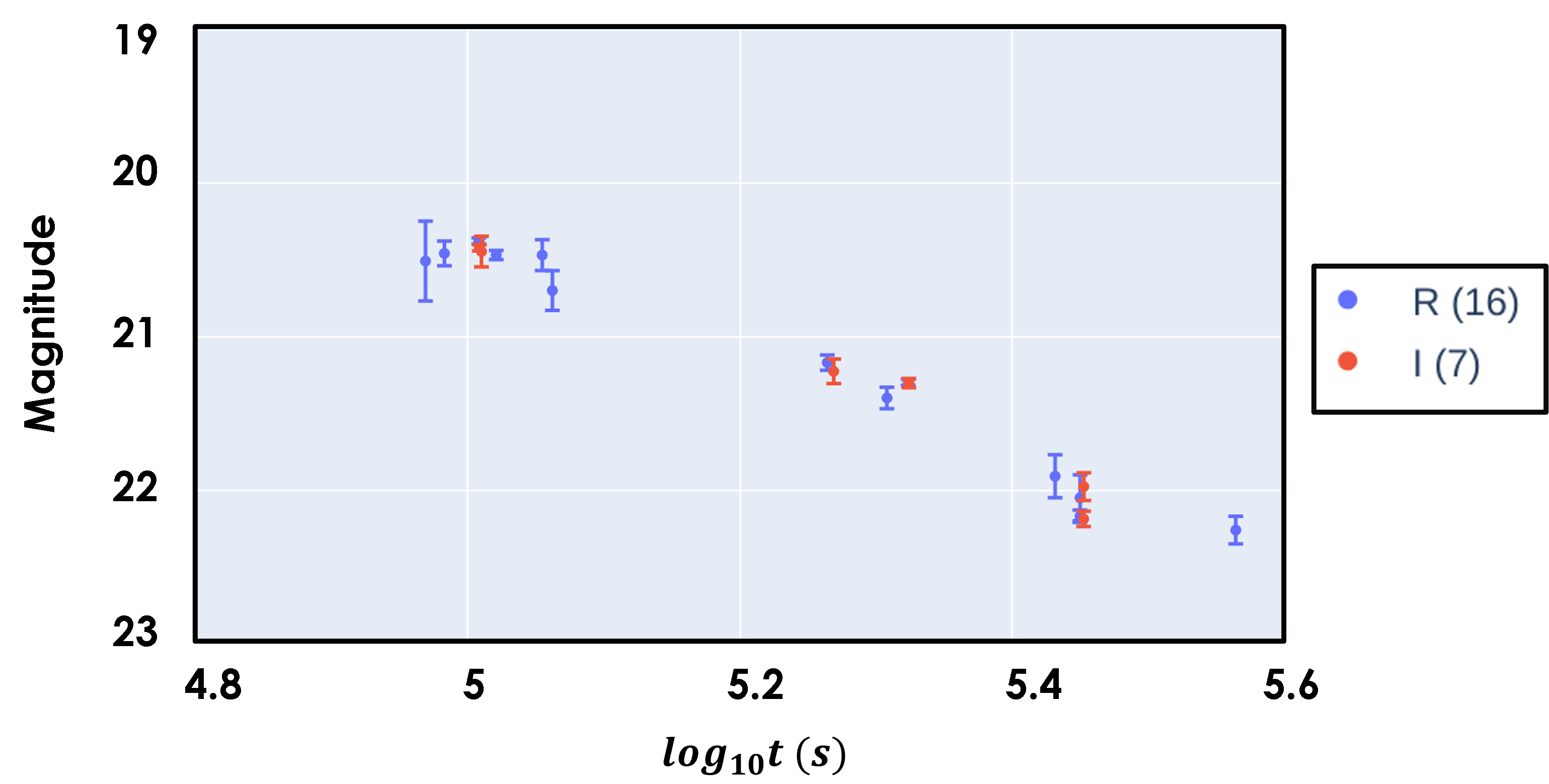}
    \caption{An example of GRB rescaling when times are duplicated, namely magnitudes co-occurring in different filters. GRB 000911A is shown before (upper left panel) and after (upper right panel) rescaling. Lower panels show a case for magnitudes that do not have time duplicates, but obey the condition in Equation \ref{eq:coincidenttimes}. In the example, GRB 030323A is shown before (lower left panel) and after (lower right panel) rescaling.}
    \label{fig:rescaleduplicates}
\end{figure*}

The rescaling factors of the $f$ filter to the most numerous inside a GRB LC (we use the $mn$ label to mark the most numerous filter) are fitted through the \textit{lmfit} package that uses the weighted Levenberg–Marquardt algorithm \citep{Levenberg1944AMF} with a linear model in the form:

\vspace{-0.4cm}
\begin{equation}
rf_{mn,f}=a*\log_{10}(t_{f})+b,
\label{eq:linmodel}
\end{equation}

where $a$ is the slope and $b$ is the normalization. Figure \ref{fig:rescaleduplicates} shows an example of magnitude rescaling in both the cases of duplicate and coincident epochs. GRB 000911A is shown before (upper left panel) and after (upper right panel) rescaling. The $I$ band is the most numerous, with 23 occurrences. Lower panels show a case (GRB 030323A) for magnitudes that are not time duplicates, but obey the criteria of $0.025$ of Equation \ref{eq:coincidenttimes}. Here, the $R$ band is the most numerous, with 16 occurrences.
The error on the rescaling factor is defined as $\Delta rf_{mn,f}=\sqrt{\Delta \mathrm{mag}_{mn}^2 + \Delta \mathrm{mag}_{f}^2}$.
At least three data points for the rescaling factors among the most numerous filter and the other filters are needed for the fitting. 
The colour evolution analysis is performed with two approaches: the $a=0$ \textit{fitting} and the \textit{variable slope} fitting. In the {$a=0$ fitting}, we fix the slope $a$ in Equation \ref{eq:linmodel} to zero and we fit the normalization $b$, estimating then the probability $P$ according to Equation \ref{eq:probability}, {the reduced $\chi^2$, and the Bayesian Information Criterion (BIC) value.} In this approach, if for the fitting we have $P \geq 0.05$ (namely, the probability for the fitting to not be drawn by chance is greater than $0.05$), then we consider the fitting as a no colour evolution case, otherwise for $P < 0.05$, we consider the fitting as a colour evolution case. In the variable slope approach, the slope $a$ of Equation \ref{eq:rescalingfactor} is fitted together with the normalization $b$ and two fitting parameters are free to vary. Also in this case we collect the values of $P$, the reduced $\chi^2$, and the BIC.
The definition adopted in this approach for the absence of colour evolution is that $a$ should be compatible with zero to within 3 $\sigma$. If this condition is not satisfied, the filter has colour evolution. We consider as undetermined cases (neither colour nor no colour evolution) the ones where the absolute value of the uncertainty of the slope $\sigma_a$ is larger than the slope $a$ itself.
Concerning the rescaling procedure, when the magnitudes in different filters at a given time are within 1 $\sigma$ of each other, the rescaling factor is not needed for those magnitudes. Even if no rescaling is needed in certain filters, some individual magnitude measurements in those filters might stand out as unusually different. If these outliers are not caused by flares or bumps, then they might be due to problems like observation difficulties or calibration errors.
We have checked the agreement of both the $a=0$ and the variable slope methods with already published results in the literature. The cases of disagreement with the literature are presented in Section \ref{sec:colorevolution}. Furthermore, in the cases where most of the filters do not agree with the literature, but at least two filters agree, these are considered as "apparent" disagreement.
In the estimation of the rescaling factors, we apply the following approximations: $u'\approx u$, $g'\approx g$, $r'\approx r$, $i'\approx i$, $z'\approx z$, $K'\approx K$, and $K_{S} \approx K$.
These approximations guarantee that we can have better coverage of the LCs for the colour evolution investigation.
The same assumption does not hold between the Johnson and Cousins filters, since approximating $I\approx I_{C}$ and $R\approx R_{C}$ would induce systematic effects in the estimation of the colour evolution.\\
The main difference between our analysis and prior approaches for estimating the colour behaviour is our use of the most numerous filter as a reference for all other filters, while in the literature, they consider a pre-determined set of colour indices. Indeed, in the literature, it is very common to find the magnitude differences among the following pairs of filters: $B-V$, $U-B$, $V-R$, $V-R$, $R-I$, $UVW2-UVM2$, $UVM2-UVW1$, $UVW1-U$, $u-g$, $g-r$, $r-i$, and $i-z$ \citep{Simon2001,Li2018a}. Nevertheless, with the script in our package, it is possible to change the reference filter for rescaling without necessarily taking the most numerous filter, allowing users to perform the colour evolution analysis using a different filter.

\begin{figure}
   \centering
   \includegraphics[scale=0.25]{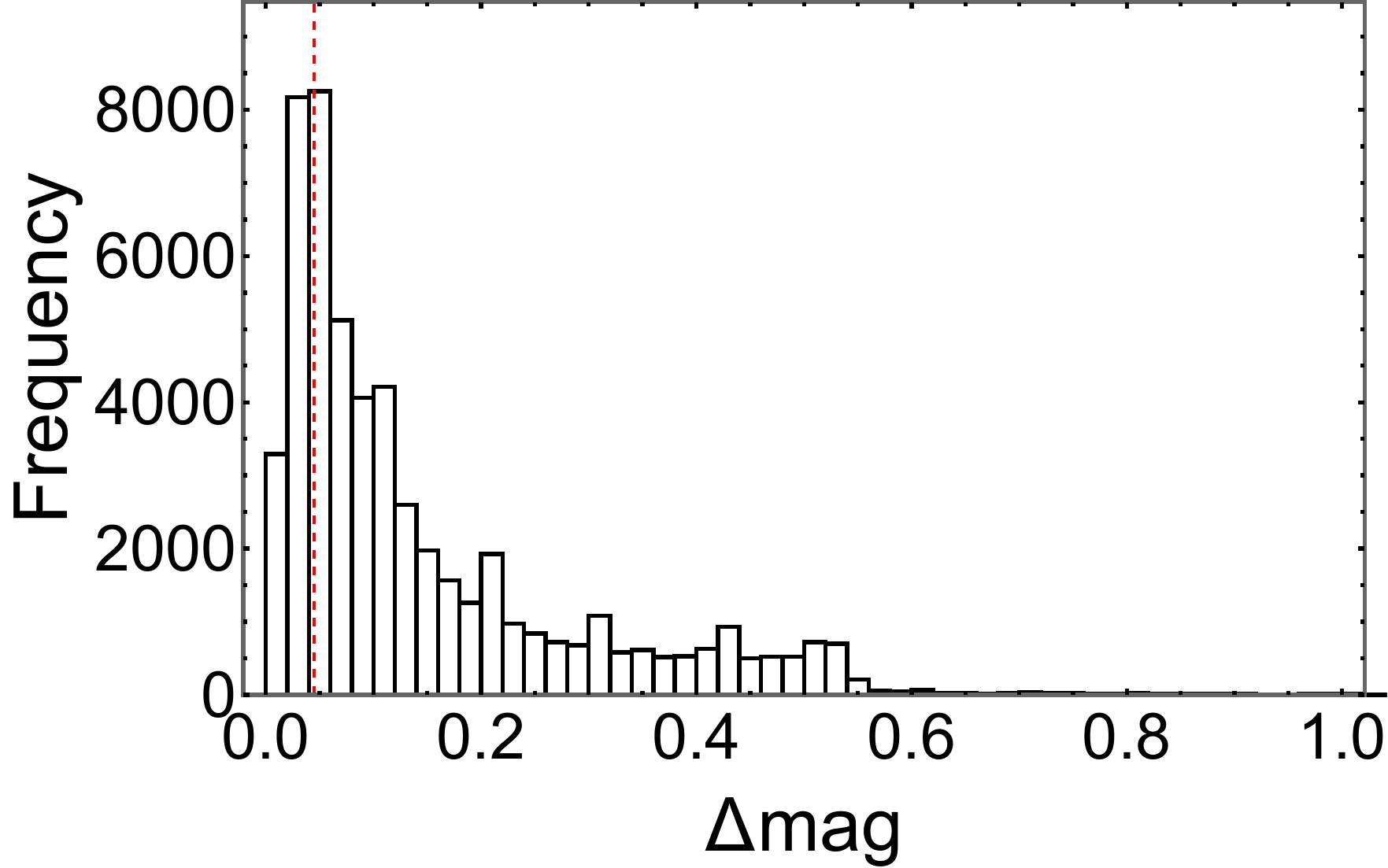}
    \caption{The distribution of the errors on magnitudes. The dashed red line marks the peak of the distribution (at around $0.05\,mag$).}
    \label{fig:errorsdist}
\end{figure}

\vspace{-0.7cm}
\subsection{The definition of the Gold and Diamond samples}
We define the "Gold" sample and the "Diamond" sample among our total GRBs. To be part of the Gold sample, a GRB must have a computed $\beta_\mathrm{opt}$ in the current analysis and fulfill one of two conditions:

A) the achromatic behaviour, characterized by colour evolution or the absence thereof, aligns with findings from existing literature.

B) if the chromatic behaviour is determined in this work due to an increased data set compared to previous analyses.

\begin{figure}
    \centering
    \includegraphics[width=7cm, height=6cm]{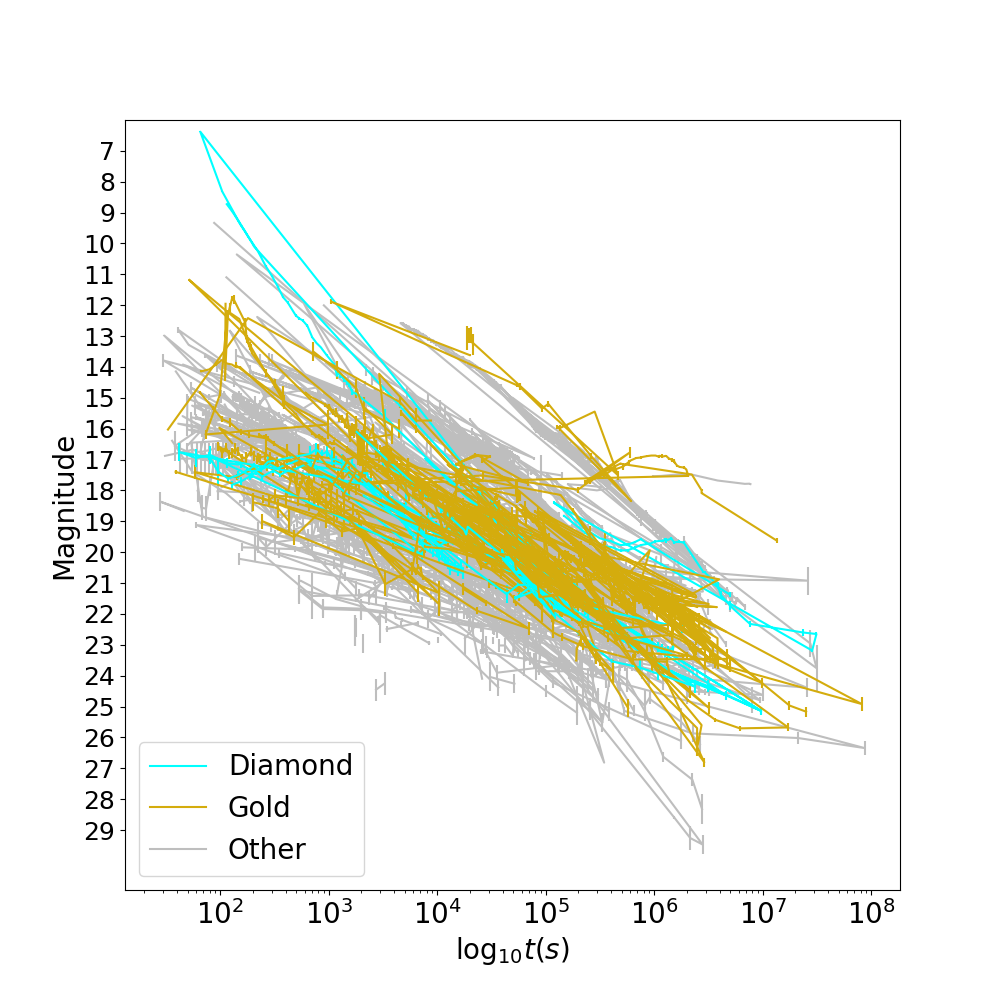}
    \caption{Magnitudes in the AB system, corrected for Galactic extinction, are shown for the Gold (in yellow), the Diamond samples (in cyan), and all the other GRBs (in grey), identified via the $a=0$ fitting. We call this plot the "Kann plot" in memory of D. A. Kann.} 
    \label{fig:kannplot}
\end{figure}

To be part of the Diamond sample, the GRB must have a computed $\beta_{\rm opt}$ and exhibit no colour evolution. It is crucial to define the Gold sample since it encompasses all the cases where we either find a confirmation of the chromatic behaviour found in the literature or we show for the first time their colour evolution analysis through the unprecedented collection of data of the same GRBs. The Diamond sample, which also includes a subset of the Gold one, is the one characterized by the possibility of rescaling the GRBs to the most numerous filter, which we remind is a necessary step to improve the fitting quality for the LCs. When $a$ is set to zero, the Gold sample comprises 92 GRBs, and the Diamond sample comprises 100 GRBs. When $a$ is variable, 92 GRBs fall in the Gold sample and 67 GRBs in the Diamond sample. To show LCs with the same filters, since the $R$ is on average the most numerous, in Figure \ref{fig:kannplot}, we present all the GRBs' $R$ band (345 cases) together with the Gold (shown in yellow) and Diamond (shown in blue) samples for the $a=0$. Furthermore, Figure \ref{fig:kannplot}, which we name the "Kann plot" in memory of D. A. Kann.

\vspace{-0.8cm}
\section{Results on the spectral analysis and the colour evolution}\label{sec:colorevolution}
In the following Subsections, we detail the spectral analysis and the colour evolution results, respectively.

\vspace{-0.3cm}
\subsection{The spectral analysis fitting results}
\begin{figure}
    \centering
    \includegraphics[scale=0.50]{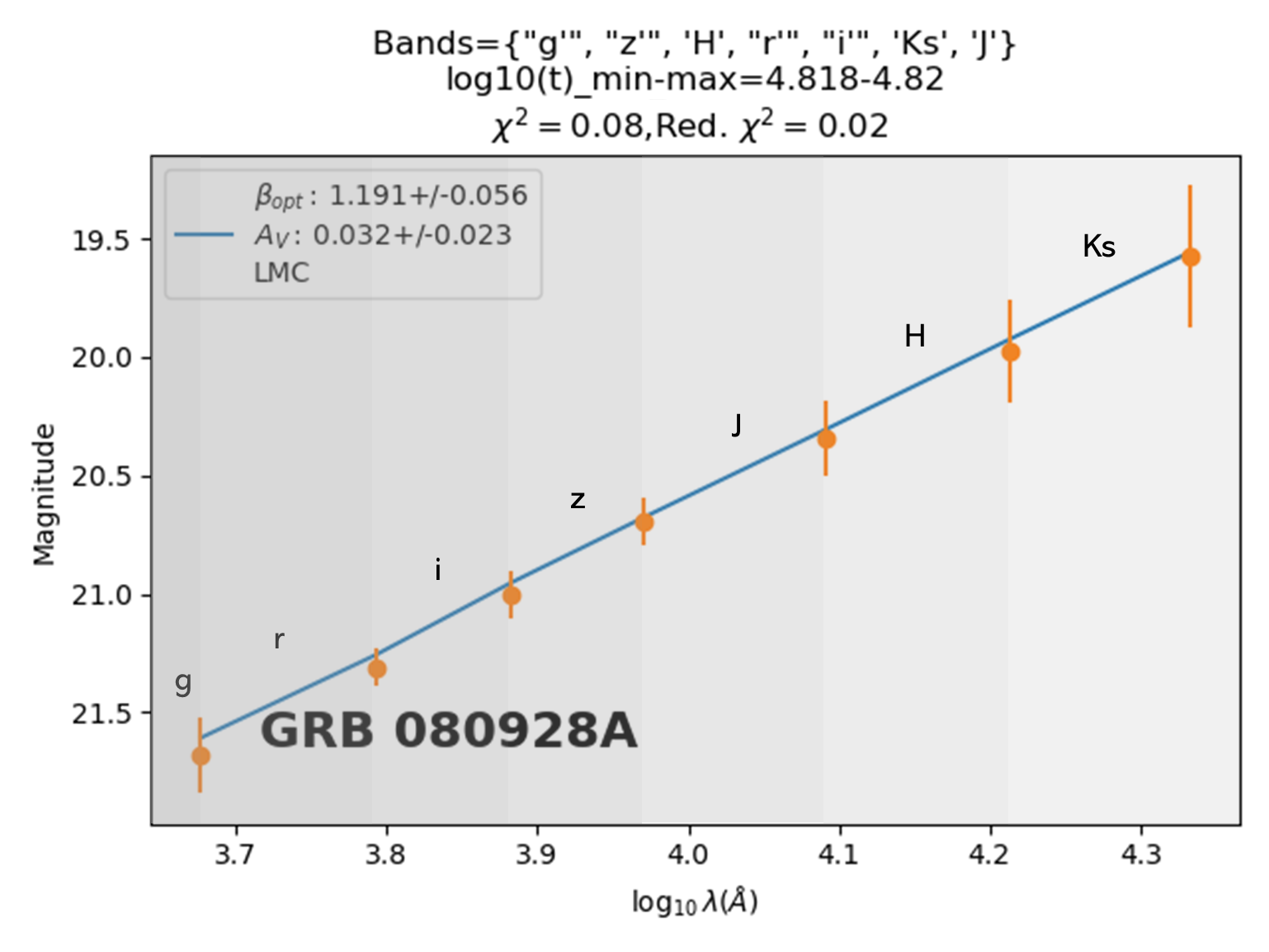}
    \caption{The SED fitting for GRB 080928A. The regression line is estimated by the Levemberg-Marquardt method. The filters, the log-time range, the $\chi^2$, and the reduced $\chi^2$ of the fit are shown in the title of the plot. The magnitudes and the wavelengths are in AB system, corrected for the Galactic extinction.} 
    \label{fig:betaexample}
\end{figure}

\begin{figure}
    \centering
    \includegraphics[scale=0.3]{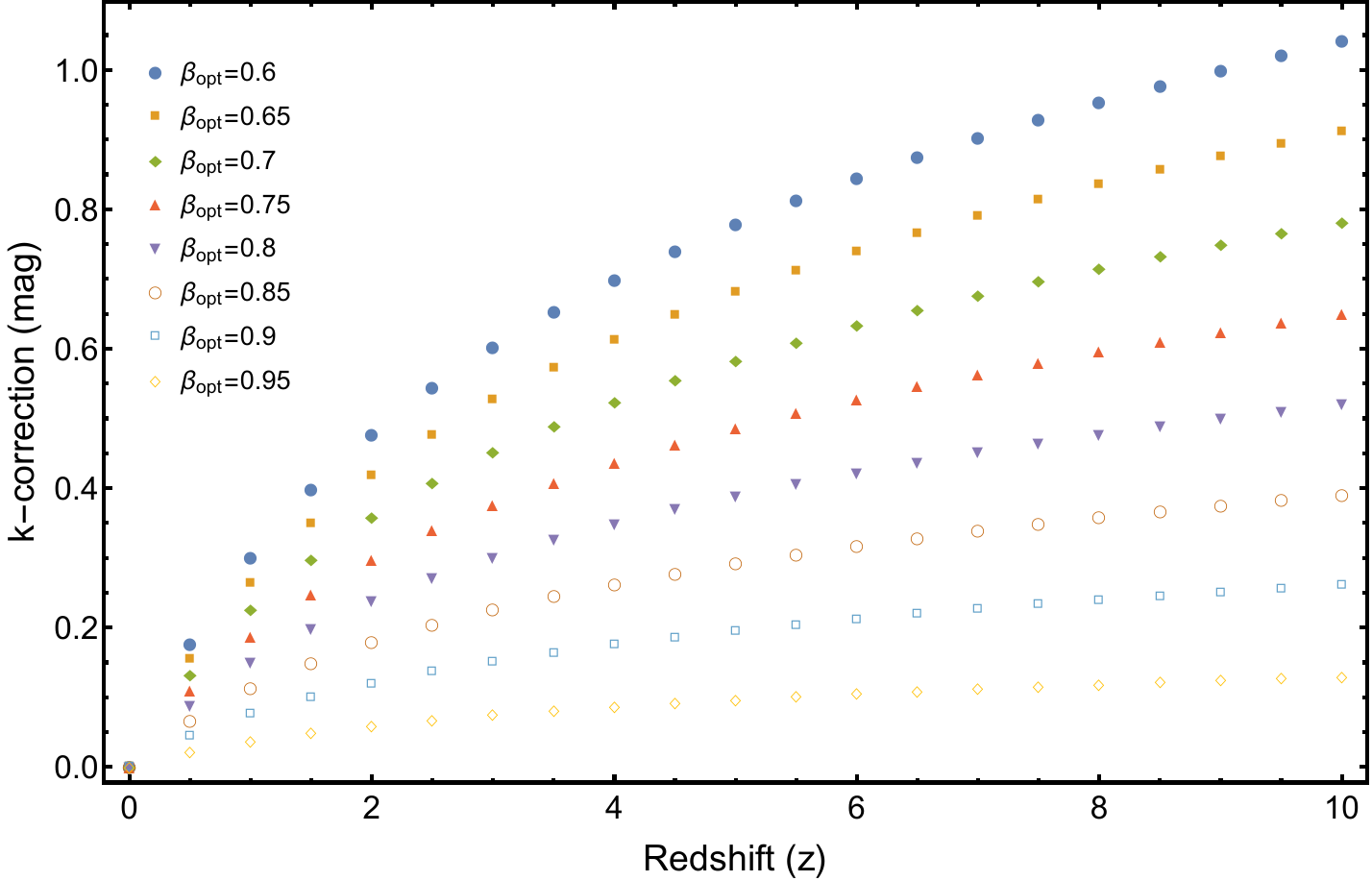}
    \caption{The plot of $k$-correction as a function of $z$ for different $\beta_\mathrm{opt}$ values.}
    \label{fig:kcorrection}
\end{figure}

Following the procedure described in Section \ref{sec:spectral}, we were able to estimate the spectral index for each epoch of the LC and obtain a total of $924$ good fits (namely, with a probability of not being drawn by chance $\geq 0.05$) for $138$ GRBs. As an example, in Figure \ref{fig:betaexample} we show the SED fitting for GRB 080928A.
While comparing spectral indices from different epoch, we need to consider how the evolution in redshift plays a substantial role.
In Figure \ref{fig:kcorrection}, we show the behaviour of the $k$-correction with redshift for possible spectral indices values in GRBs, color-coded and with different symbols: 0.5 (blue circle), 0.6 (blue filled circles), 0.65 (yellow filled squares), 0.70 (green filled diamonds), 0.80 (purple downward triangles), 0.85 (red empty circles), 0.90 (blue empty squares), and 0.95 (yellow empty diamonds) at the redshifts from 0.1 to 10. The most relevant contribution given by the $k$-correction may arise from spectral indices that are $\sim 0.6$ at a redshift $z \sim 10$ and impact the observations by $1\,mag$. 
Given that our sample does not extend to such high redshifts, the GRBs in our sample where the $k$-correction is not computed do not suffer from major systematic effects.\\ 

\vspace{-0.8cm}

\subsection{The colour evolution analysis results}
\begin{figure}
    \centering
    \includegraphics[scale=0.25]{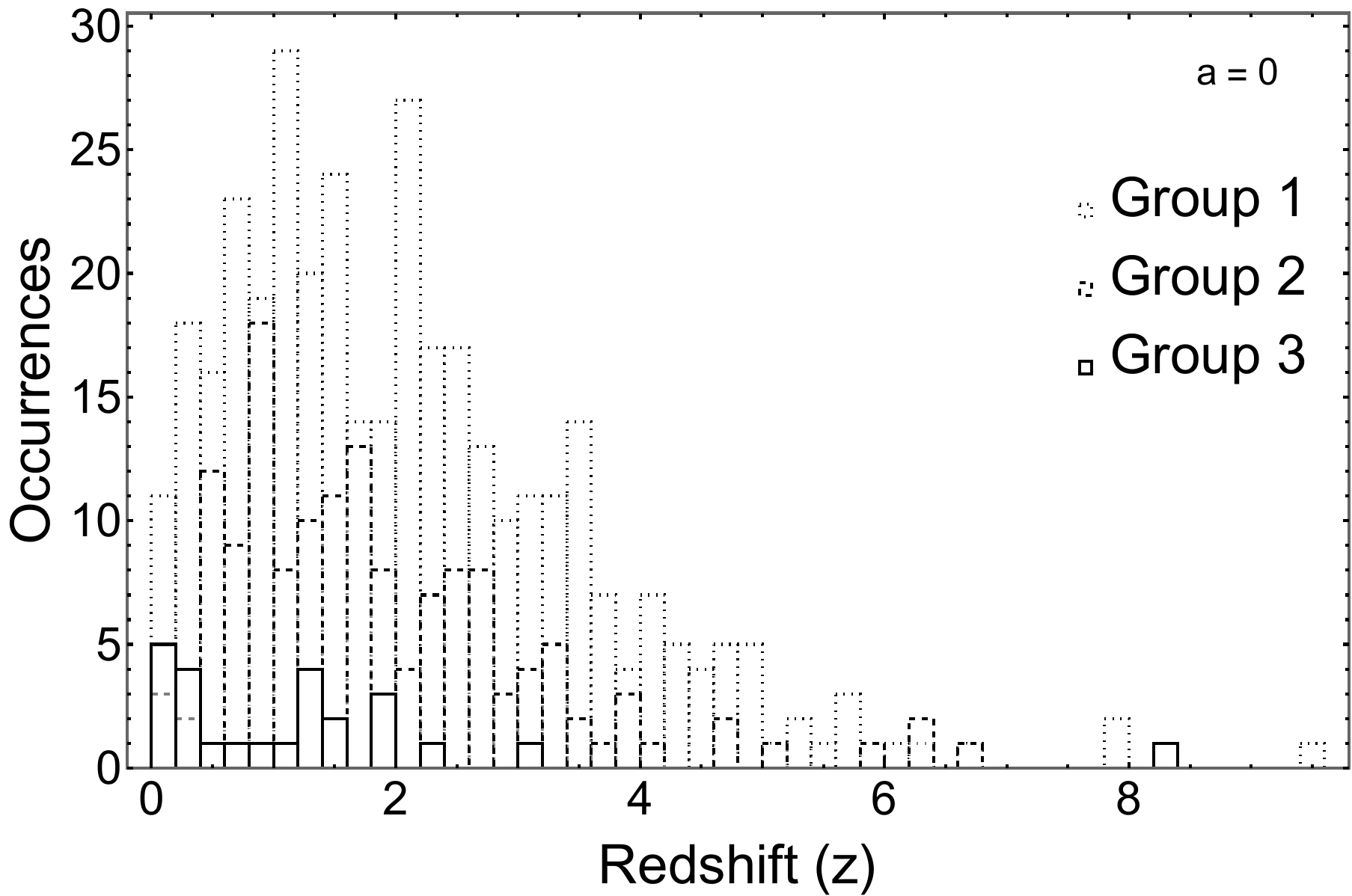}
    \caption{Redshift histograms for Groups 1 (dotted lines), 2 (dashed lines), and 3 (continuous lines) defined through the results of the colour evolution with $a=0$.} 
    \label{fig:redshifthistogram}
\end{figure}

\begin{figure}
    \centering
    \includegraphics[scale=0.255]{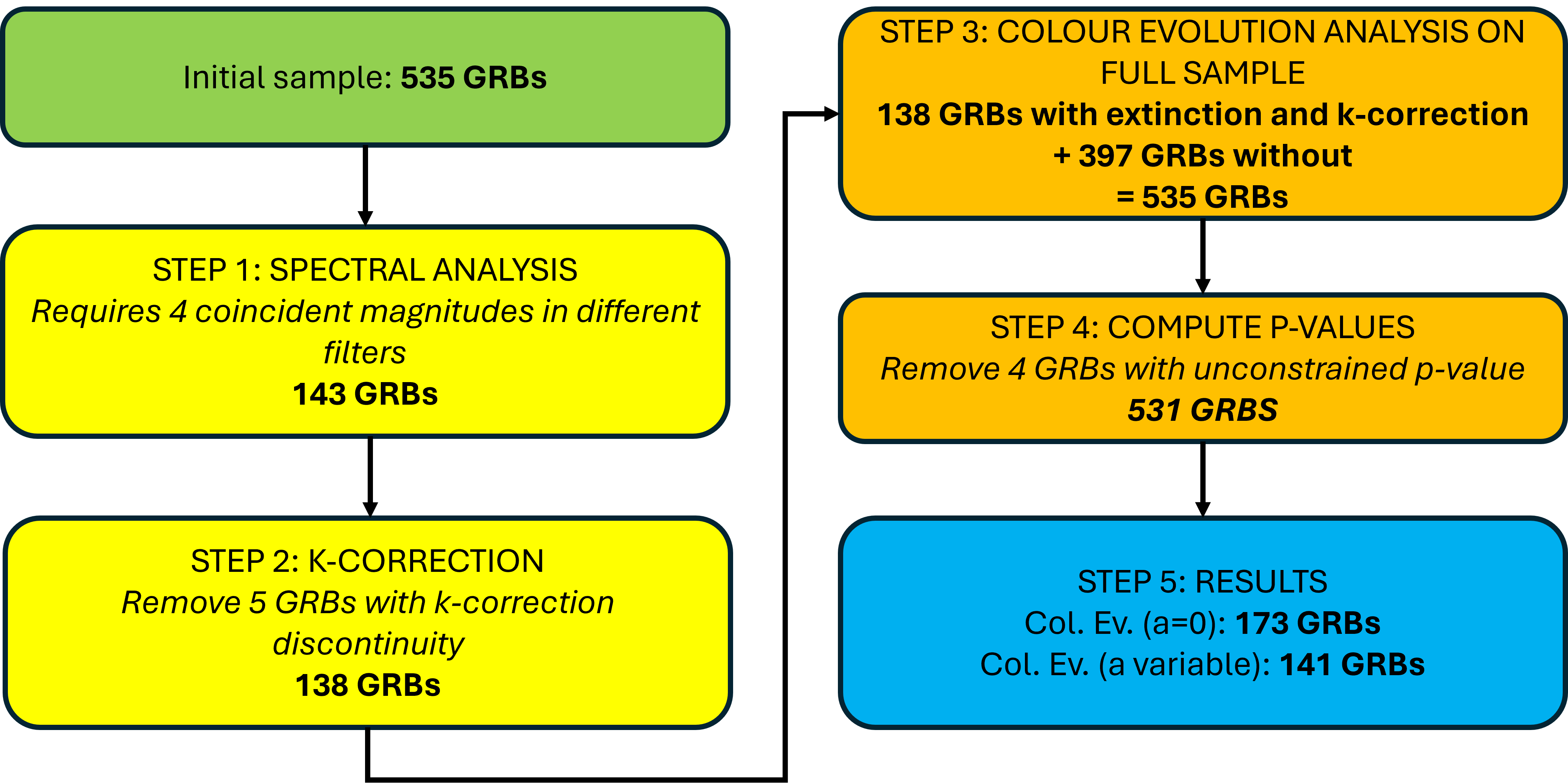}
    \caption{The flowchart of the GRB analysis.}
    \label{fig:flowchart}
\end{figure}

In the following three Subsections, we show the results of the colour evolution analysis performed by fixing $a=0$, then letting $a$ be free to vary, and their comparison with the literature. In both treatments, we define three major groups listed below:
\begin{itemize}
    \item GROUP 1: GRBs without colour evolution;
    \item GROUP 2: GRBs with colour evolution;
    \item GROUP 3: GRBs undetermined by us.
\end{itemize}
All 535 GRBs and their classification into groups are reported in the Table 4 of the Online Materials.
In order to understand if these groups have a trend in their redshifts, we have plotted histograms of redshifts for these three groups in Figure \ref{fig:redshifthistogram} based on the analysis with the slope $a=0$. There appears to be no clear dependence of the groups' population on the redshift, with the distributions being clustered at $z<4$. The same holds for the variable slope case (reported in the Online Materials). The flowchart in Figure \ref{fig:flowchart} summarizes the number of GRBs retained at each step, from the $k$-correction and the host extinction correction, to the colour evolution analysis.
Due to the restrictive criteria on the epoch, we retain fewer GRBs than the literature, but we highlight that our procedure is uniform across all cases.

\vspace{-0.5cm}
\subsubsection{The colour evolution: fitting with the fixed $a=0$}
Following the methodology described in Subsection \ref{sec:scaling}, we fit the rescaling factors with a slope $a=0$ and determined the colour evolution for a given GRB on the compatibility in majority of the filters. We determined this compatibility for 173 GRBs, excluding the 5 GRBs discussed in \ref{sec:spectral}.
There are only 4 GRBs (030329A, 100814A, 130427A, and 141220A) in both methods ($a=0$ and variable slope) that, despite having enough data points, do not have a constrained value of the probability for the fitting. The results of this analysis are schematized in the left panel of Figure \ref{fig:diagram}.
The comparison with the literature of the $a=0$ cases is detailed below:
\begin{itemize}
    \item agreement with the literature: 84 cases (75 cases with no colour evolution, and 9 with colour evolution).
    \item disagreement with the literature: 20 cases (6 cases reported colour evolution in our analysis, but the literature suggests no colour evolution, and 14 cases where our analysis indicates no coour evolution, but the literature suggests colour evolution).
    \item apparent disagreement with the literature: 3 cases. As defined in Section \ref{sec:scaling}, the apparent disagreement cases are the ones where most filters do not agree with the literature, but at least two filters do agree (1 case with colour evolution in our analysis, but no colour evolution in the literature, and 2 cases with no colour evolution in our analysis, but colour evolution in the literature). The disagreement and apparent disagreement GRBs are listed in Table \ref{tab:disagreement}.
    \item determined in our analysis, but not in the literature: 66 cases. 9 of these have colour evolution, while 57 show no colour evolution.
    \item determined in the literature, but not by us: 70 cases (46 cases with no colour evolution, and 24 with colour evolution). We stress that the cases that are undetermined by us are the GRBs for which we do not find at least 3 rescaling factors in at least one filter.
\end{itemize}
According to our colour evolution analysis with $a=0$, we define the following groups (summarized in the upper panel of Figure \ref{fig:flow}):
\begin{itemize}
    \item GROUP 1: GRBs without colour evolution (148, among which 100 have a measured value of $\beta_\mathrm{opt}$). In this Group, 75 agree with the literature (61 of which have $\beta_\mathrm{opt}$), 14 disagree with the literature (10 of which have $\beta_\mathrm{opt}$), 2 have an apparent disagreement with literature and have $\beta_\mathrm{opt}$, and 57 are determined by us, but undetermined in the literature (27 have $\beta_\mathrm{opt}$).
    \item GROUP 2: GRBs with colour evolution (25, and 4 of them with $\beta_\mathrm{opt}$). In this Group, 9 agree with the literature (3 with $\beta_\mathrm{opt}$), 6 disagree with the literature and none of these has $\beta_\mathrm{opt}$, 1 has an apparent disagreement (without $\beta_\mathrm{opt}$), and 9 are determined by us, but undetermined in the literature (1 with $\beta_\mathrm{opt}$).
    \item GROUP 3: GRBs undetermined by us (358 with $\beta_\mathrm{opt}$ available in 34 cases). In this Group, 45 are undetermined for us but determined in the literature as cases with no colour evolution, 8 them with $\beta_\mathrm{opt}$. 21 GRBs are undetermined by us, but are determined in the literature as cases with colour evolution, 4 with $\beta_\mathrm{opt}$. Finally, 292 are both undetermined by us and in the literature (for those, 22 have $\beta_\mathrm{opt}$).
\end{itemize}

The advantage of fixing the slope $a=0$ is that it reduces the number of GRBs for which we have the same number of filters that exhibit colour evolution and not. Thus, we have a reduced number of undetermined cases. As a consequence, this increases the number of GRBs for which there is an agreement with the literature. In Figure \ref{fig:appdisagreea0}, we report the colour evolution analysis for the 3 apparent disagreement cases for $a=0$ fitting (080109A, 080319B, and 130702A).
\vspace{-0.5cm}
\subsubsection{The colour evolution: fitting with the linear model (variable slope)}
We also fit the linear model of Equation \ref{eq:linmodel} allowing the slope $a$ to vary. Thus, the slope obtained describes the time evolution of the rescaling factors, if present. We determined the chromatic behaviour of 141 GRBs, 18.5\% less than the slope $a=0$ analysis. This variation is due to the low probability for a linear fitting with variable $a$.
We note that the 4 GRBs for which the fitting probability $P$ is not constrained in the $a=0$ fitting (030329A, 100814A, 130427A, and 141220A) have an unconstrained $P$ also in the variable slope fitting and consequently, they are excluded from the groups and the following discussion.
The results of the comparison between the variable slope case and the literature are the following:
\begin{itemize}
    \item agreement with the literature: 66 cases (57 cases with no colour evolution, and 9 with colour evolution).
    \item disagreement with the literature: 20 cases (10 cases reported colour evolution in the literature, but our analysis suggests no colour evolution and 10 cases where the literature suggests no colour evolution, but our analysis indicates colour evolution).
    \item apparent disagreement with the literature: 4 cases (3 cases with colour evolution in our analysis, but no colour evolution in the literature, and 1 case with no colour evolution in our analysis, but colour evolution reported in the literature).
    \item determined in our analysis, but not in the literature: 51 cases. 16 of these have colour evolution, while 35 show no colour evolution.
    \item determined in the literature, but not by us: 87 cases (58 cases with no colour evolution, and 29 with colour evolution). In the variable $a$ fitting, the cases that are undetermined by us are the GRBs for which either we do not find at least 3 rescaling factors in at least one filter or all the filters with at least 3 rescaling factors have $\sigma_{a}>a$.
\end{itemize}
According to our colour evolution analysis with variable slope instead, we define the following groups (reported in the lower panel of Figure \ref{fig:flow}):
\begin{itemize}
    \item GROUP 1: GRBs without colour evolution (103, among which 68 have a measured value of $\beta_\mathrm{opt}$). In this Group, 56 agree with the literature (43 of which have $\beta_\mathrm{opt}$), 10 disagree with the literature (6 of which have $\beta_\mathrm{opt}$), 1 has an apparent disagreement with literature and has $\beta_\mathrm{opt}$, and 36 are determined by us, but undetermined in the literature (18 have $\beta_\mathrm{opt}$).
    \item GROUP 2: GRBs with colour evolution (38, and 19 of them with $\beta_\mathrm{opt}$). In this Group, 9 agree with the literature (5 with $\beta_\mathrm{opt}$), 11 disagree with the literature and 7 of them have $\beta_\mathrm{opt}$, 2 have an apparent disagreement (both with $\beta_\mathrm{opt}$), and 16 are determined by us, but undetermined in the literature (5 $\beta_\mathrm{opt}$).
    \item GROUP 3: GRBs undetermined by us (389, with $\beta_\mathrm{opt}$ available in 51 cases). In this Group, 57 are undetermined for us but determined in the literature as cases with no colour evolution, 16 of them with $\beta_\mathrm{opt}$. 27 are undetermined by us but are determined in the literature as cases with colour evolution, 7 with $\beta_\mathrm{opt}$. Finally, 305 are both undetermined by us and in the literature (for those, 28 have $\beta_\mathrm{opt}$).
\end{itemize}

\vspace{-0.5cm}
\subsubsection{Comparison between the $a=0$ and the variable slope methods}
\begin{figure*}
    \centering
    \includegraphics[scale=0.35]{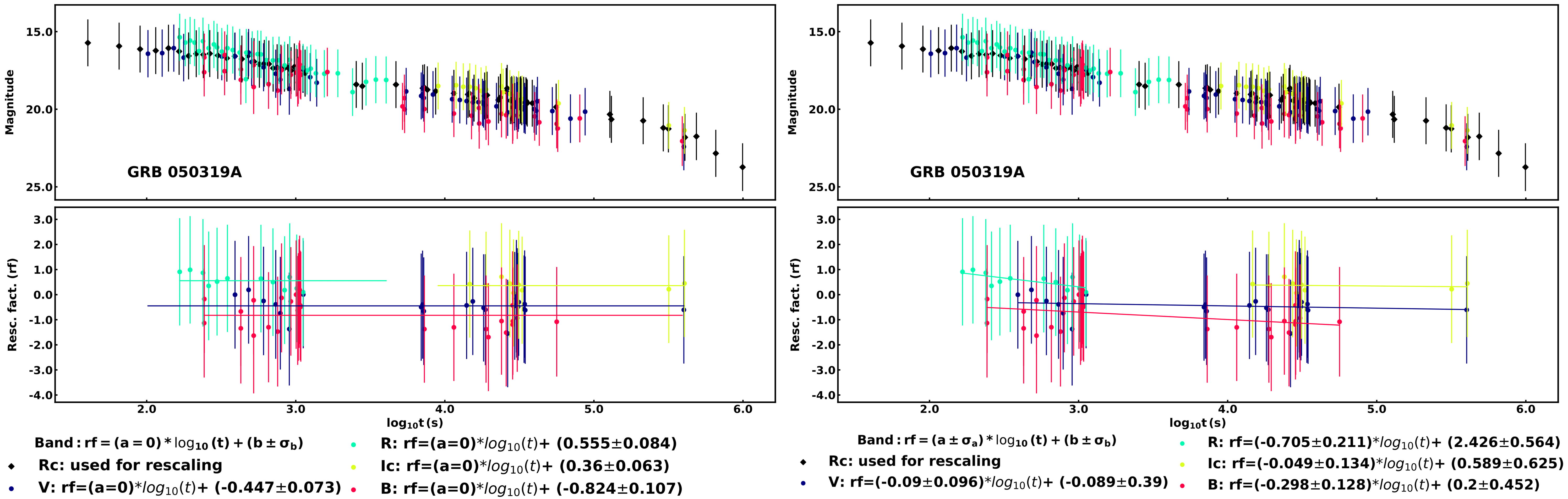}
    \caption{A comparison of the $a=0$ fitting (left panel) and the variable slope fitting (right panel) for GRB 050319A. For both panels, the upper half shows magnitudes versus times, while in the lower half we report the rescaling factors vs time with their fitting functions.}
    \label{fig:a0vslinear}
\end{figure*}

We compare the results of the slope $a=0$ and the variable slope methods. Setting the slope $a=0$, the number of GRBs with determined chromatic behaviour is increased by 23\% (173 versus 141) compared to the variable slope approach; $a=0$ fitting also somewhat reduced the disagreement cases with the literature (17\% versus 13\%) and slightly increased agreement (47\% versus 49\%) compared to the variable $a$ fitting. Overall, while there is a slight increase in agreement with the fixed slope model, the overall level of agreement between the two models remains similar and thus, leads to a similar comparison with the literature. As an example, in Figure \ref{fig:a0vslinear}, we compare the $a=0$ fit with the variable slope fit for GRB 050319A. We see that for the filters $B,I_C,V$ the variable slope value $a$ is compatible with zero in less than 3 $\sigma$ ($a_{B}=-0.298 \pm 0.128,a_{I_C}=-0.049 \pm 0.134,a_V=-0.090 \pm 0.096$), in agreement with what we observe in the $a=0$ fitting. The cases of disagreement with the literature and the 5 $\sigma$ outliers detailed in the Sec. \ref{sec:outliers3sigma} for both methods are listed on the Online Materials.

We evaluated BIC for 136 GRBs for which both fitting were determined. To compare the two fitting methods, we follow the procedure described by \citet{dainotti2021c}. For each filter, we find the minimum of the BIC between the variable $a$ and the $a=0$ fittings, namely $BIC_{min}=min(BIC_{var.a},BIC_{a=0})$. Then, for the two models, we evaluate the \textit{relative likelihood} $p_i=W_i/(\sum_{i}W_i)$, where $W_i$ is the weight of the $i$-th model, defined as $W_i=exp((BIC_{min}-BIC_{i})/2)$. If, for the $i$-th model, the $p_i>0.95$ then that model is preferred to the other. If both of the $p_i$ values are $<0.95$ then there is no statistical difference between the two models. When the majority of the filters in a given GRB reveal a model that is favoured compared to the other according to its $p_i$, then the $i$ model is preferred.

\begin{figure*}
    \centering
    \includegraphics[scale=0.35]{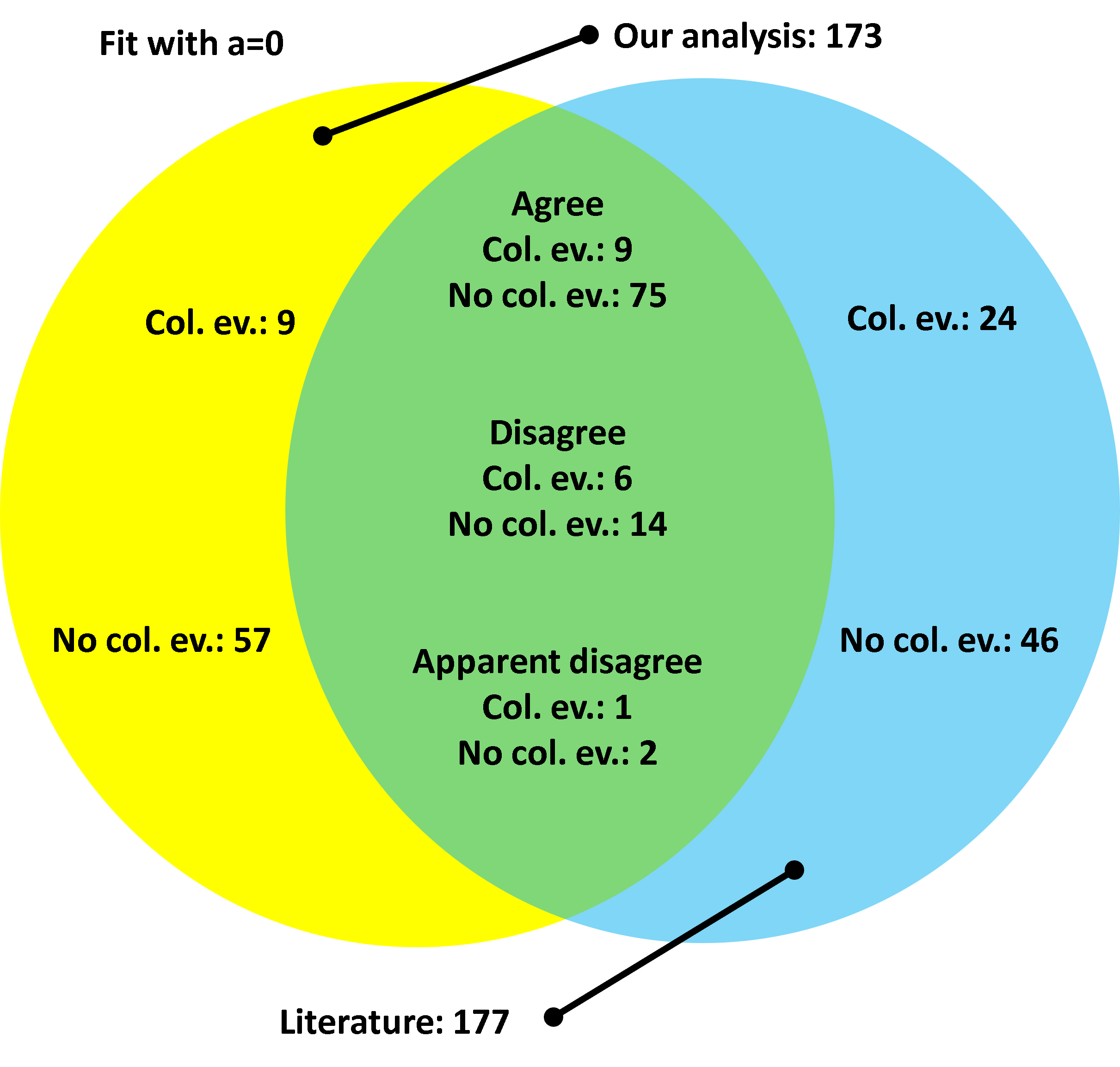}
    \hspace{2ex}
    \includegraphics[scale=0.35]{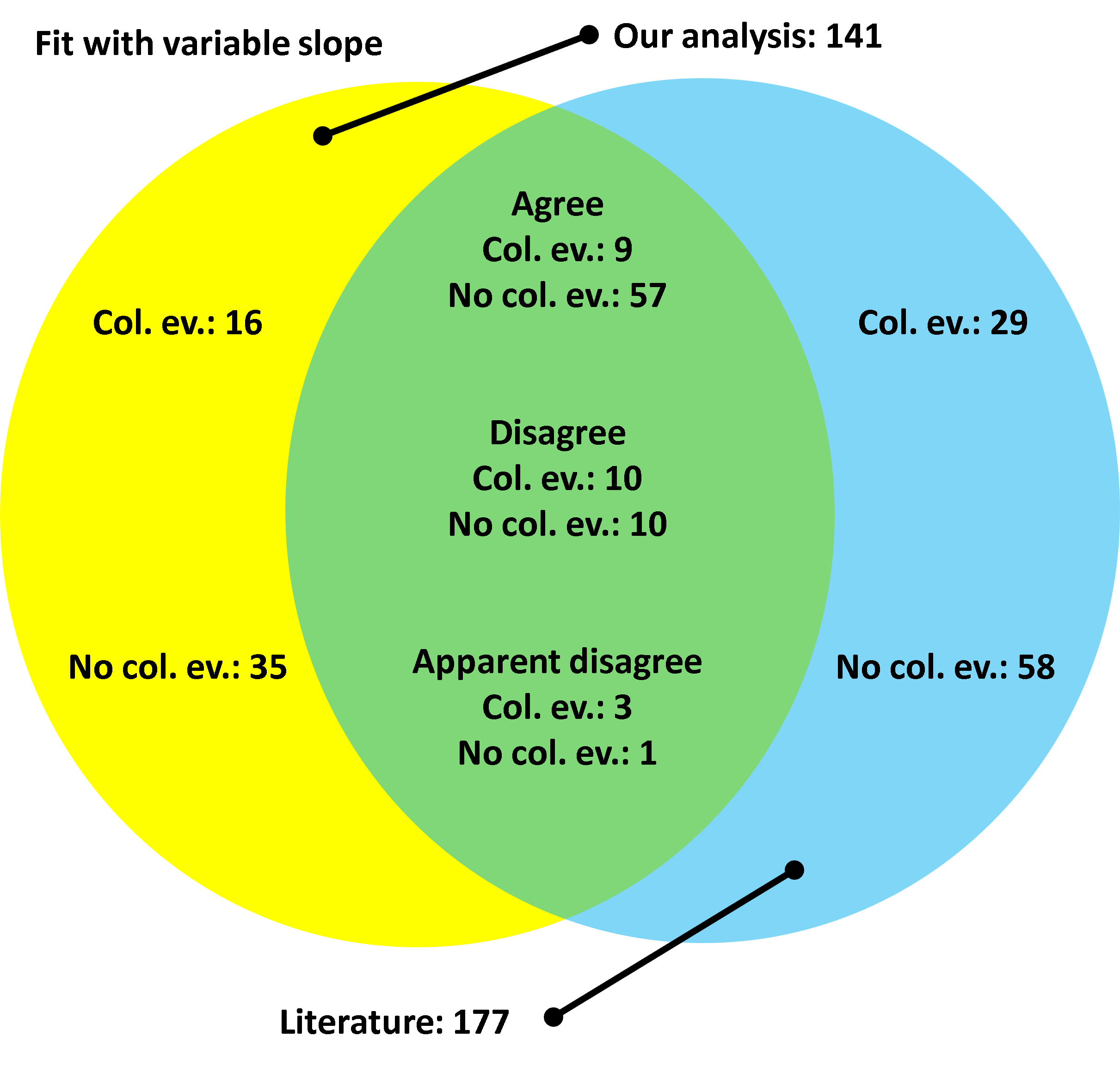}
    \includegraphics[scale=0.45]{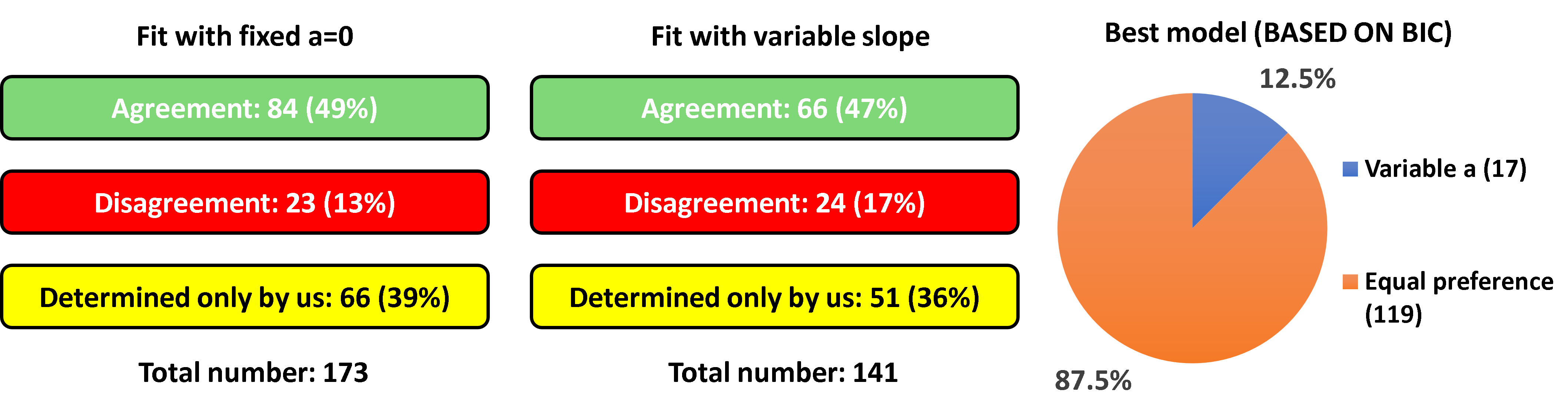}
    \caption{Visual representation of the results of our colour evolution analysis. \textbf{Upper panel.} Venn diagrams detailing agreement with literature for the two models: \textit{Right} - A linear model with $a=0$, \textit{Left} - A linear model with variable slope. \textbf{Lower left panel.} The comparison between the two models. Note that in this last panel the disagreement and apparent disagreement cases have been counted together for simplicity and have been reported in the red cell. \textbf{Lower right panel.} The best model based on BIC. In the pie chart, Blue represents the linear model with $a=0$, Orange is the linear model with variasble $a$ and Grey represents the cases where both models yielded equal BIC, hence equal preference.}
    \label{fig:diagram}
\end{figure*}

Of the 136 GRBs, for 119 GRBs (87.5\%), the $p_i$ is $<95\%$ for both models, making them statistically indistinguishable. In the remaining 17 GRBs, the $p_i$ favours the variable slope over the fixed $a=0$ slope. Furthermore, all these 17 cases have colour evolution according to the variable slope analysis, showing the model's effectiveness in capturing colour evolution. The pie chart in lower right panel of Figure \ref{fig:diagram} provides a summary of the BIC analysis' results. 

To summarize, comparing the results of the spectral index evolution with the colour evolution results through the two fitting methods, we find the following cases: 6 GRBs for which both $a=0$ and variable $a$ show no colour evolution similarly to the results obtained with the SED; 6 GRBs for which $a=0$ shows no colour evolution, variable $a$ has colour evolution, and the SED shows in 3 cases evolution and in other 3 cases no evolution; 15 GRBs for which both $a=0$ and variable $a$ show no colour evolution but the SED shows evolution; 3 GRBs for which either the $a=0$ or variable $a$ are undetermined but the SED shows evolution.

\vspace{-0.5cm}
\subsubsection{The outliers}
\label{sec:outliers3sigma}
\begin{figure*}
    \centering
    \includegraphics[scale=0.40]{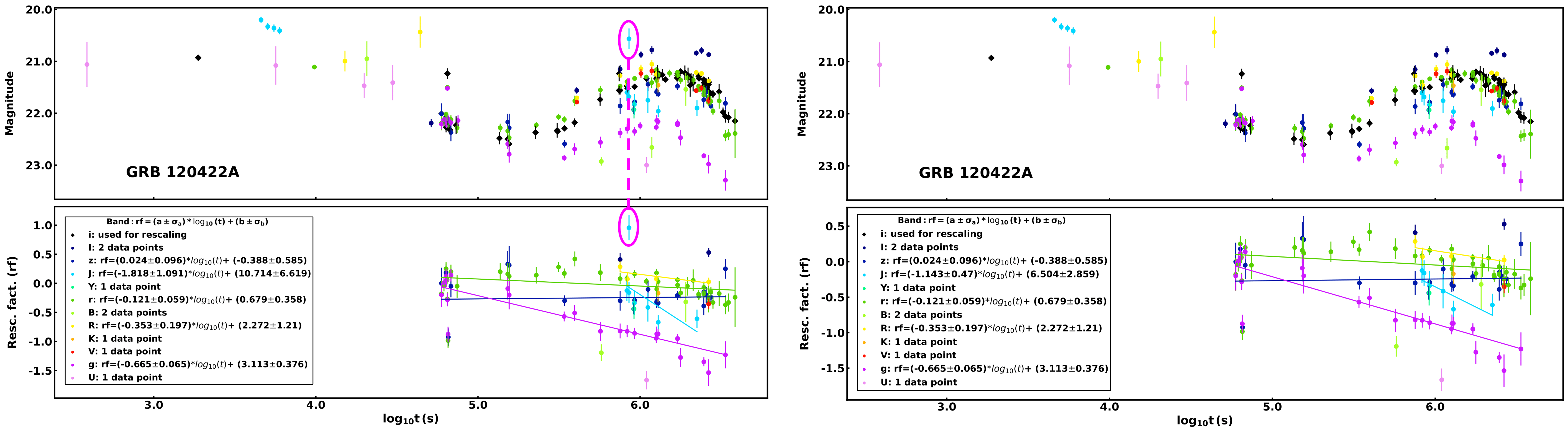}
    \caption{\textbf{Left panel.} The GRB 120422A where the outlier for the $J$ band (in the magenta circles) is included and the variable slope fitting is performed. The most numerous filter is shown with black rhombuses, while the other filters are shown with coloured dots. \textbf{Right panel.} The same case of the upper panel but with the removal of the $J$ band outlier. The same convention on the colours of left panel is applied here.}
    \label{fig:outlier}
\end{figure*}

In this Section, we discuss the outliers found in the catalogue and their impact on the colour evolution analysis. We classify the outliers as follows: the \textit{bad photometry points} for which the photometry is not reliable according to their sources (for example, in the case of some GCNs), marked with $X$; the points whose magnitudes are separated by at least 5 $\sigma$ from the other points at coincident epochs marked with $Y$; the non-simultaneous outliers, for which it is not possible to apply the criterion of coincident epochs defined in Equation \ref{eq:coincidenttimes} and their magnitudes deviate more than 5 $\sigma$ from the other data points that are the closest according to their time epoch, marked with $Z$. According to our definition, there are many outliers with mixed properties, for example, they can be $X$ and $Y$ or $X$ and $Z$. In total, there are 1492 outliers in this catalogue: 1022 are of the $X$-type, 801 are $Y$-type, and 656 are $Z$-type. These outliers are in total a fraction of $1492/64813=2\%$. We have also added to clarify that the points marked $Y$ {\it are} outliers, which will provide issues during the fits, while the points marked $Z$ {\it may be} outliers. A conservative approach, similar to the one we used, would reject both the $Y$ and $Z$ points, while a less cautious approach focused on looking into small effects may require the inclusion of the $Z$ points in the fits. In the following, we focused only on the conservative approach and thus rejected points labeled $X$, $Y$, and $Z$. However, we here highlight that the users can also investigate and use these data points for their own data analysis.
To evaluate the impact of the outliers with respect to the linear fit of the rescaling factors, we also performed the fitting of the rescaling factors after removing these outliers when their photometry is not fully reliable (like in some cases of the GCNs). This analysis has been performed for both the $a=0$ and the variable slope fitting.
In the $a=0$ approach, there are 49 GRBs for which there are one or more outliers. Among these 49, 12 have outliers caused by non-reliable photometry of GCNs.
In the cases of 060218A, 100621A, 120422A, and 210619B, the removal of the outliers does not change the colour behaviour of the GRBs. For 080430A, 081028A, 110503A, 141028A, 150323A, 160121A, 181010A, and 210610B, instead, the removal of the outliers from GCNs prevents the analysis from finding a determined colour behaviour given the lack of sufficient rescaling factors. Concerning GRB 160131A, before the removal, the $V$ band has no colour evolution, but it shows colour evolution after the removal of GCN 18960 \citep{2016GCN.18960....1H}.
In the variable slope approach, among all the cases that show outliers (41 in total), in 10 GRBs (060218A, 110503A, 111228A, 120422A, 150323A, 160131A, 180325A, 181010A, 210619B, and 230812B) the photometry may not be reliable since the data points that constitute outliers are taken from GCNs.
In detail, in the case of 060218A and 120422A, the removal of the outliers from GCN 4831 \citep{2006GCN..4831....1R} and GCN 13279 \citep{2012GCN.13279....1P} respectively, does not change the results. Concerning 160131A, removing data from GCN 18960 \citep{2016GCN.18960....1H} in the $V$ filter changes the behaviour from no colour evolution to colour evolution. 
In all the 7 remaining cases, the removal of the outliers led to undetermined colour evolution due to the lack of sufficient data.
The case below (in Figure \ref{fig:outlier}) shows GRB 120422A where there is no difference between removing and leaving in the outliers. 
In the left upper panel, the LC shows the fit highlighting the outlier in the J band with the magenta circles, with the respective rescaling factors in the lower panel. The J filter has rescaling factors slope $a=-1.818 \pm 1.091$ (shown in the bottom panel), which indicates no colour evolution. In the right panel, the LC is presented without the outlier in the J band: the J filter has rescaling factors slope $a=-1.143 \pm 0.470$, which still indicates no colour evolution.

\vspace{-0.5cm}
\subsubsection{Disagreement of our results with the literature}
The main reasons for disagreement are summarized below with the label corresponding to Table 4 of the Online Materials:

\begin{enumerate}
    \item GRBs may exhibit achromatic (labeled as A.1) or chromatic behaviour (labeled as A.2) across different wavelengths, but the comparison between us and the literature is performed in different bands.
    \item Different methodologies in data analysis (labeled with B).
    \item Inconsistent data availability at coincident times with certain filters (C.1) or lack of sufficient data (labeled with C.2).
    \item rescaling in different filters between us and the literature (D.1 labels cases of agreement with the literature when changing the filter for rescaling; D.2 indicates GRBs when only a few filters agree with the literature after changing the reference filter; D.3 labels GRBs when we have the same number of filters with and without colour evolution).
    \end{enumerate}
For the $a=0$ fitting, there are 20 GRBs whose colour evolution is in disagreement with the ones reported in the literature and 3 GRBs for which there is an apparent disagreement behaviour. These cases are listed in the left half of Table \ref{tab:disagreement}. There are additional 70 cases for which we are not able to discern any colour  behaviour but they are discussed in the literature.
For the variable slope analysis, we find 21 GRBs whose colour evolution is in disagreement with the ones reported in previous work and 4 GRBs for which there is an apparent disagreement. For 87 cases, we are not able to discern any colour behaviour while the literature discusses them. These events are listed in the right half of Table \ref{tab:disagreement}.
The results of this scheme are reported in the 4th column of Table 4 on the Online Materials.

\begin{figure}
    \centering
    \includegraphics[scale=0.39]{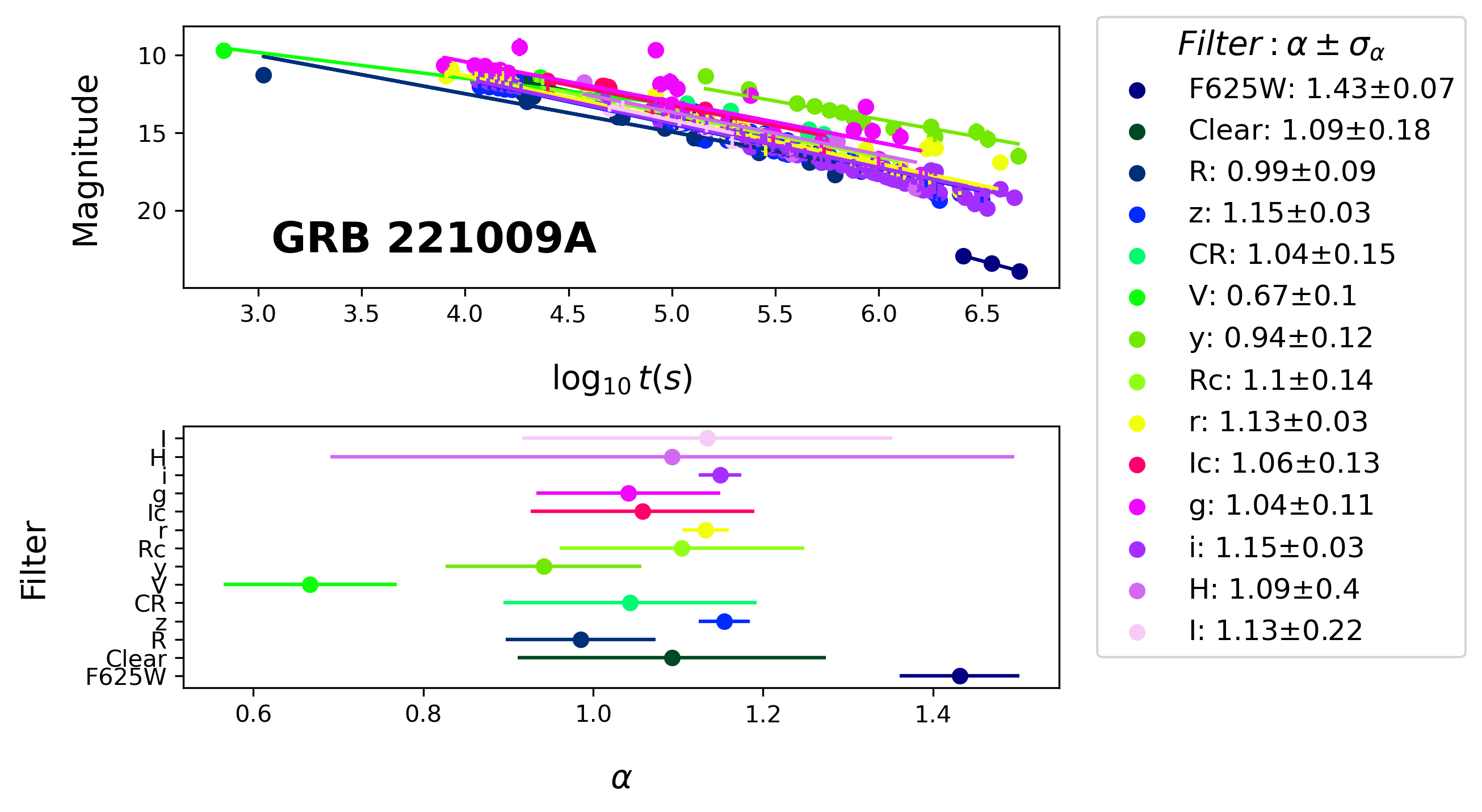}
    \caption{The monochromatic LCs fitting for the GRB 221009A: the plot shows the magnitudes vs the $\log_{10}$ of time after the trigger (in seconds) together with the best fit curves estimated with the simple power-law model. In the legend, the $\alpha$ values for each filter together with the 1 $\sigma$ values.}
    \label{fig:221009Aalpha}
\end{figure}

\begin{figure*}
    \centering
    \includegraphics[scale=0.35]{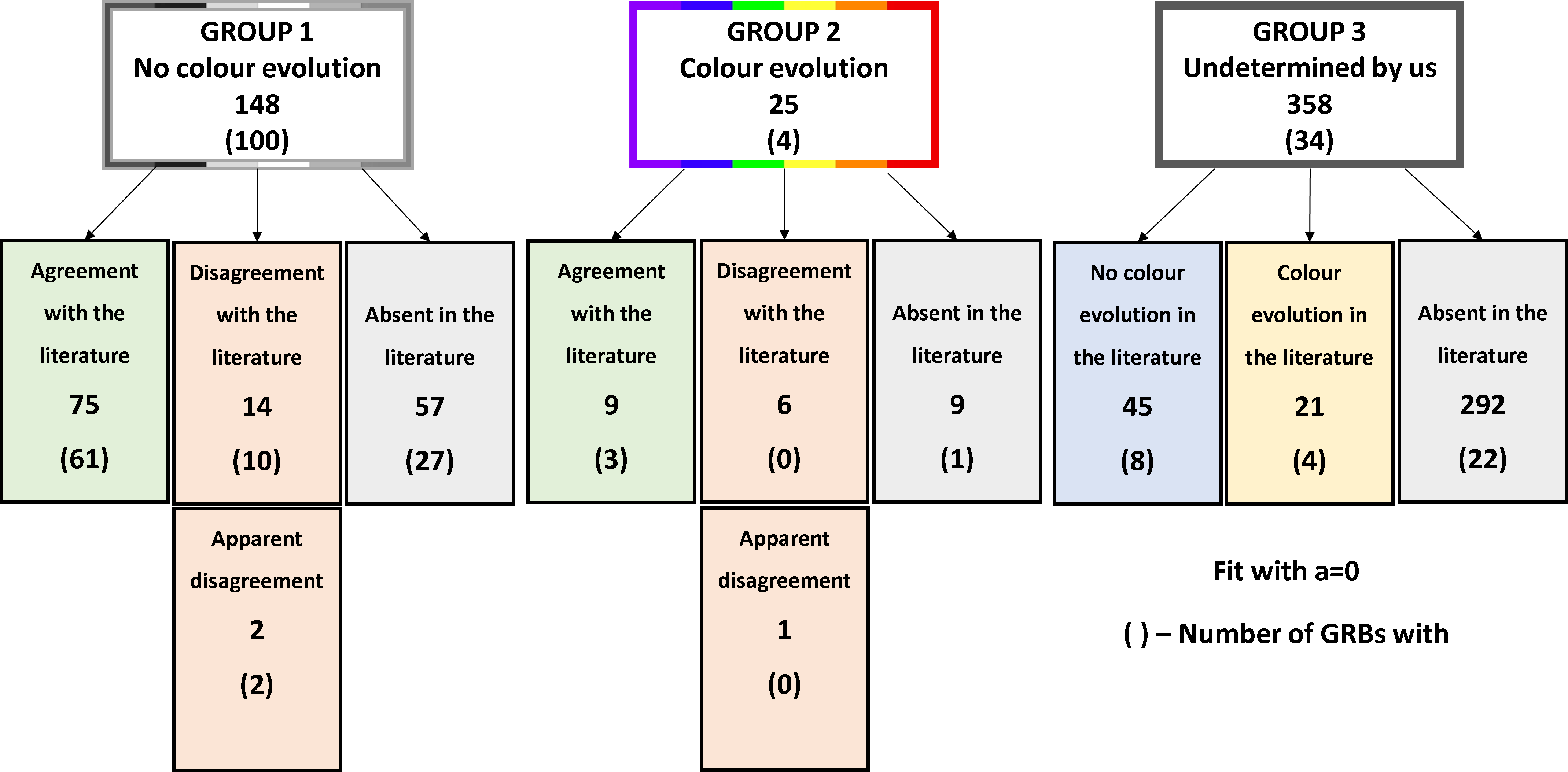}
    \includegraphics[scale=0.35]{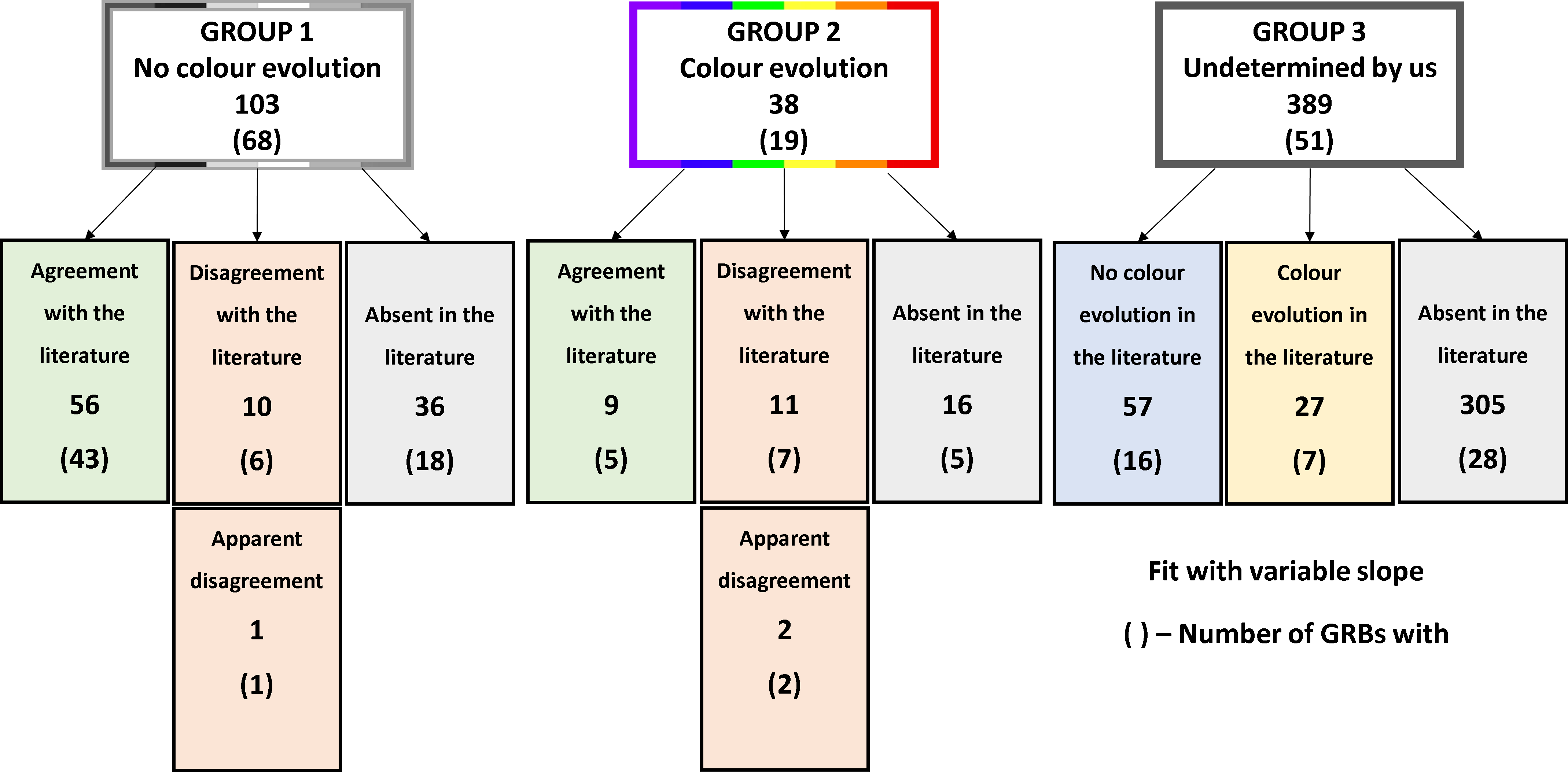}
    \caption{The flowchart of the Groups division. \textbf{Upper panel.} Groups division based on the colour evolution with a linear fit where $a=0$. \textbf{Lower panel.} Same groups for a fit with variable slope. In round brackets, the number of GRBs with known $\beta_\mathrm{opt}$.}
    \label{fig:flow}
\end{figure*}

\begin{table*}
    \centering
    \setlength{\tabcolsep}{5pt}
    \begin{tabular}{ccc|ccc}
    \hline
    DISAGREEMENT & CASES & & DISAGREEMENT & CASES & \\
    \hline
        GRB($a=0$) & Col.ev.(lit) & Col.ev.(us) & GRB(var.slope) & Col.ev.(lit) & Col.ev.(us) \\
        \hline
        000911A &  Yes & No  &  990510A & No & Yes \\
        011121A &  Yes & No &  991208A & No & Yes \\
        041006A	&	Yes & No	& 	011121A &  Yes & No  \\
        060124A	&	No & Yes	&	041006A &	Yes & No   \\
        061126A	&	Yes & No	&  060124A & No & Yes  \\
        071003A &  Yes & No  &  061126A &  Yes & No \\
        071031A &  Yes & No  &	080109A &	No & Yes \\
        100418A	&	Yes & No	&	090313A &	No & Yes	  \\
        100316D	& 	No & Yes	&	090926A &	No & Yes	  \\
        101225A	& 	Yes & No	&	091018A &	No & Yes	  \\
        120119A	& Yes	& No	&	101219B &	No & Yes	  \\
        140606B	&	Yes & No	&	111228A &	No & Yes	  \\
        141004A	&	Yes & No	&	120422A &	Yes & No	  \\
        141221A	&	No & Yes	&  150910A &  Yes & No	  \\
        150910A & Yes  &  No  &  160821B & Yes & No \\
        160131A & No  & Yes  &  161017A & Yes & No  \\
        161017A &  Yes &  No &	161023A &	No & Yes \\
        180418A	&	No & Yes	&	170202A &	No & Yes	  \\
        201015A	&	Yes & No	&	200829A &	Yes & No	  \\
        210731A	&	No & Yes	& 210619B &	Yes & No	  \\ 
        -	&	- & -	& 230812B &	Yes & No	 \\
    \hline
    APPARENT & DISAGREEMENT & CASES & APPARENT & DISAGREEMENT & CASES \\
    \hline
    GRB($a=0$) & Col.ev.(lit) & Col.ev.(us) & GRB(var.slope) & Col.ev.(lit) & Col.ev.(us) \\
    \hline
    080109A & No & Yes  & 080330A  & No & Yes \\  
    080319B &  Yes & No  & 080413B  & No & Yes \\
    130702A & Yes & No  & 081029A  & No & Yes \\  
    - & - & - &  120119A & Yes & No \\  
     \hline
    \end{tabular}
    \caption{The disagreement and apparent disagreement cases for both the $a=0$ and the variable slope fittings. For each of the subtables the structure is the following: first column = the GRB is specified; second column = the colour evolution (Yes) or no colour evolution (No) behaviour from literature; third column = the colour or no colour evolution of the GRB from our analysis.}
    \label{tab:disagreement}
\end{table*}

\subsubsection{The approach of the monochromatic light curve fitting}
Finally, we briefly discuss the monochromatic LC fitting approach to estimate the colour evolution. It might be argued that our approach could be less accurate than the monochromatic LC fitting. We do infer the evolutionary behaviour by comparing the time-decay power-law index parameter ($\alpha$) estimated in each filter for a given GRB.
The advantage of this method is that we can determine it for multiple segments of the LCs assuming our $\delta_t/t=0.025$. If we fit all LCs we can have evolution in some segments which would be masked out if we had not considered this $\delta_t/t$ condition. The two-step process is now automated and the code will be made publicly available so anyone can change the condition on $\delta_t/t$ from $0.025$ to the desired value. In such a process, the LCs features will not be washed out since the criterion for $\delta_t /t$ is pretty restrictive, so all features, if they exist, must be preserved.

Fitting the monochromatic LCs and estimating their slopes is a less restrictive assumption than our method where we investigate the colour evolution only in the time ranges where multiple filters are available.
For completeness of the treatment, we tested the monochromatic method. We consider here as an example the case of GRB 221009A, shown in Figure \ref{fig:221009Aalpha}, where we can observe the fitting through a power-law model defined below. We fitted the monochromatic magnitude values versus $\log_{10}(t)$ with a simple power-law model following \citet{Li2018a}. In this model, $\alpha$ is the slope value and all the slopes are compatible with the ones in $H$ filter to 1$\sigma$, indicating no colour evolution. The power-law model is the following: $\mathrm{mag}_{corr}(t)=2.5\alpha_{X} \log_{10}(t) + \mathrm{mag}_{0}$, where the corrections for the Galactic and host extinctions are already included in the $\mathrm{mag}_{AB}$ value on the left side, the subscript $X$ is the filter, and $\mathrm{mag}_{0}$ is the intercept of the fitting.
Applying this model to GRB 221009A and fitting only the filters for which at least 3 data points are available, we obtain the following values of $\alpha$: $\alpha_{Clear}=1.09\pm0.18, \alpha_{g}=1.04\pm0.11, \alpha_{y}=0.94\pm0.12, \alpha_{i}=1.15\pm0.03, \alpha_{R}=0.99\pm0.09, \alpha_{r}=1.13\pm0.03, \alpha_{I}=1.13\pm0.22, \alpha_{I_C}=1.06\pm0.13, \alpha_{H}=1.09\pm0.40, \alpha_{V}=0.67\pm0.10, \alpha_{z}=1.15\pm0.03, \alpha_{R_C}=1.10\pm0.14, \alpha_{CR}=1.04\pm0.15,$ and $\alpha_{F625W}=1.43\pm0.07$. These values are all compatible in 1$\sigma$ and this is the same outcome that we obtain with our approach of fitting the rescaling factors as a function of time. We note that the average value of the slope of GRB 221009 is different in some filters due to the different number of data points, however, the results are compatible within the uncertainties.

\normalfont
\vspace{-0.8cm}
\section{Optical Repository Webpage}\label{sec:website}
Here, we introduce the framework for the dedicated webpage developed to facilitate the analysis of optical LCs. Following a similar approach to the repositories established by \citet{Evans2009}, our webpage is designed as a powerful and efficient resource for the community. It enables users to view and interact with LCs and provides valuable insights into their colour evolution by utilizing modules developed as integral components of the \textit{grbLC} package (see Section \ref{sec:grblc}).
Currently, the repository comprises 535 LCs, encompassing the magnitude files corrected for Galactic extinction and transformed into the AB system, and the initially gathered raw data, which users can conveniently download.
The webpage infrastructure is constructed using \textit{streamlit}, a {\sc python}-based framework for generating web-based data visualization applications with minimal reliance on web technologies. The seamless integration of \textit{streamlit} with other {\sc python} packages has streamlined the development of an interactive web application based on \textit{grbLC}, allowing a fully functional online interactive user experience. Furthermore, the expandable and customizable nature of \textit{streamlit} enables us to iterate between improved versions of the webpage readily.
Shortly, our webpage is set to expand its scope by incorporating additional observed optical LCs from the literature and private sources that may not be present currently. Interested users can contact the corresponding author and contribute their materials - be they codes or data - as part of a potential follow-up to this catalogue.

\subsection{The \textit{grb}\textit{LC} package}
\label{sec:grblc}
The \textit{grbLC} package is an open-source {\sc python}-based program designed to handle photometric data from various telescopes worldwide. The package associates the telescope information in the GRB magnitude files with our comprehensive table of telescope calibration data. This data encompasses details such as effective wavelengths, zero point flux densities of filters, and conversion values for magnitude system transformations. The package is comprised of three modules for scraping GCNs, conversion in uniform units, and colour evolution, respectively.
Currently, \textit{grbLC} can be manually built from the source, accessible on GitHub at the link \url{https://github.com/SLAC-Gamma-Rays/grbLC}, by Mac, Linux and Windows users. We provide the users with detailed documentation on GitHub to walk them through the package, including the installation instructions.

\begin{figure*}
    \centering
    \includegraphics[scale=0.40]{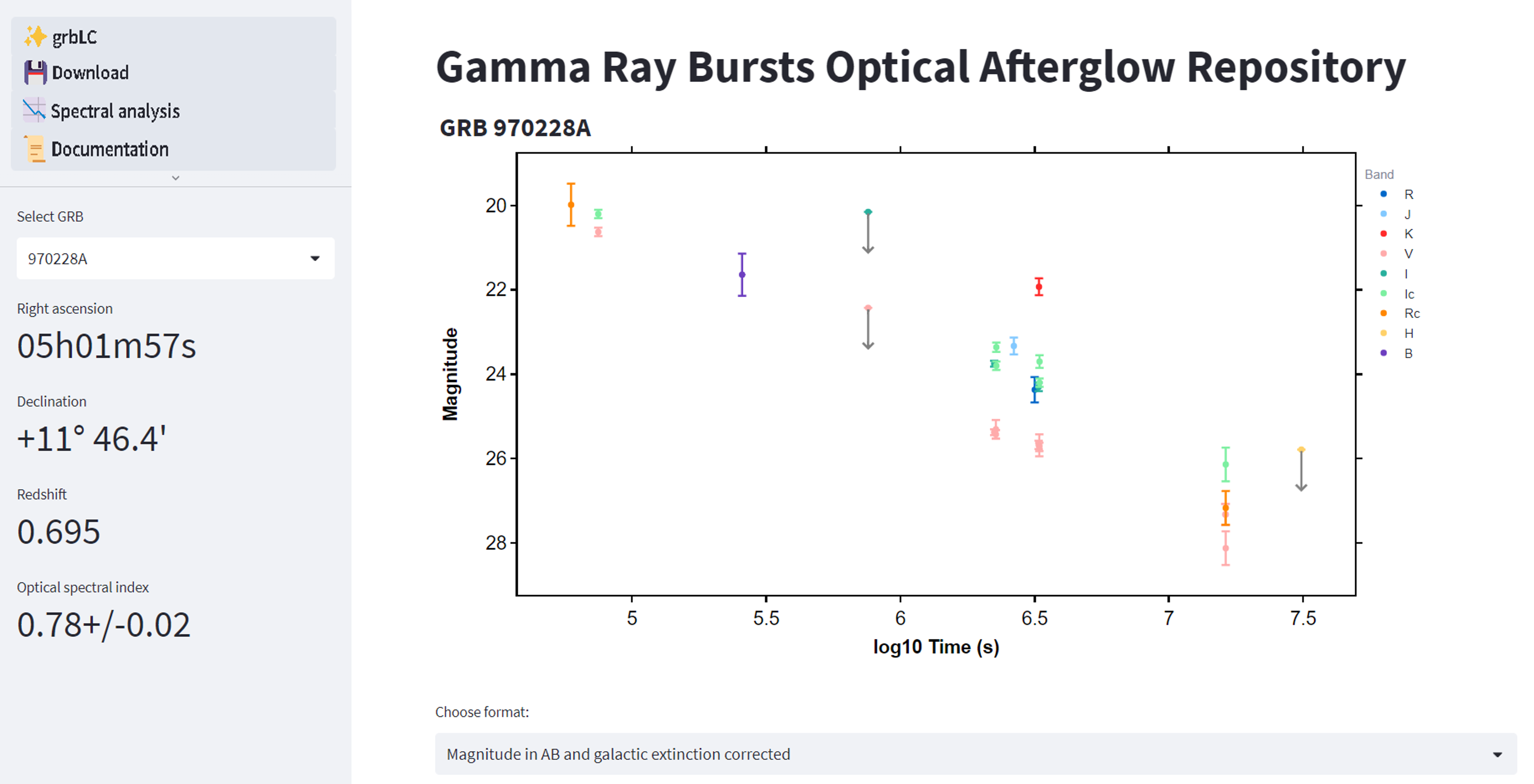}
    \caption{An example of the \textit{grbLC} main page for GRB 970228A. The plot shows the LC in magnitudes versus the $\log_{10}$ of the midtime of the observations (in seconds) after the trigger.
    The points with error bars indicate the magnitudes, while the points with downward grey arrows indicate the limiting magnitudes. The GRB can be selected from the drop-down menu on the left. The GRBs' right ascension, declination, redshift, and spectral index are displayed on the left. In the Colour Evolution menu, it is possible to visualize the filter fitting parameters and GRB rescaling factors. GCN and ADS searches enable searches on the NASA ADS website, providing access to the results about the required GRB. The magnitude file for the shown GRB can be downloaded as a {\tt .txt} file and stored using Download. The magnitudes in the original format or those in AB corrected for galactic extinction can be switched between using the Choose format. The filters to be shown can be chosen by clicking on them.}
    \label{fig:webappmainpage}
\end{figure*}

In addition to hosting our catalogue, the webpage provides an alternative user-friendly way to interact with the \textit{grbLC} package. The primary page provides essential information about a given GRB, including its RA, DEC, redshift ($z$), optical spectral index ($\beta_\mathrm{opt}$), host extinction correction ($A^{host}_V$) along with the dust model, and plots of magnitudes. Users can toggle between plots of the original magnitudes collected from literature and the homogenised version where all magnitudes are shifted to the AB system and corrected for Galactic extinction. By using the user-friendly interface menu, users can explore the colour evolution behaviour of a GRB and download all magnitude values in a machine-readable \textit{.txt} file. Figure~\ref{fig:webappmainpage} depicts an example of this interface.  A demonstration of the web repository can be viewed at \url{https://youtu.be/3n1jtJH4uGU}

\subsection{The GRB data scraper}\label{sec:scraper}
To streamline the compilation of our data sample, we created a web scraper that automates the GCN search process for pertinent information about a specific GRB. A major part of this module is developed with the {\sc Julia} Programming Language-based for efficient data scrapping.
This web scraper employs reading techniques to parse and segregate data for our sample automatically. 
Like a "scavenger hunt," the process encompasses two primary phases: manual data collection and coding automation for future endeavours.The manual data collection phase involves crawling GCNs and published data sources to gather relevant information. This phase helps us find GRB identifiers and keywords related to each telescope to help automation. Based on this information, our data scraper can conduct searches by employing regular expression (regex) packages to filter GCNs containing the specified keyword. By inputting the GRB name, the web scraper compiles all associated GCNs and stores them in a user-accessible \textit{.txt} file.
Furthermore, the data scraper includes a feature that extends its search to the NASA Astrophysics Data System (ADS) catalogue. This functionality allows users to search for papers related to a designated GRB and keyword. The GCN scraping process is conducted on the local machine, working on already downloaded \textit{.txt} files, while the relevant papers are efficiently downloaded in \textit{.pdf} and compressed into \textit{.gz} format to the user's computer, facilitating a swifter and more streamlined access to published data. In the current version, the data scraper can identify more than 32500 optical data points from the following telescopes and networks: UVOT, MASTER, RATIR, GROND, KAIT, IKI-FuN GRBs. We stress that the development of this tool is an ongoing project and the plan is to include all the optical telescopes reported in the literature.

\subsection{The magnitude conversion}\label{sec:conversion}
Standardization becomes crucial when dealing with a diverse data sample from various sources to accommodate the different reporting formats. The primary variables requiring attention are the correction for Galactic extinction and the magnitude system. This standardization process is the core function of the conversion module.
The Galactic extinction correction is implemented based on Equation~\ref{eq:magABextcorr}. The module utilizes the {\sc python} package \textit{dustmaps} to retrieve the colour excess ($E_{B-V}$) from \cite{Schlegel1998}. The conversion coefficient ($g_{X}$) for a specific band is sourced from \cite{Schlafly2011} or the Asiago Database on Photometric Systems \citep[ADPS,][]{AsiagoADPS} if the former lacks the band information.
The package also facilitates the conversion between Vega and AB magnitude systems. After the Vega$\rightarrow$AB shift, the module allows for the application of $k$-correction (utilizing the optical spectral index from literature) and the host galaxy extinction correction, following Equation \ref{eq:maghostkcorr}. The extinction maps presented in \citet{Pei1992} for the three dust models (SMC, LMC, and MW) are used alongside the measured $A^{host}_V$ value from the literature. These conversions can be executed using the \texttt{convertGRB} function.
\vspace{-0.3cm}
\subsection{The colour evolution}\label{sec:colorevolutionmodule}
The evolution module is dedicated to facilitating our colour evolution studies. Once a \texttt{Lightcurve} object is defined, users can visually explore the LC using the \texttt{displayGRB} function. This function generates an interactive plot using \textit{Plotly}, featuring a colour map based on the observation bands.
For a comprehensive colour evolution analysis, the \texttt{colorevolGRB} function comes into play. As detailed in Section \ref{sec:scaling}, this function utilizes a specific methodology and provides users with a \textit{Pandas} data frame. The data frame encompasses crucial information about the fit results of rescaling factor slopes for each filter and indications of whether colour evolution is detected. Additionally, the function generates a \textit{Matplotlib} plot illustrating the rescaling factor slope.
A significant aspect of this module is its ability to list the names of GRBs where colour evolution is identified or absent. 

\vspace{-0.7cm}
\section{Conclusions}\label{sec:conclusions}
We introduce a compilation of 535 optical LCs with known redshifts, including our publicly accessible web-based LC repository. The Online Materials show each singular LC and spectral indices. This compilation encompasses 64813 data points gathered from literature, circular notices, and private communications.
We computed the optical spectral index for 138 GRBs and investigated the presence of colour evolution. This analysis lays the groundwork for the subsequent rescaling of GRB LCs, a procedure feasible only in cases without colour evolution. Our results reveal that the colour evolution scenario, whether present or absent, has been determined for 173 GRBs. To reach this conclusion, we have provided a new homogeneous analysis for all LCs in the repository, which comprises the analysis for the colour evolution and spectral indices. We also compared our analysis and results to the ones provided in the literature.

In particular, we identified 84 cases of agreement with previous studies and 23 instances of disagreement by fitting the rescaling factors with fixed slope $a=0$. Among the disagreement cases, it is worth noting that the disparity could arise from differences in the choice of the reference filter for estimating rescaling factors, where the literature may not necessarily consider the most numerous filter within the GRB as the reference. Another potential difference lies in assuming a $0.025$ precision level for two epochs to be considered coincident. Besides the comprehensive analysis of the colour evolution, we have developed a user-friendly software suite for analyzing GRB LCs, offering various capabilities. The software can efficiently convert magnitudes reported in the literature to the AB system, correct for Galactic extinction, and incorporate $k$-correction and host galaxy correction, where applicable. Additionally, we have created a web scraper tool that can search through GCNs and the NASA ADS to extract relevant data related to specific GRBs or keywords. This powerful tool streamlines data gathering for studies involving extensive datasets.
We also provide detailed documentation of the functions included in the \textit{grbLC} package in this study. The web scraper tool is also available within this webpage, alongside the option to conveniently download available data for specific LCs.
Given the wide variety of formats in the available data, this painstaking data gathering procedure started with this open-source project, and it is still ongoing as additional refinements need to be added.
We hope the community will join the effort and contribute via the GitHub repository, enhancing this package and its future use.

\vspace{-0.4cm}
\section{Data availability}\label{sec:usage}
The data gathered in this catalogue was obtained either by public sources, in which case the published work is cited; GCN circulars, which have their number cited; or private communications. 
The open-source compiled catalogue and its data are available in \textit{grbLC} at \url{https://github.com/SLAC-Gamma-Rays/grbLC}.
The \textit{grbLC} package and web-based repository are designed to be open-access and available to all community members. 
The data will be available after the acceptance of this manuscript. All are welcome to use our software, though we ask that if the provided data or software is used in any publication, the authors cite this paper and include the following statement in the acknowledgements: 
"Data used in our work is taken from the catalogue Dainotti et al. (2024), and the original data sources are cited within." All the photometric data for the 535 GRBs weigh $5.64$ MB, while the literature information about GRB properties is stored in a table of $100$ KB.

\vspace{-0.8cm}
\section{Acknowledgements}\label{sec:acknowledgements}
We acknowledge Elena Zaninoni for kindly providing part of the data. We acknowledge the support of Shubham Bhardwaj, Angana Chakraborty, Sai Bhargavi Gangula, Gowri Govindaraj, Andrea Boria Denis, Reinaldo Aponte, Swati Shreya, Swapnil Singh, Snehadeep Kumar, Suman Sahu, Sukhjit Singh, and Deeshani Mitra in the data check and collection. We acknowledge the contribution of Sam Young to the building of the web tools. We acknowledge the contribution of Agnieszka Pollo and Aditya Narendra to the discussion about the colour evolution analysis. We are grateful to Massimiliano De Pasquale for the precious suggestions about the analysis of UVOT data and to Alice Breweld and Paul Kuin for the in-depth discussion about colour evolution. We are grateful to Elena Pian for her contribution to the discussion of the spectral analysis. NF acknowledges financial support from UNAM-DGAPA-PAPIIT  through grant IN106521. BG acknowledges support from the Australian Research Council Centre of Excellence for Gravitational Wave Discovery (OzGrav), under project number CE170100004. AP, AV, SB acknowledge support by the Russian Science Foundation (RSF) grant 23-12-00220. RLB acknowledges support from the CONAHCyT postdoctoral fellowship.


\vspace{-0.7cm}
\bibliographystyle{mnras}
\bibliography{webpage} 

\begin{thebibliography}{}
\makeatletter
\relax
\def\mn@urlcharsother{\let\do\@makeother \do\$\do\&\do\#\do\^\do\_\do\%\do\~}
\def\mn@doi{\begingroup\mn@urlcharsother \@ifnextchar [ {\mn@doi@} {\mn@doi@[]}}
\def\mn@doi@[#1]#2{\def\@tempa{#1}\ifx\@tempa\@empty \href {http://dx.doi.org/#2} {doi:#2}\else \href {http://dx.doi.org/#2} {#1}\fi \endgroup}
\def\mn@eprint#1#2{\mn@eprint@#1:#2::\@nil}
\def\mn@eprint@arXiv#1{\href {http://arxiv.org/abs/#1} {{\tt arXiv:#1}}}
\def\mn@eprint@dblp#1{\href {http://dblp.uni-trier.de/rec/bibtex/#1.xml} {dblp:#1}}
\def\mn@eprint@#1:#2:#3:#4\@nil{\def\@tempa {#1}\def\@tempb {#2}\def\@tempc {#3}\ifx \@tempc \@empty \let \@tempc \@tempb \let \@tempb \@tempa \fi \ifx \@tempb \@empty \def\@tempb {arXiv}\fi \@ifundefined {mn@eprint@\@tempb}{\@tempb:\@tempc}{\expandafter \expandafter \csname mn@eprint@\@tempb\endcsname \expandafter{\@tempc}}}

\bibitem[\protect\citeauthoryear{{Ackermann} et~al.,}{{Ackermann} et~al.}{2010}]{2010ApJ...716.1178A}
{Ackermann} M.,  et~al., 2010, \mn@doi [\apj] {10.1088/0004-637X/716/2/1178}, \href {https://ui.adsabs.harvard.edu/abs/2010ApJ...716.1178A} {716, 1178}

\bibitem[\protect\citeauthoryear{{Adkins} et~al.,}{{Adkins} et~al.}{2014}]{2014SPIE.9147E..03A}
{Adkins} S.~M.,  et~al., 2014, in {Ramsay} S.~K.,  {McLean} I.~S.,   {Takami} H.,  eds,  SPIE Conference Series Vol. 9147, Ground-based and Airborne Instrumentation for Astronomy V. p. 914703, \mn@doi{10.1117/12.2056971}

\bibitem[\protect\citeauthoryear{{Agayeva} et~al.,}{{Agayeva} et~al.}{2021}]{GRANDMA2021RMxAC..53..198A}
{Agayeva} S.,  et~al., 2021, in RMxAA Conference Series. pp 198--205 (\mn@eprint {arXiv} {2008.03962}), \mn@doi{10.22201/ia.14052059p.2021.53.39}

\bibitem[\protect\citeauthoryear{{Amati}, {Frontera}  \& {Guidorzi}}{{Amati} et~al.}{2009}]{2009A&A...508..173A}
{Amati} L.,  {Frontera} F.,   {Guidorzi} C.,  2009, \mn@doi [\aap] {10.1051/0004-6361/200912788}, \href {https://ui.adsabs.harvard.edu/abs/2009A&A...508..173A} {508, 173}

\bibitem[\protect\citeauthoryear{{Arnold}, {Steele}, {Bates}, {Mottram}  \& {Smith}}{{Arnold} et~al.}{2012}]{2012SPIE.8446E..2JA}
{Arnold} D.~M.,  {Steele} I.~A.,  {Bates} S.~D.,  {Mottram} C.~J.,   {Smith} R.~J.,  2012, in {McLean} I.~S.,  {Ramsay} S.~K.,   {Takami} H.,  eds,  SPIE Conference Series Vol. 8446, Ground-based and Airborne Instrumentation for Astronomy IV. p. 84462J, \mn@doi{10.1117/12.927000}

\bibitem[\protect\citeauthoryear{{Becerra} et~al.,}{{Becerra} et~al.}{2023a}]{Becerra2023}
{Becerra} R.~L.,  et~al., 2023a, \mn@doi [\mnras] {10.1093/mnras/stad1372}, \href {https://ui.adsabs.harvard.edu/abs/2023MNRAS.522.5204B} {522, 5204}

\bibitem[\protect\citeauthoryear{{Becerra} et~al.,}{{Becerra} et~al.}{2023b}]{Becerra2023b}
{Becerra} R.~L.,  et~al., 2023b, \mn@doi [\mnras] {10.1093/mnras/stad2513}, \href {https://ui.adsabs.harvard.edu/abs/2023MNRAS.525.3262B} {525, 3262}

\bibitem[\protect\citeauthoryear{{Beskin}, {Karpov}, {Bondar}, {Greco}, {Guarnieri}, {Bartolini}  \& {Piccioni}}{{Beskin} et~al.}{2010}]{Beskin2010ApJ}
{Beskin} G.,  {Karpov} S.,  {Bondar} S.,  {Greco} G.,  {Guarnieri} A.,  {Bartolini} C.,   {Piccioni} A.,  2010, \mn@doi [\apjl] {10.1088/2041-8205/719/1/L10}, \href {https://ui.adsabs.harvard.edu/abs/2010ApJ...719L..10B} {719, L10}

\bibitem[\protect\citeauthoryear{{Blagorodnova} et~al.,}{{Blagorodnova} et~al.}{2018}]{2018PASP..130c5003B}
{Blagorodnova} N.,  et~al., 2018, \mn@doi [\pasp] {10.1088/1538-3873/aaa53f}, \href {https://ui.adsabs.harvard.edu/abs/2018PASP..130c5003B} {130, 035003}

\bibitem[\protect\citeauthoryear{{Blake} et~al.,}{{Blake} et~al.}{2005}]{Blake2005Natur}
{Blake} C.~H.,  et~al., 2005, \mn@doi [\nat] {10.1038/nature03520}, \href {https://ui.adsabs.harvard.edu/abs/2005Natur.435..181B} {435, 181}

\bibitem[\protect\citeauthoryear{Blanton \& Roweis}{Blanton \& Roweis}{2007}]{Blanton2007}
Blanton M.~R.,  Roweis S.,  2007, \mn@doi [\aj] {10.1086/510127}, 133, 734

\bibitem[\protect\citeauthoryear{{Bloom}, {Starr}, {Blake}, {Skrutskie}  \& {Falco}}{{Bloom} et~al.}{2006}]{bloom2005autonomous}
{Bloom} J.~S.,  {Starr} D.~L.,  {Blake} C.~H.,  {Skrutskie} M.~F.,   {Falco} E.~E.,  2006, in {Gabriel} C.,  {Arviset} C.,  {Ponz} D.,   {Enrique} S.,  eds,  ASP Conference Series Vol. 351, Astronomical Data Analysis Software and Systems XV. p.~751 (\mn@eprint {arXiv} {astro-ph/0511842}), \mn@doi{10.48550/arXiv.astro-ph/0511842}

\bibitem[\protect\citeauthoryear{{Boer} et~al.,}{{Boer} et~al.}{2003}]{2003Msngr.113...45B}
{Boer} M.,  et~al., 2003, The Messenger, \href {https://ui.adsabs.harvard.edu/abs/2003Msngr.113...45B} {113, 45}

\bibitem[\protect\citeauthoryear{{Butler} et~al.,}{{Butler} et~al.}{2012}]{2012SPIE.8446E..10B}
{Butler} N.,  et~al., 2012, in {McLean} I.~S.,  {Ramsay} S.~K.,   {Takami} H.,  eds,  SPIE Conference Series Vol. 8446, Ground-based and Airborne Instrumentation for Astronomy IV. p. 844610, \mn@doi{10.1117/12.926471}

\bibitem[\protect\citeauthoryear{{Cano}, {Wang}, {Dai}  \& {Wu}}{{Cano} et~al.}{2017}]{Cano2017}
{Cano} Z.,  {Wang} S.-Q.,  {Dai} Z.-G.,   {Wu} X.-F.,  2017, \mn@doi [Advances in Astronomy] {10.1155/2017/8929054}, \href {https://ui.adsabs.harvard.edu/abs/2017AdAst2017E...5C} {2017, 8929054}

\bibitem[\protect\citeauthoryear{{Castro-Tirado} et~al.,}{{Castro-Tirado} et~al.}{1997}]{1997IAUC.6657....1C}
{Castro-Tirado} A.~J.,  et~al., 1997, \iaucirc, \href {https://ui.adsabs.harvard.edu/abs/1997IAUC.6657....1C} {6657, 1}

\bibitem[\protect\citeauthoryear{{Cenko} et~al.,}{{Cenko} et~al.}{2009}]{2009ApJ...693.1484C}
{Cenko} S.~B.,  et~al., 2009, \mn@doi [\apj] {10.1088/0004-637X/693/2/1484}, \href {https://ui.adsabs.harvard.edu/abs/2009ApJ...693.1484C} {693, 1484}

\bibitem[\protect\citeauthoryear{{Chuprakov}, {Eselevich}  \& {Korobtsev}}{{Chuprakov} et~al.}{2018}]{Chuprakov18}
{Chuprakov} S.~A.,  {Eselevich} M.~V.,   {Korobtsev} I.~V.,  2018, \mn@doi [Journal of Astronomical Telescopes, Instruments, and Systems] {10.1117/1.JATIS.4.2.024002}, \href {https://ui.adsabs.harvard.edu/abs/2018JATIS...4b4002C/abstract} {4}

\bibitem[\protect\citeauthoryear{{Costa} et~al.,}{{Costa} et~al.}{1997}]{1997Natur.387..783C}
{Costa} E.,  et~al., 1997, \mn@doi [\nat] {10.1038/42885}, \href {https://ui.adsabs.harvard.edu/abs/1997Natur.387..783C} {387, 783}

\bibitem[\protect\citeauthoryear{{Covino} et~al.,}{{Covino} et~al.}{2013}]{2013MNRAS.432.1231C}
{Covino} S.,  et~al., 2013, \mn@doi [\mnras] {10.1093/mnras/stt540}, \href {https://ui.adsabs.harvard.edu/abs/2013MNRAS.432.1231C} {432, 1231}

\bibitem[\protect\citeauthoryear{{Dainotti}, {Cardone}  \& {Capozziello}}{{Dainotti} et~al.}{2008}]{Dainotti2008}
{Dainotti} M.~G.,  {Cardone} V.~F.,   {Capozziello} S.,  2008, \mn@doi [\mnras] {10.1111/j.1745-3933.2008.00560.x}, \href {https://ui.adsabs.harvard.edu/abs/2008MNRAS.391L..79D} {391, L79}

\bibitem[\protect\citeauthoryear{{Dainotti}, {Willingale}, {Capozziello}, {Fabrizio Cardone}  \& {Ostrowski}}{{Dainotti} et~al.}{2010}]{Dainotti2010}
{Dainotti} M.~G.,  {Willingale} R.,  {Capozziello} S.,  {Fabrizio Cardone} V.,   {Ostrowski} M.,  2010, \mn@doi [\apjl] {10.1088/2041-8205/722/2/L215}, \href {https://ui.adsabs.harvard.edu/abs/2010ApJ...722L.215D} {722, L215}

\bibitem[\protect\citeauthoryear{{Dainotti}, {Fabrizio Cardone}, {Capozziello}, {Ostrowski}  \& {Willingale}}{{Dainotti} et~al.}{2011}]{Dainotti2011}
{Dainotti} M.~G.,  {Fabrizio Cardone} V.,  {Capozziello} S.,  {Ostrowski} M.,   {Willingale} R.,  2011, \mn@doi [\apj] {10.1088/0004-637X/730/2/135}, \href {https://ui.adsabs.harvard.edu/abs/2011ApJ...730..135D} {730, 135}

\bibitem[\protect\citeauthoryear{{Dainotti}, {Petrosian}, {Singal}  \& {Ostrowski}}{{Dainotti} et~al.}{2013}]{Dainotti2013}
{Dainotti} M.~G.,  {Petrosian} V.,  {Singal} J.,   {Ostrowski} M.,  2013, \mn@doi [\apj] {10.1088/0004-637X/774/2/157}, \href {https://ui.adsabs.harvard.edu/abs/2013ApJ...774..157D} {774, 157}

\bibitem[\protect\citeauthoryear{{Dainotti}, {Del Vecchio}, {Shigehiro}  \& {Capozziello}}{{Dainotti} et~al.}{2015}]{dainotti2015}
{Dainotti} M.~G.,  {Del Vecchio} R.,  {Shigehiro} N.,   {Capozziello} S.,  2015, \mn@doi [\apj] {10.1088/0004-637X/800/1/31}, \href {https://ui.adsabs.harvard.edu/abs/2015ApJ...800...31D} {800, 31}

\bibitem[\protect\citeauthoryear{{Dainotti}, {Postnikov}, {Hernandez}  \& {Ostrowski}}{{Dainotti} et~al.}{2016}]{Dainotti2016}
{Dainotti} M.~G.,  {Postnikov} S.,  {Hernandez} X.,   {Ostrowski} M.,  2016, \mn@doi [\apjl] {10.3847/2041-8205/825/2/L20}, \href {https://ui.adsabs.harvard.edu/abs/2016ApJ...825L..20D} {825, L20}

\bibitem[\protect\citeauthoryear{{Dainotti}, {Lenart}, {Sarracino}, {Nagataki}, {Capozziello}  \& {Fraija}}{{Dainotti} et~al.}{2020a}]{2020ApJ...904...97D}
{Dainotti} M.~G.,  {Lenart} A.~{\L}.,  {Sarracino} G.,  {Nagataki} S.,  {Capozziello} S.,   {Fraija} N.,  2020a, \mn@doi [\apj] {10.3847/1538-4357/abbe8a}, \href {https://ui.adsabs.harvard.edu/abs/2020ApJ...904...97D} {904, 97}

\bibitem[\protect\citeauthoryear{{Dainotti} et~al.,}{{Dainotti} et~al.}{2020b}]{2020ApJ...905L..26D}
{Dainotti} M.~G.,  et~al., 2020b, \mn@doi [\apjl] {10.3847/2041-8213/abcda9}, \href {https://ui.adsabs.harvard.edu/abs/2020ApJ...905L..26D} {905, L26}

\bibitem[\protect\citeauthoryear{{Dainotti} et~al.,}{{Dainotti} et~al.}{2021}]{dainotti2021c}
{Dainotti} M.~G.,  et~al., 2021, \mn@doi [\apjs] {10.3847/1538-4365/abfe17}, \href {https://ui.adsabs.harvard.edu/abs/2021ApJS..255...13D} {255, 13}

\bibitem[\protect\citeauthoryear{{Dainotti} et~al.,}{{Dainotti} et~al.}{2022a}]{2022ApJS..261...25D}
{Dainotti} M.~G.,  et~al., 2022a, \mn@doi [\apjs] {10.3847/1538-4365/ac7c64}, \href {https://ui.adsabs.harvard.edu/abs/2022ApJS..261...25D} {261, 25}

\bibitem[\protect\citeauthoryear{{Dainotti} et~al.,}{{Dainotti} et~al.}{2022b}]{2022arXiv220312908D}
{Dainotti} M.~G.,  et~al., 2022b, \mn@doi [\apjs] {10.3847/1538-4365/ac7c64}, \href {https://ui.adsabs.harvard.edu/abs/2022ApJS..261...25D} {261, 25}

\bibitem[\protect\citeauthoryear{{Davilla}, {Wood}, {Atcheson}, {Saunders}, {Sullivan}, {Vaughan}  \& {Saisse}}{{Davilla} et~al.}{1993}]{1993ApOpt..32.1775D}
{Davilla} P.,  {Wood} H.~J.,  {Atcheson} P.~D.,  {Saunders} R.,  {Sullivan} J.,  {Vaughan} A.~H.,   {Saisse} M.,  1993, \mn@doi [\ao] {10.1364/AO.32.001775}, \href {https://ui.adsabs.harvard.edu/abs/1993ApOpt..32.1775D} {32, 1775}

\bibitem[\protect\citeauthoryear{{Ehgamberdiev}, {Baijumanov}, {Ilyasov}, {Sarazin}, {Tillayev}, {Tokovinin}  \& {Ziad}}{{Ehgamberdiev} et~al.}{2000}]{Ehgamberdiev00}
{Ehgamberdiev} S.~A.,  {Baijumanov} A.~K.,  {Ilyasov} S.~P.,  {Sarazin} M.,  {Tillayev} Y.~A.,  {Tokovinin} A.~A.,   {Ziad} A.,  2000, \mn@doi [Astronomy and Astrophysics Supplement] {10.1051/aas:2000244}, \href {https://ui.adsabs.harvard.edu/abs/2000A%26AS..145..293E/abstract} {145, 293}

\bibitem[\protect\citeauthoryear{{Eichler}, {Livio}, {Piran}  \& {Schramm}}{{Eichler} et~al.}{1989}]{Eichler1989}
{Eichler} D.,  {Livio} M.,  {Piran} T.,   {Schramm} D.~N.,  1989, \mn@doi [\nat] {10.1038/340126a0}, \href {https://ui.adsabs.harvard.edu/abs/1989Natur.340..126E} {340, 126}

\bibitem[\protect\citeauthoryear{{Elenin}, {Kusakin}  \& {Molotov}}{{Elenin} et~al.}{2015}]{Elenin2015}
{Elenin} L.,  {Kusakin} A.,   {Molotov} I.,  2015, The Minor Planet Bulletin (ISSN 1052-8091). Bulletin of the Minor Planets Section of the Association of Lunar and Planetary Observers, \href {https://ui.adsabs.harvard.edu/abs/2015MPBu...42...27E/abstract} {42, 27}

\bibitem[\protect\citeauthoryear{{Evans}, {Beardmore}, {Goad}, {Osborne}, {Burrows}  \& {Gehrels}}{{Evans} et~al.}{2008}]{2008AIPC.1000..539E}
{Evans} P.~A.,  {Beardmore} A.~P.,  {Goad} M.~R.,  {Osborne} J.~P.,  {Burrows} D.~N.,   {Gehrels} N.,  2008, in {Galassi} M.,  {Palmer} D.,   {Fenimore} E.,  eds,  AIP Conference Series Vol. 1000, Gamma-ray Bursts 2007. pp 539--542 (\mn@eprint {arXiv} {0801.4462}), \mn@doi{10.1063/1.2943526}

\bibitem[\protect\citeauthoryear{{Evans} et~al.,}{{Evans} et~al.}{2009}]{Evans2009}
{Evans} P.~A.,  et~al., 2009, \mn@doi [\mnras] {10.1111/j.1365-2966.2009.14913.x}, \href {https://ui.adsabs.harvard.edu/abs/2009MNRAS.397.1177E} {397, 1177}

\bibitem[\protect\citeauthoryear{{Fenimore} \& {Ramirez-Ruiz}}{{Fenimore} \& {Ramirez-Ruiz}}{2000}]{2000astro.ph..4176F}
{Fenimore} E.~E.,  {Ramirez-Ruiz} E.,  2000, \mn@doi [arXiv e-prints] {10.48550/arXiv.astro-ph/0004176}, \href {https://ui.adsabs.harvard.edu/abs/2000astro.ph..4176F} {pp astro--ph/0004176}

\bibitem[\protect\citeauthoryear{{Fiorucci, M.} \& {Munari, U.}}{{Fiorucci, M.} \& {Munari, U.}}{2003}]{AsiagoADPS}
{Fiorucci, M.} {Munari, U.} 2003, \mn@doi [A\&A] {10.1051/0004-6361:20030075}, 401, 781

\bibitem[\protect\citeauthoryear{{Fraija}, {Laskar}, {Dichiara}, {Beniamini}, {Duran}, {Dainotti}  \& {Becerra}}{{Fraija} et~al.}{2020}]{2020ApJ...905..112F}
{Fraija} N.,  {Laskar} T.,  {Dichiara} S.,  {Beniamini} P.,  {Duran} R.~B.,  {Dainotti} M.~G.,   {Becerra} R.~L.,  2020, \mn@doi [\apj] {10.3847/1538-4357/abc41a}, \href {https://ui.adsabs.harvard.edu/abs/2020ApJ...905..112F} {905, 112}

\bibitem[\protect\citeauthoryear{{Fraija}, {Dainotti}, {Ugale}, {Jyoti}  \& {Warren}}{{Fraija} et~al.}{2022}]{2022ApJ...934..188F}
{Fraija} N.,  {Dainotti} M.~G.,  {Ugale} S.,  {Jyoti} D.,   {Warren} D.~C.,  2022, \mn@doi [\apj] {10.3847/1538-4357/ac7a9c}, \href {https://ui.adsabs.harvard.edu/abs/2022ApJ...934..188F} {934, 188}

\bibitem[\protect\citeauthoryear{{Fraija}, {Dainotti}, {Kamenetskaia}, {Levine}  \& {Galvan-Gamez}}{{Fraija} et~al.}{2023}]{2023MNRAS.tmp.2216F}
{Fraija} N.,  {Dainotti} M.~G.,  {Kamenetskaia} B.~B.,  {Levine} D.,   {Galvan-Gamez} A.,  2023, \mn@doi [\mnras] {10.1093/mnras/stad2236}, \href {https://ui.adsabs.harvard.edu/abs/2023MNRAS.tmp.2216F} {}

\bibitem[\protect\citeauthoryear{{Gehrels} et~al.,}{{Gehrels} et~al.}{2004}]{2004ApJ...611.1005G}
{Gehrels} N.,  et~al., 2004, \mn@doi [\apj] {10.1086/422091}, \href {https://ui.adsabs.harvard.edu/abs/2004ApJ...611.1005G} {611, 1005}

\bibitem[\protect\citeauthoryear{{Gehrels}, {Ramirez-Ruiz}  \& {Fox}}{{Gehrels} et~al.}{2009}]{Gehrels2009ARA&A}
{Gehrels} N.,  {Ramirez-Ruiz} E.,   {Fox} D.~B.,  2009, \mn@doi [\araa] {10.1146/annurev.astro.46.060407.145147}, \href {https://ui.adsabs.harvard.edu/abs/2009ARA&A..47..567G} {47, 567}

\bibitem[\protect\citeauthoryear{{Ghirlanda}, {Ghisellini}  \& {Lazzati}}{{Ghirlanda} et~al.}{2004}]{2004ApJ...616..331G}
{Ghirlanda} G.,  {Ghisellini} G.,   {Lazzati} D.,  2004, \mn@doi [\apj] {10.1086/424913}, \href {https://ui.adsabs.harvard.edu/abs/2004ApJ...616..331G} {616, 331}

\bibitem[\protect\citeauthoryear{{Ghirlanda}, {Ghisellini}  \& {Firmani}}{{Ghirlanda} et~al.}{2006}]{2006NJPh....8..123G}
{Ghirlanda} G.,  {Ghisellini} G.,   {Firmani} C.,  2006, \mn@doi [New Journal of Physics] {10.1088/1367-2630/8/7/123}, \href {https://ui.adsabs.harvard.edu/abs/2006NJPh....8..123G} {8, 123}

\bibitem[\protect\citeauthoryear{{Goad} et~al.,}{{Goad} et~al.}{2007}]{2007A&A...476.1401G}
{Goad} M.~R.,  et~al., 2007, \mn@doi [\aap] {10.1051/0004-6361:20078436}, \href {https://ui.adsabs.harvard.edu/abs/2007A&A...476.1401G} {476, 1401}

\bibitem[\protect\citeauthoryear{{Greiner} et~al.,}{{Greiner} et~al.}{2008}]{2008PASP..120..405G}
{Greiner} J.,  et~al., 2008, \mn@doi [\pasp] {10.1086/587032}, \href {https://ui.adsabs.harvard.edu/abs/2008PASP..120..405G} {120, 405}

\bibitem[\protect\citeauthoryear{{Greiner} et~al.,}{{Greiner} et~al.}{2015}]{Greiner2015}
{Greiner} J.,  et~al., 2015, \mn@doi [\apj] {10.1088/0004-637X/809/1/76}, \href {https://ui.adsabs.harvard.edu/abs/2015ApJ...809...76G} {809, 76}

\bibitem[\protect\citeauthoryear{{Hentunen}, {Nissinen}  \& {Salmi}}{{Hentunen} et~al.}{2016}]{2016GCN.18960....1H}
{Hentunen} V.~P.,  {Nissinen} M.,   {Salmi} T.,  2016, GRB Coordinates Network, \href {https://ui.adsabs.harvard.edu/abs/2016GCN.18960....1H} {18960, 1}

\bibitem[\protect\citeauthoryear{{Hoang}, {Giang}  \& {Tram}}{{Hoang} et~al.}{2020}]{2020ApJ...895...16H}
{Hoang} T.,  {Giang} N.~C.,   {Tram} L.~N.,  2020, \mn@doi [\apj] {10.3847/1538-4357/ab8ae1}, \href {https://ui.adsabs.harvard.edu/abs/2020ApJ...895...16H} {895, 16}

\bibitem[\protect\citeauthoryear{{Hu} et~al.,}{{Hu} et~al.}{2023}]{Youdong2023}
{Hu} Y.~D.,  et~al., 2023, \mn@doi [Frontiers in Astronomy and Space Sciences] {10.3389/fspas.2023.952887}, \href {https://ui.adsabs.harvard.edu/abs/2023FrASS..10.2887H} {10, 952887}

\bibitem[\protect\citeauthoryear{{Izzo} et~al.,}{{Izzo} et~al.}{2019}]{2019Natur.565..324I}
{Izzo} L.,  et~al., 2019, \mn@doi [\nat] {10.1038/s41586-018-0826-3}, \href {https://ui.adsabs.harvard.edu/abs/2019Natur.565..324I} {565, 324}

\bibitem[\protect\citeauthoryear{{Jakobsson}, {Hjorth}, {Fynbo}, {Watson}, {Pedersen}, {Bj{\"o}rnsson}  \& {Gorosabel}}{{Jakobsson} et~al.}{2004}]{Jakobsson2004}
{Jakobsson} P.,  {Hjorth} J.,  {Fynbo} J.~P.~U.,  {Watson} D.,  {Pedersen} K.,  {Bj{\"o}rnsson} G.,   {Gorosabel} J.,  2004, \mn@doi [\apjl] {10.1086/427089}, \href {https://ui.adsabs.harvard.edu/abs/2004ApJ...617L..21J} {617, L21}

\bibitem[\protect\citeauthoryear{{Jel{\'\i}nek} et~al.,}{{Jel{\'\i}nek} et~al.}{2016}]{jelinekBOOTES}
{Jel{\'\i}nek} M.,  et~al., 2016, \mn@doi [Advances in Astronomy] {10.1155/2016/1928465}, \href {https://ui.adsabs.harvard.edu/abs/2016AdAst2016E..12J} {2016, 192846}

\bibitem[\protect\citeauthoryear{{Kann}, {Klose}  \& {Zeh}}{{Kann} et~al.}{2006}]{kann2006signatures}
{Kann} D.~A.,  {Klose} S.,   {Zeh} A.,  2006, \mn@doi [\apj] {10.1086/500652}, \href {https://ui.adsabs.harvard.edu/abs/2006ApJ...641..993K} {641, 993}

\bibitem[\protect\citeauthoryear{{Kann} et~al.,}{{Kann} et~al.}{2010}]{2010ApJ...720.1513K}
{Kann} D.~A.,  et~al., 2010, \mn@doi [\apj] {10.1088/0004-637X/720/2/1513}, \href {https://ui.adsabs.harvard.edu/abs/2010ApJ...720.1513K} {720, 1513}

\bibitem[\protect\citeauthoryear{{Kann} et~al.,}{{Kann} et~al.}{2011}]{2011ApJ...734...96K}
{Kann} D.~A.,  et~al., 2011, \mn@doi [\apj] {10.1088/0004-637X/734/2/96}, \href {https://ui.adsabs.harvard.edu/abs/2011ApJ...734...96K} {734, 96}

\bibitem[\protect\citeauthoryear{Kann et~al.,}{Kann et~al.}{2023}]{Kann2023BOAT}
Kann D.~A.,  et~al., 2023, \mn@doi [\apjl] {10.3847/2041-8213/acc8d0}, 948, L12

\bibitem[\protect\citeauthoryear{{Kann} et~al.,}{{Kann} et~al.}{2024}]{2024arXiv240300101K}
{Kann} D.~A.,  et~al., 2024, \mn@doi [arXiv e-prints] {10.48550/arXiv.2403.00101}, \href {https://ui.adsabs.harvard.edu/abs/2024arXiv240300101K} {p. arXiv:2403.00101}

\bibitem[\protect\citeauthoryear{{Khetsuriani}}{{Khetsuriani}}{1967}]{Khetsuriani67}
{Khetsuriani} T.~S.,  1967, \mn@doi [Solar Physics] {10.1007/BF00155927}, \href {https://ui.adsabs.harvard.edu/abs/1967SoPh....2..237K/abstract} {2, 237}

\bibitem[\protect\citeauthoryear{{Kochanek} \& {Piran}}{{Kochanek} \& {Piran}}{1993}]{Kochanek93}
{Kochanek} C.~S.,  {Piran} T.,  1993, \mn@doi [\apjl] {10.1086/187083}, \href {https://ui.adsabs.harvard.edu/abs/1993ApJ...417L..17K} {417, L17}

\bibitem[\protect\citeauthoryear{{Kodaira}}{{Kodaira}}{1992}]{1992ESOC...42...43K}
{Kodaira} K.,  1992, in ESO Conference and Workshop Proceedings. p.~43

\bibitem[\protect\citeauthoryear{{Kohama} et~al.,}{{Kohama} et~al.}{2001}]{Kohama2002}
{Kohama} M.,  et~al., 2001, in {Inoue} H.,  {Kunieda} H.,  eds,  ASP Conference Series Vol. 251, New Century of X-ray Astronomy. p.~558

\bibitem[\protect\citeauthoryear{{Kouveliotou}, {Meegan}, {Fishman}, {Bhat}, {Briggs}, {Koshut}, {Paciesas}  \& {Pendleton}}{{Kouveliotou} et~al.}{1993}]{kouveliotou1993identification}
{Kouveliotou} C.,  {Meegan} C.~A.,  {Fishman} G.~J.,  {Bhat} N.~P.,  {Briggs} M.~S.,  {Koshut} T.~M.,  {Paciesas} W.~S.,   {Pendleton} G.~N.,  1993, \mn@doi [\apjl] {10.1086/186969}, \href {https://ui.adsabs.harvard.edu/abs/1993ApJ...413L.101K} {413, L101}

\bibitem[\protect\citeauthoryear{{Kuin}, {Sbarufatti}, {Marshall}  \& {Schady}}{{Kuin} et~al.}{2008}]{2008GCN..7844....1K}
{Kuin} N.~P.~M.,  {Sbarufatti} B.,  {Marshall} F.,   {Schady} P.,  2008, GRB Coordinates Network, \href {https://ui.adsabs.harvard.edu/abs/2008GCN..7844....1K} {7844, 1}

\bibitem[\protect\citeauthoryear{Levenberg}{Levenberg}{1944}]{Levenberg1944AMF}
Levenberg K.,  1944, Quarterly of Applied Mathematics, 2, 164

\bibitem[\protect\citeauthoryear{{Levine}, {Dainotti}, {Zvonarek}, {Fraija}, {Warren}, {Chandra}  \& {Lloyd-Ronning}}{{Levine} et~al.}{2022}]{2022ApJ...925...15L}
{Levine} D.,  {Dainotti} M.,  {Zvonarek} K.~J.,  {Fraija} N.,  {Warren} D.~C.,  {Chandra} P.,   {Lloyd-Ronning} N.,  2022, \mn@doi [\apj] {10.3847/1538-4357/ac4221}, \href {https://ui.adsabs.harvard.edu/abs/2022ApJ...925...15L} {925, 15}

\bibitem[\protect\citeauthoryear{{Li} et~al.,}{{Li} et~al.}{2012}]{Li2012}
{Li} L.,  et~al., 2012, \mn@doi [\apj] {10.1088/0004-637X/758/1/27}, \href {https://ui.adsabs.harvard.edu/abs/2012ApJ...758...27L} {758, 27}

\bibitem[\protect\citeauthoryear{{Li} et~al.,}{{Li} et~al.}{2015}]{Li2015}
{Li} L.,  et~al., 2015, \mn@doi [\apj] {10.1088/0004-637X/805/1/13}, \href {https://ui.adsabs.harvard.edu/abs/2015ApJ...805...13L} {805, 13}

\bibitem[\protect\citeauthoryear{{Li}, {Wang}, {Shao}, {Wu}, {Huang}, {Zhang}, {Ryde}  \& {Yu}}{{Li} et~al.}{2018}]{Li2018a}
{Li} L.,  {Wang} Y.,  {Shao} L.,  {Wu} X.-F.,  {Huang} Y.-F.,  {Zhang} B.,  {Ryde} F.,   {Yu} H.-F.,  2018, \mn@doi [\apjs] {10.3847/1538-4365/aaa02a}, \href {https://ui.adsabs.harvard.edu/abs/2018ApJS..234...26L} {234, 26}

\bibitem[\protect\citeauthoryear{{Lian}, {Wang}, {Gan}, {Li}  \& {Liang}}{{Lian} et~al.}{2022}]{2022ApJ...931...90L}
{Lian} J.-S.,  {Wang} S.-Q.,  {Gan} W.-P.,  {Li} J.-Y.,   {Liang} E.-W.,  2022, \mn@doi [\apj] {10.3847/1538-4357/ac69db}, \href {https://ui.adsabs.harvard.edu/abs/2022ApJ...931...90L} {931, 90}

\bibitem[\protect\citeauthoryear{{Lloyd} \& {Petrosian}}{{Lloyd} \& {Petrosian}}{1999}]{1999ApJ...511..550L}
{Lloyd} N.~M.,  {Petrosian} V.,  1999, \mn@doi [\apj] {10.1086/306719}, \href {https://ui.adsabs.harvard.edu/abs/1999ApJ...511..550L} {511, 550}

\bibitem[\protect\citeauthoryear{{Maksut} \& {Grossan}}{{Maksut} \& {Grossan}}{2021}]{2021RMxAC..53..169M}
{Maksut} Z.,  {Grossan} B.,  2021, in RMxAA Conference Series. pp 169--173, \mn@doi{10.22201/ia.14052059p.2021.53.35}

\bibitem[\protect\citeauthoryear{Mazets et~al.,}{Mazets et~al.}{1981}]{mazets1981catalog}
Mazets E.,  et~al., 1981, Astrophysics and Space Science, 80, 3

\bibitem[\protect\citeauthoryear{{Melandri} et~al.,}{{Melandri} et~al.}{2019}]{2019MNRAS.490.5366M}
{Melandri} A.,  et~al., 2019, \mn@doi [\mnras] {10.1093/mnras/stz2900}, \href {https://ui.adsabs.harvard.edu/abs/2019MNRAS.490.5366M} {490, 5366}

\bibitem[\protect\citeauthoryear{{Metzger}, {Giannios}, {Thompson}, {Bucciantini}  \& {Quataert}}{{Metzger} et~al.}{2011}]{2011MNRAS.413.2031M}
{Metzger} B.~D.,  {Giannios} D.,  {Thompson} T.~A.,  {Bucciantini} N.,   {Quataert} E.,  2011, \mn@doi [\mnras] {10.1111/j.1365-2966.2011.18280.x}, \href {https://ui.adsabs.harvard.edu/abs/2011MNRAS.413.2031M} {413, 2031}

\bibitem[\protect\citeauthoryear{{Minaev} \& {Pozanenko}}{{Minaev} \& {Pozanenko}}{2020}]{Minaev2020}
{Minaev} P.~Y.,  {Pozanenko} A.~S.,  2020, \mn@doi [\mnras] {10.1093/mnras/stz3611}, \href {https://ui.adsabs.harvard.edu/abs/2020MNRAS.492.1919M} {492, 1919}

\bibitem[\protect\citeauthoryear{Morgan et~al.,}{Morgan et~al.}{2014}]{Morgan_2014}
Morgan A.~N.,  et~al., 2014, \mn@doi [\mnras] {10.1093/mnras/stu344}, 440, 1810

\bibitem[\protect\citeauthoryear{{Morris} et~al.,}{{Morris} et~al.}{1997}]{1997IAUC.6666....1M}
{Morris} M.,  et~al., 1997, \iaucirc, \href {https://ui.adsabs.harvard.edu/abs/1997IAUC.6666....1M} {6666, 1}

\bibitem[\protect\citeauthoryear{{Morris}, {Gendre}, {Orange}, {Cucchiara}, {Giblin}, {Klotz}  \& {Thierry}}{{Morris} et~al.}{2018}]{MorrisVIRT}
{Morris} D.~C.,  {Gendre} B.,  {Orange} N.~B.,  {Cucchiara} A.,  {Giblin} T.~W.,  {Klotz} A.,   {Thierry} P.,  2018, in AAS Meeting Abstracts \#231. p. 442.07

\bibitem[\protect\citeauthoryear{Murdin}{Murdin}{2000}]{Abastumani}
Murdin P.,  2000, \mn@doi [Encyclopedia of Astronomy and Astrophysics] {10.1888/0333750888/4219}

\bibitem[\protect\citeauthoryear{{Narayan}, {Paczynski}  \& {Piran}}{{Narayan} et~al.}{1992}]{1992ApJ...395L..83N}
{Narayan} R.,  {Paczynski} B.,   {Piran} T.,  1992, \mn@doi [\apjl] {10.1086/186493}, \href {http://adsabs.harvard.edu/abs/1992ApJ...395L..83N} {395, L83}

\bibitem[\protect\citeauthoryear{{Norris}, {Marani}  \& {Bonnell}}{{Norris} et~al.}{2000}]{2000ApJ...534..248N}
{Norris} J.~P.,  {Marani} G.~F.,   {Bonnell} J.~T.,  2000, \mn@doi [\apj] {10.1086/308725}, \href {https://ui.adsabs.harvard.edu/abs/2000ApJ...534..248N} {534, 248}

\bibitem[\protect\citeauthoryear{{O'Brien} et~al.,}{{O'Brien} et~al.}{2006}]{2006ApJ...647.1213O}
{O'Brien} P.~T.,  et~al., 2006, \mn@doi [\apj] {10.1086/505457}, \href {https://ui.adsabs.harvard.edu/abs/2006ApJ...647.1213O} {647, 1213}

\bibitem[\protect\citeauthoryear{{Oates} et~al.,}{{Oates} et~al.}{2009}]{Oates2009}
{Oates} S.~R.,  et~al., 2009, \mn@doi [\mnras] {10.1111/j.1365-2966.2009.14544.x}, \href {https://ui.adsabs.harvard.edu/abs/2009MNRAS.395..490O} {395, 490}

\bibitem[\protect\citeauthoryear{{Oates}, {Page}, {De Pasquale}, {Schady}, {Breeveld}, {Holland}, {Kuin}  \& {Marshall}}{{Oates} et~al.}{2012}]{Oates2012}
{Oates} S.~R.,  {Page} M.~J.,  {De Pasquale} M.,  {Schady} P.,  {Breeveld} A.~A.,  {Holland} S.~T.,  {Kuin} N.~P.~M.,   {Marshall} F.~E.,  2012, \mn@doi [\mnras] {10.1111/j.1745-3933.2012.01331.x}, \href {https://ui.adsabs.harvard.edu/abs/2012MNRAS.426L..86O} {426, L86}

\bibitem[\protect\citeauthoryear{{Oke} \& {Sandage}}{{Oke} \& {Sandage}}{1968}]{Oke1968}
{Oke} J.~B.,  {Sandage} A.,  1968, \mn@doi [\apj] {10.1086/149737}, \href {https://ui.adsabs.harvard.edu/abs/1968ApJ...154...21O} {154, 21}

\bibitem[\protect\citeauthoryear{{Ozaki}}{{Ozaki}}{2005}]{2005ARNis..15....6O}
{Ozaki} S.,  2005, Annual Report of the Nishi-Harima Astronomical Observatory (ISSN 0917-6926), \href {https://ui.adsabs.harvard.edu/abs/2005ARNis..15....6O} {15, 6}

\bibitem[\protect\citeauthoryear{{Paczy{\'n}ski}}{{Paczy{\'n}ski}}{1998}]{1998ApJ...494L..45P}
{Paczy{\'n}ski} B.,  1998, \mn@doi [\apjl] {10.1086/311148}, \href {https://ui.adsabs.harvard.edu/abs/1998ApJ...494L..45P} {494, L45}

\bibitem[\protect\citeauthoryear{{Pei}}{{Pei}}{1992}]{Pei1992}
{Pei} Y.~C.,  1992, \mn@doi [\apj] {10.1086/171637}, \href {https://ui.adsabs.harvard.edu/abs/1992ApJ...395..130P} {395, 130}

\bibitem[\protect\citeauthoryear{{Perley}, {Jones}  \& {Ellis}}{{Perley} et~al.}{2012}]{2012GCN.13279....1P}
{Perley} D.~A.,  {Jones} T.,   {Ellis} R.,  2012, GRB Coordinates Network, \href {https://ui.adsabs.harvard.edu/abs/2012GCN.13279....1P} {13279, 1}

\bibitem[\protect\citeauthoryear{{Perna}, {Lazzati}  \& {Fiore}}{{Perna} et~al.}{2003}]{2003ApJ...585..775P}
{Perna} R.,  {Lazzati} D.,   {Fiore} F.,  2003, \mn@doi [\apj] {10.1086/346109}, \href {https://ui.adsabs.harvard.edu/abs/2003ApJ...585..775P} {585, 775}

\bibitem[\protect\citeauthoryear{Peterson}{Peterson}{1997}]{Peterson1997}
Peterson B.~M.,  1997, An Introduction to Active Galactic Nuclei.
Cambridge University Press

\bibitem[\protect\citeauthoryear{{Piran}}{{Piran}}{1999}]{1999PhR...314..575P}
{Piran} T.,  1999, \mn@doi [\physrep] {10.1016/S0370-1573(98)00127-6}, \href {https://ui.adsabs.harvard.edu/abs/1999PhR...314..575P} {314, 575}

\bibitem[\protect\citeauthoryear{{Piro} et~al.,}{{Piro} et~al.}{1998}]{1998A&A...331L..41P}
{Piro} L.,  et~al., 1998, \mn@doi [\aap] {10.48550/arXiv.astro-ph/9710355}, \href {https://ui.adsabs.harvard.edu/abs/1998A&A...331L..41P} {331, L41}

\bibitem[\protect\citeauthoryear{{Rastinejad} et~al.,}{{Rastinejad} et~al.}{2022}]{2022Natur.612..223R}
{Rastinejad} J.~C.,  et~al., 2022, \mn@doi [\nat] {10.1038/s41586-022-05390-w}, \href {https://ui.adsabs.harvard.edu/abs/2022Natur.612..223R} {612, 223}

\bibitem[\protect\citeauthoryear{{Rodgers}, {Allen}, {Barlow}, {Garcia}, {Pierce}  \& {Canterna}}{{Rodgers} et~al.}{2006}]{2006GCN..4831....1R}
{Rodgers} C.,  {Allen} D.,  {Barlow} B.,  {Garcia} C.,  {Pierce} M.,   {Canterna} R.,  2006, GRB Coordinates Network, \href {https://ui.adsabs.harvard.edu/abs/2006GCN..4831....1R} {4831, 1}

\bibitem[\protect\citeauthoryear{{Roming} et~al.,}{{Roming} et~al.}{2009}]{Roming2009}
{Roming} P.~W.~A.,  et~al., 2009, \mn@doi [\apj] {10.1088/0004-637X/690/1/163}, \href {https://ui.adsabs.harvard.edu/abs/2009ApJ...690..163R} {690, 163}

\bibitem[\protect\citeauthoryear{{Roming} et~al.,}{{Roming} et~al.}{2017}]{roming2017large}
{Roming} P. W.~A.,  et~al., 2017, \mn@doi [\apjs] {10.3847/1538-4365/228/2/13}, \href {https://ui.adsabs.harvard.edu/abs/2017ApJS..228...13R} {228, 13}

\bibitem[\protect\citeauthoryear{{Rossi} et~al.,}{{Rossi} et~al.}{2020}]{rossi2020comparison}
{Rossi} A.,  et~al., 2020, \mn@doi [\mnras] {10.1093/mnras/staa479}, \href {https://ui.adsabs.harvard.edu/abs/2020MNRAS.493.3379R} {493, 3379}

\bibitem[\protect\citeauthoryear{{Rumyantsev}, {Kiselev}  \& {Ivanova}}{{Rumyantsev} et~al.}{2019}]{Rumyantsev19}
{Rumyantsev} V.~V.,  {Kiselev} N.~N.,   {Ivanova} A.~V.,  2019, \mn@doi [Solar System Research] {10.1134/S0038094619020060}, \href {https://ui.adsabs.harvard.edu/abs/2019SoSyR..53...91R/abstract} {53, 91}

\bibitem[\protect\citeauthoryear{{Sahu} et~al.,}{{Sahu} et~al.}{1997}]{1997ApJ...489L.127S}
{Sahu} K.~C.,  et~al., 1997, \mn@doi [\apjl] {10.1086/316786}, \href {https://ui.adsabs.harvard.edu/abs/1997ApJ...489L.127S} {489, L127}

\bibitem[\protect\citeauthoryear{{Sakamoto} et~al.,}{{Sakamoto} et~al.}{2007}]{2007ApJ...669.1115S}
{Sakamoto} T.,  et~al., 2007, \mn@doi [\apj] {10.1086/521640}, \href {https://ui.adsabs.harvard.edu/abs/2007ApJ...669.1115S} {669, 1115}

\bibitem[\protect\citeauthoryear{{Sako} et~al.,}{{Sako} et~al.}{2018}]{Sako2018}
{Sako} S.,  et~al., 2018, in {Evans} C.~J.,  {Simard} L.,   {Takami} H.,  eds,  SPIE Conference Series Vol. 10702, Ground-based and Airborne Instrumentation for Astronomy VII. p. 107020J, \mn@doi{10.1117/12.2310049}

\bibitem[\protect\citeauthoryear{{Sari}, {Piran}  \& {Narayan}}{{Sari} et~al.}{1998}]{Sari1998}
{Sari} R.,  {Piran} T.,   {Narayan} R.,  1998, \mn@doi [\apjl] {10.1086/311269}, \href {https://ui.adsabs.harvard.edu/abs/1998ApJ...497L..17S} {497, L17}

\bibitem[\protect\citeauthoryear{{Sbarufatti} et~al.,}{{Sbarufatti} et~al.}{2008}]{2008GCN..7828....1S}
{Sbarufatti} B.,  et~al., 2008, GRB Coordinates Network, \href {https://ui.adsabs.harvard.edu/abs/2008GCN..7828....1S} {7828, 1}

\bibitem[\protect\citeauthoryear{{Schlafly} \& {Finkbeiner}}{{Schlafly} \& {Finkbeiner}}{2011}]{Schlafly2011}
{Schlafly} E.~F.,  {Finkbeiner} D.~P.,  2011, \mn@doi [\apj] {10.1088/0004-637X/737/2/103}, \href {https://ui.adsabs.harvard.edu/abs/2011ApJ...737..103S} {737, 103}

\bibitem[\protect\citeauthoryear{{Schlegel}, {Finkbeiner}  \& {Davis}}{{Schlegel} et~al.}{1998}]{Schlegel1998}
{Schlegel} D.~J.,  {Finkbeiner} D.~P.,   {Davis} M.,  1998, \mn@doi [\apj] {10.1086/305772}, \href {https://ui.adsabs.harvard.edu/abs/1998ApJ...500..525S} {500, 525}

\bibitem[\protect\citeauthoryear{{Serebryanskiy}, {Krugov}, {Komarov}, {Usol'tseva}  \& {Akniyazov}}{{Serebryanskiy} et~al.}{2018}]{Serebryanskiy18}
{Serebryanskiy} A.~V.,  {Krugov} M.~A.,  {Komarov} A.~A.,  {Usol'tseva} L.~A.,   {Akniyazov} C.~B.,  2018, \mn@doi [Solar System Research] {10.1134/S0038094618040044}, \href {https://ui.adsabs.harvard.edu/abs/2018SoSyR..52..320S/abstract} {52, 320}

\bibitem[\protect\citeauthoryear{{Severny}}{{Severny}}{1955}]{Severny1955}
{Severny} A.~B.,  1955, \mn@doi [Vistas in Astronomy] {10.1016/0083-6656(55)90086-6}, \href {https://ui.adsabs.harvard.edu/abs/1955VA......1..701S/abstract} {1, 701}

\bibitem[\protect\citeauthoryear{{Si} et~al.,}{{Si} et~al.}{2018}]{Si2018}
{Si} S.-K.,  et~al., 2018, \mn@doi [\apj] {10.3847/1538-4357/aad08a}, \href {https://ui.adsabs.harvard.edu/abs/2018ApJ...863...50S} {863, 50}

\bibitem[\protect\citeauthoryear{{Simon, V.}, {Hudec, R.}, {Pizzichini, G.}  \& {Masetti, N.}}{{Simon, V.} et~al.}{2001}]{Simon2001}
{Simon, V.} {Hudec, R.} {Pizzichini, G.}  {Masetti, N.} 2001, \mn@doi [A\&A] {10.1051/0004-6361:20011158}, 377, 450

\bibitem[\protect\citeauthoryear{{Smith} \& {Shetrone}}{{Smith} \& {Shetrone}}{2020}]{2020PASP..132l5002S}
{Smith} G.~H.,  {Shetrone} M.,  2020, \mn@doi [\pasp] {10.1088/1538-3873/abc022}, \href {https://ui.adsabs.harvard.edu/abs/2020PASP..132l5002S} {132, 125002}

\bibitem[\protect\citeauthoryear{{Srinivasaragavan}, {Dainotti}, {Fraija}, {Hernandez}, {Nagataki}, {Lenart}, {Bowden}  \& {Wagner}}{{Srinivasaragavan} et~al.}{2020}]{Dainotti2020a}
{Srinivasaragavan} G.~P.,  {Dainotti} M.~G.,  {Fraija} N.,  {Hernandez} X.,  {Nagataki} S.,  {Lenart} A.,  {Bowden} L.,   {Wagner} R.,  2020, \mn@doi [\apj] {10.3847/1538-4357/abb702}, \href {https://ui.adsabs.harvard.edu/abs/2020ApJ...903...18S} {903, 18}

\bibitem[\protect\citeauthoryear{{Srinivasaragavan} et~al.,}{{Srinivasaragavan} et~al.}{2024}]{2024ApJ...960L..18S}
{Srinivasaragavan} G.~P.,  et~al., 2024, \mn@doi [\apjl] {10.3847/2041-8213/ad16e7}, \href {https://ui.adsabs.harvard.edu/abs/2024ApJ...960L..18S} {960, L18}

\bibitem[\protect\citeauthoryear{{Takase}, {Ishida}, {Shimizu}, {Maehara}, {Hamajima}, {Noguchi}  \& {Ohashi}}{{Takase} et~al.}{1977}]{1977AnTok..16...74T}
{Takase} B.,  {Ishida} K.,  {Shimizu} M.,  {Maehara} H.,  {Hamajima} K.,  {Noguchi} T.,   {Ohashi} M.,  1977, Annals of the Tokyo Astronomical Observatory, \href {https://ui.adsabs.harvard.edu/abs/1977AnTok..16...74T} {16, 74}

\bibitem[\protect\citeauthoryear{{Torii} et~al.,}{{Torii} et~al.}{2003}]{Torii2003}
{Torii} K.,  et~al., 2003, \mn@doi [\apjl] {10.1086/379846}, \href {https://ui.adsabs.harvard.edu/abs/2003ApJ...597L.101T} {597, L101}

\bibitem[\protect\citeauthoryear{{Torii}, {Hasuike}  \& {Tsunemi}}{{Torii} et~al.}{2006}]{2006sgrb.confE..61T}
{Torii} K.,  {Hasuike} K.,   {Tsunemi} H.,  2006, in {Chevalier} R.,  {Hwang} U.,   {Laming} M.,  eds, KITP Conference: Supernova and Gamma-Ray Burst Remnants. p.~61

\bibitem[\protect\citeauthoryear{{Uemura} \& {Kanata Team}}{{Uemura} \& {Kanata Team}}{2009}]{2009ASPC..404...69U}
{Uemura} M.,  {Kanata Team} 2009, in {Murphy} S.~J.,  {Bessell} M.~S.,  eds,  ASP Conference Series Vol. 404, The Eighth Pacific Rim Conference on Stellar Astrophysics: A Tribute to Kam-Ching Leung. p.~69

\bibitem[\protect\citeauthoryear{{Vestrand} et~al.,}{{Vestrand} et~al.}{2005}]{Vestrand2005Natur}
{Vestrand} W.~T.,  et~al., 2005, \mn@doi [\nat] {10.1038/nature03515}, \href {https://ui.adsabs.harvard.edu/abs/2005Natur.435..178V} {435, 178}

\bibitem[\protect\citeauthoryear{{Volnova}, {Pozanenko}, {Mazaeva}, {Belkin}, {Molotov}, {Elenin}, {Tungalag}  \& {Buckley}}{{Volnova} et~al.}{2021}]{Volnova21}
{Volnova} A.,  {Pozanenko} A.,  {Mazaeva} E.,  {Belkin} S.,  {Molotov} I.,  {Elenin} L.,  {Tungalag} N.,   {Buckley} D.,  2021, \mn@doi [Anais da Academia Brasileira de Ciencias] {/10.1590/0001-3765202120200883}, 93

\bibitem[\protect\citeauthoryear{{Wang} et~al.,}{{Wang} et~al.}{2015}]{Wang2015}
{Wang} X.-G.,  et~al., 2015, \mn@doi [\apjs] {10.1088/0067-0049/219/1/9}, \href {https://ui.adsabs.harvard.edu/abs/2015ApJS..219....9W} {219, 9}

\bibitem[\protect\citeauthoryear{{Watson} et~al.,}{{Watson} et~al.}{2016}]{2016SPIE.9910E..0GW}
{Watson} A.~M.,  et~al., 2016, in {Peck} A.~B.,  {Seaman} R.~L.,   {Benn} C.~R.,  eds,  SPIE Conference Series Vol. 9910, Observatory Operations: Strategies, Processes, and Systems VI. p. 99100G (\mn@eprint {arXiv} {1606.00695}), \mn@doi{10.1117/12.2232898}

\bibitem[\protect\citeauthoryear{{Woosley}}{{Woosley}}{1993}]{1993ApJ...405..273W}
{Woosley} S.~E.,  1993, \mn@doi [\apj] {10.1086/172359}, \href {https://ui.adsabs.harvard.edu/abs/1993ApJ...405..273W} {405, 273}

\bibitem[\protect\citeauthoryear{{Woosley} \& {Bloom}}{{Woosley} \& {Bloom}}{2006}]{Woosley2006ARA&A}
{Woosley} S.~E.,  {Bloom} J.~S.,  2006, \mn@doi [\araa] {10.1146/annurev.astro.43.072103.150558}, \href {https://ui.adsabs.harvard.edu/abs/2006ARA&A..44..507W} {44, 507}

\bibitem[\protect\citeauthoryear{Woźniak, Vestrand, Panaitescu, Wren, Davis  \& White}{Woźniak et~al.}{2009}]{Wozniak2009}
Woźniak P.~R.,  Vestrand W.~T.,  Panaitescu A.~D.,  Wren J.~A.,  Davis H.~R.,   White R.~R.,  2009, \mn@doi [\apj] {10.1088/0004-637X/691/1/495}, 691, 495

\bibitem[\protect\citeauthoryear{{Yamaoka}, {Kato}, {Uemura}, {Ishioka}  \& {Naito}}{{Yamaoka} et~al.}{2002}]{Yamaoka2002}
{Yamaoka} H.,  {Kato} T.,  {Uemura} M.,  {Ishioka} R.,   {Naito} H.,  2002, in {Ikeuchi} S.,  {Hearnshaw} J.,   {Hanawa} T.,  eds, 8th Asian-Pacific Regional Meeting, Volume II. pp 363--364

\bibitem[\protect\citeauthoryear{{Yatsu} et~al.,}{{Yatsu} et~al.}{2007}]{2007PhyE...40..434Y}
{Yatsu} Y.,  et~al., 2007, \mn@doi [Physica E Low-Dimensional Systems and Nanostructures] {10.1016/j.physe.2007.06.050}, \href {https://ui.adsabs.harvard.edu/abs/2007PhyE...40..434Y} {40, 434}

\bibitem[\protect\citeauthoryear{{Yonetoku}, {Murakami}, {Nakamura}, {Yamazaki}, {Inoue}  \& {Ioka}}{{Yonetoku} et~al.}{2004}]{2004ApJ...609..935Y}
{Yonetoku} D.,  {Murakami} T.,  {Nakamura} T.,  {Yamazaki} R.,  {Inoue} A.~K.,   {Ioka} K.,  2004, \mn@doi [\apj] {10.1086/421285}, \href {https://ui.adsabs.harvard.edu/abs/2004ApJ...609..935Y} {609, 935}

\bibitem[\protect\citeauthoryear{{Zaninoni}}{{Zaninoni}}{2013}]{2013PhDT.......144Z}
{Zaninoni} E.,  2013, PhD thesis, University of Padua, Italy

\bibitem[\protect\citeauthoryear{{Zaninoni}, {Bernardini}, {Margutti}, {Oates}  \& {Chincarini}}{{Zaninoni} et~al.}{2013}]{Zaninoni2013}
{Zaninoni} E.,  {Bernardini} M.~G.,  {Margutti} R.,  {Oates} S.,   {Chincarini} G.,  2013, \mn@doi [\aap] {10.1051/0004-6361/201321221}, \href {https://ui.adsabs.harvard.edu/abs/2013A&A...557A..12Z} {557, A12}

\bibitem[\protect\citeauthoryear{{de Ugarte Postigo} et~al.,}{{de Ugarte Postigo} et~al.}{2005}]{2005AA...443..841D}
{de Ugarte Postigo} A.,  et~al., 2005, \mn@doi [\aap] {10.1051/0004-6361:20052898}, \href {https://ui.adsabs.harvard.edu/abs/2005A&A...443..841D} {443, 841}

\bibitem[\protect\citeauthoryear{{van Paradijs} et~al.,}{{van Paradijs} et~al.}{1997}]{1997Natur.386..686V}
{van Paradijs} J.,  et~al., 1997, \mn@doi [\nat] {10.1038/386686a0}, \href {https://ui.adsabs.harvard.edu/abs/1997Natur.386..686V} {386, 686}

\makeatother
\end{thebibliography}


\vspace{-0.7cm}
\appendix
\section{Instruments in this sample}\label{sec:instruments}
In this overview, we highlight the contributions of the telescopes included in our sample, which provide our data, both privately shared and publicly available. This overview is structured based on the number of data points gathered by each telescope, presented in descending order, along with the corresponding percentage of these data points compared to the total data gathered within this catalogue.

The Swift satellite primarily observed the GRBs in our sample, with additional data from over 570 instruments on both ground-based telescopes and satellites. 

The most recurring instrument is the UVOT aboard the Swift satellite. Until 20 October 2023, Swift has observed 1608 GRBs. Of these, 420 have measured redshifts, about 26\% of the total sample. Up to 20 October 2023, 1356 have UVOT observations, 496 UVOT detections, 783 GRBs have ground detections, and 451 have both UVOT and ground detections. In our sample, UVOT has observed 469 GRBs, the $88\%$ of the catalogue: 130 have only limiting magnitudes, while 339 either have both observed magnitudes and limiting magnitudes or have only observed magnitudes. UVOT provides 19786 data points in the 170-650 nm range. The UVOT data represents 30.53$\%$ of the total (64813), of which 14708 are observed magnitude values, while 5078 are limiting magnitudes. When a new GRB triggers the Swift Burst Alert Telescope (Swift/BAT), the satellite directs the XRT and UVOT to the burst's location. Source localization using UVOT is more accurate than XRT localization (like GRB 080605A, see \citealt{2008GCN..7828....1S,2008GCN..7844....1K}) since the latter has a systematic error of 3.5", while the former allows reducing the uncertainty of the X-ray position down to 1.5" \citep{2008AIPC.1000..539E} after the alignment with UVOT. Considering one UVOT image, the residual mean uncertainty in the position of sources can be reduced to 0.5-1.0" after applying the appropriate astrometric corrections \citep{2007A&A...476.1401G}. Optical ground-based telescopes follow the BAT trigger, re-pointing if the new UVOT location is available.
UVOT provides observations in the following bands: $U,B,V,UVW1,UVW2,UVM2,white$.

The second most recurring instrument is the Gamma-ray Burst Optical/Near-infrared Detector Telescope 
(GROND, \citealt{2008PASP..120..405G}) on the MPG/ESO 2.2-meter telescope, located at La Silla. GROND has observed 8272 data points (12.76$\%$ of the total) in the 0.45–2.1 micrometer range, with 7509 magnitude values and 763 limiting magnitudes. GROND provides contemporary data in the $u,g,r, i,z,J,H,K$ bands.

The third is the Reionization and Transients Infrared/Optical Project (RATIR), an instrument mounted on the 1.5-meter Harold Johnson telescope at the Observatorio Astronómico Nacional located in San Pedro Mártir, Mexico, until summer 2022. RATIR was designed for the follow-up observations of the optical/NIR afterglows triggered by the Swift/BAT instrument. RATIR has provided 3180 points for this catalogue, or 4.91$\%$ of the data sample, obtaining simultaneous data in $r,i,Z,J$ or $r,i,Y,H$. For the 
commissioning and first light, see \citet{2012SPIE.8446E..10B}. 

The fourth is the Spectral Energy Distribution Machine (SEDM, \citealt{2018PASP..130c5003B}), mounted on the 60-inch telescope at the Palomar Observatory near San Diego, California. This instrument observed 1667 data points (2.57$\%$) in the 350-950 nm range. The available filters for this instrument are $u,g,r,i,U,B,V,R_C,I_C$.

The fifth is the Peters Automated Infrared Imaging Telescope (PAIRITEL, \citealt{bloom2005autonomous}), a 1.3-meter telescope located at the Fred Lawrence Whipple Observatory, Arizona, with 1299 data points, 2$\%$. It observes the sky in the $J,H,K_S$ bands previously used in the 2MASS survey. The sixth is the RINGO3 \citep{2012SPIE.8446E..2JA} instrument of the 2-meter Liverpool Telescope in the Roque de Los Muchachos Observatory. RINGO3, decommissioned in summer 2020, was a polarimeter with a 3-camera system that covered roughly the range $350-1000{\rm nm}$ and has contributed with 1013 data points in the $I, R,V$ bands, for the 1.56$\%$. 

\begin{figure}
    \centering
    \includegraphics[scale=0.38]{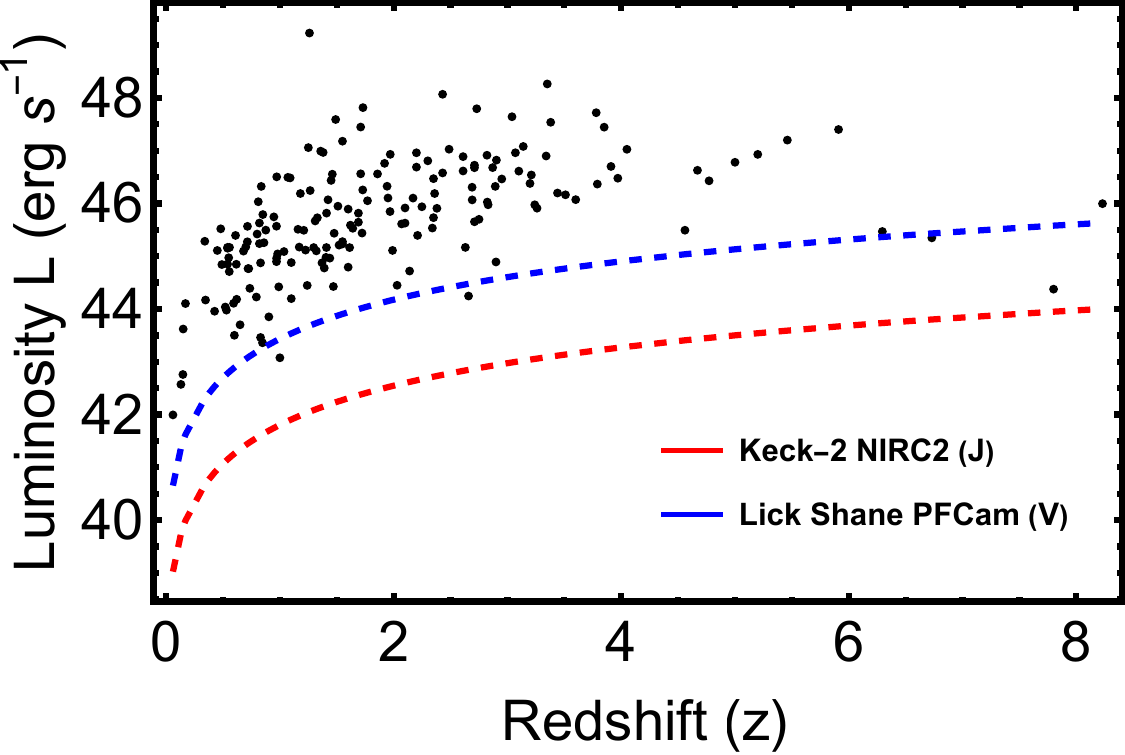}
    \caption{The plateau luminosities for the 179 optical GRBs sample reported in \citet{2022arXiv220312908D} plotted against the limiting luminosities of the 10-meter telescope Keck-2 NIRC2 instrument (in $J$ band) and the PFCam mounted on the 3-meter Shane telescope of the Lick Observatory (in $V$ band).}
    \label{fig:limlumKeckLick}
\end{figure}

The sixth are represented by the Lick \citep{2020PASP..132l5002S} and the Keck \citep{2014SPIE.9147E..03A} Observatories. The Lick provided a total of 1283 data points from the 0.8-meter Katzman Automatic Imaging Telescope (KAIT, in the $U,B,V,R,I_C$), the 1.10-meter Nickel telescope ($U,B,V,R,I,Z,g,r,i,z$), and the 3-meter Shane telescope (in the $I,g$ bands). The Keck has released a total of 200 data points, 165 from the 10-meter Keck-1 telescope ($B, V, R, I, Z, Y, J, H, K$) and 35 from the 10-meter Keck-2 telescope ($B, V, R, I, Z,g, J, H, K$), having provided information for 53 GRBs. 11 GRBs from the Keck Observatory have $z>2$, while 6 of these have $z>3$.
Furthermore, we illustrate the concept of limiting luminosity for a given telescope, using the Lick and Keck observatories as an example. Figure \ref{fig:limlumKeckLick} contains the end-of-plateau luminosity in optical bands for the 179 GRBs sample taken from \citet{2022arXiv220312908D}, together with the limiting luminosities for the Near Infrared Camera 2 (NIRC2) instrument mounted on the 10-meter Keck-2 telescope (in $J$ band) and the Primary Focus Camera (PFCam) mounted on the 3-meter Shane telescope of the Lick Observatory (in $V$ band). The corresponding limiting fluxes are $1.22*10^{-16} \rm{erg\,cm^{-2}\,s^{-1}}$ and $5.20*10^{-15} \rm{erg\,cm^{-2}\,s^{-1}}$ for the NIRC2 $J$-band (for a $300\,s$ exposure) and the PFCam $V$-band (with an exposure time in the order of $3*10^6\,s$), respectively. The plot proves the capability of these two observing facilities in detecting the majority of the optical plateaus.

Next, we present a few other relevant ground-based telescopes, for which we have contributions from GCNs and private communications. 
Part of the observations (787 data points of 27 GRBs) were received from telescopes of the IKI Gamma-Ray Burst Follow-up Network (IKI GRB-FuN,~\citealt{Volnova21}). Considering the total catalogue, the IKI GRB-FuN telescopes contributed through the following: the ZTSh telescope~\citep{Rumyantsev19} of Crimean Astrophysical Observatory~\citep[CrAO,][]{Severny1955} with 422 data points, the AZT-33IK telescope of the Sayan observatory~\citep[Mondy,][]{Chuprakov18} with 238 data points, the AS-32 telescope of Abastumani Astrophysical Observatory~\citep[AbAO,][]{Khetsuriani67} with 93 data points, the Zeiss-1000 telescope of Tien Shan Astronomical Observatory~\citep[TSHAO,][]{Elenin2015} with 191 data points, the AZT-22 telescope of Maidanak Observatory~\citep[MAO,][]{Ehgamberdiev00} with 342 observations, and the AZT-20 1.5-meter telescope of Assy-Turgen Observatory~\citep[ATO,][]{Serebryanskiy18} with 54 data points.
Another relevant observational site is the Assy-Turgen Observatory (317 data points in $U,B,V,R,g,r,i,z,J,K_S$) with the 0.7-meter Nazarbayev University Transient Telescope (NUTTelA-TAO, \citealt{2021RMxAC..53..169M}) that provides 263 data points (0.4$\%$). 

We also include some telescopes that form the Global Rapid Advanced Network Devoted to the Multi-messenger Addicts network (GRANDMA, \citealt{GRANDMA2021RMxAC..53..198A, Kann2023BOAT}), which gathered 772 data points, or 1.2$\%$ of the total. The GRANDMA network telescopes provide data in the $U,B,V,R_C,I_C,u,g,r,i,z,clear$ bands. These telescopes are: the Télescope à Action Rapide pour les Objets Transitoires (TAROT) telescopes, \citep[0.25-meter in Calern and La Silla and the 0.18-meter at Les Makes Observatory,][]{2003Msngr.113...45B}, which have gathered 640 data points, 1$\%$; the Abastumani telescope \citep{Abastumani}, 0.7-meter, located in Georgia, with 96 data points, 0.1$\%$ of the total; the Montarrenti Observatory with the 0.53-meter telescope, located in Siena, with 20 data points; the T-CAT 0.40-meter telescope at the Crous des Gats Observatory, with 11 data points; the Meade SCT telescope, 0.28-meter, located at the McDonald observatory (Fort Davis), with 9 data points; the Pic du Midi telescope, 1-meter, located in Sers, France, with seven data points. Each of the following telescopes gathers two data points: the Astrolab-IRIS, 0.53-meter, in Zillebeke, Belgium; the Atlas telescope, 0.30-meter; the Burke–Gaffney Observatory telescope, 0.61-meter; the CO-K26 telescope; the EHEA-WL, 0.20-meter; the LCO, 0.4-meter, telescope at the McDonald Observatory; the Ste-Sophie, in Montreal; the OMEGON-200F5. The following telescopes gather one data point: the GPO telescope; the Hidden Valley Observatory telescope; the New Mexico Skies telescope; the Sierra Remote Observatory; Auberry; the T-BRO telescope, and the Parent telescope. 

There are also the BOOTES telescopes \citep{Youdong2023,jelinekBOOTES} that have gathered 575 data points, or $0.88\%$ of the total. This telescope network provides data in $u,g,r, i,Z, Y,clear$ filters. The BOOTES network is composed of telescopes that include the 0.6-meter in Malaga with 164 data points, the 0.3-meter telescope in Mazagon with 152, and additional telescopes in Lijiang with 138, in Otago with 84, in Sierra Nevada with 29, and at the Observatorio Astronómico Nacional in Mexico with 8. 

We now detail the telescopes that belong to the Japanese community based on the number of data points gathered by each telescope, presented in descending order, along with the corresponding percentage of occurrences in the total.
The Multicolor Imaging Telescopes for Survey and Monstrous Explosions (MITSuME, \citealt{2007PhyE...40..434Y}), 0.5-meter telescopes are automated telescopes designed to respond to GCN alerts within a few seconds to a few minutes. MITSuME observes in the optical bands ($g, R_C,I_C$) and can detect roughly 4-5 GRB afterglows annually. MITSuME has two locations: Akeno, Yamanashi Prefecture, gathering 140 data points (0.22$\%$ of the total catalog), and Okayama Astrophysical Observatory (OAO), gathering 502 data points (0.77$\%$ of the total). The data points collected from the two telescopes are 642, 0.99$\%$ of the total, making MITSuME the most prolific Japanese telescope in our catalog. 

The second most numerous observations from the Japanese contributors are given by the pair of 0.25-meter and 0.30-meter telescopes owned and located in Kyoto University \citep{Yamaoka2002}. Together, they have gathered 377 data points in the $R$ band, namely, the 0.58$\%$ of the total. The third telescope per number of observations is the Yatsugatake telescope \citep{Torii2003}, built with cameras of focal length 3.5-mm (mounting the $R_C$ filter), which has gathered 277 data points, 0.43$\%$ of the total. The fourth position for the number of data points is occupied by the Kanata 1.5-meter telescope \citep{2009ASPC..404...69U}, which has gathered 77 data points ($B, V,R, I,K$ bands), 0.12$\%$.

The fifth most recurrent Japanese telescope is the Subaru \citep{1992ESOC...42...43K}, an 8.2-meter telescope located in Hawaii and operated by the National Astronomical Observatory of Japan (NAOJ). The Subaru telescope comprises five cameras spanning optical and infrared wavelengths, including the Hyper-Suprime Cam (HSC), which is capable of wide-field observations. Subaru has observed 39 data points ($g,r, i, Z$), namely, 0.06$\%$ of the total.

In the sixth position, we have the 0.3-meter RIMOTS telescope \citep{Kohama2002}, located at the University of Miyazaki, that has observed 37 unfiltered data points, the 0.06$\%$. The NAYUTA 2-meter telescope occupies the seventh position \citep{2005ARNis..15....6O}, in Sayo Town, with 12 data points in the $J, H,K_S$ bands, 0.02$\%$, while the Gunma Observatory \citep{Torii2003} with the 1.5-meter and 0.25-meter telescopes has provided 13 data points, 0.02$\%$, in the $J, H, K, R_C$ wavelengths and occupies the eighth position. In the ninth position we have another relevant Japanese facility: the wide-field CMOS camera, Tomo-e Gozen \citep{Sako2018}, mounted on the 1.05-meter Schmidt telescope at the Kiso Observatory \citep{1977AnTok..16...74T} located in Nagano prefecture. The GRB optical follow-up with Tomo-e Gozen was initiated by some of us (Dainotti, Niino, Watson, Kalinowski) in January 2022 and has contributed with six data points with the clear filter. In the tenth place, the RIBOTS 0.3-meter telescope \citep{Kohama2002}, located in Bisei, Okayama, has provided six $R$-band data points.
Finally, the eleventh Japanese telescope in order of contributions is the ART-3a 0.35-meter \citep{2006sgrb.confE..61T} telescope in Toyonaka has contributed to our catalogue with four $I_C, J$ data.

Besides the Swift satellite, we have stored data from another space satellite: the Hubble Space Telescope (HST, see \citealt{1993ApOpt..32.1775D} for a review on the optical design), with 292 data points in the broad-band filters like the $F435W,F555W,F814W$, for the 0.46$\%$.
Other important telescopes are the 0.5-meter Virgin Islands Robotic Telescope (VIRT, \citealt{MorrisVIRT}) in the $U,B,V,R,I_C$ with 16 data points, and the 6x0.28-meter Deca-Degree Optical Transient Imager (DDOTI, \citealt{2016SPIE.9910E..0GW}) with four $white$ data points, located in Mexico, which was designed to locate optical transients.

The details of all the telescopes and instruments in the repository and their occurrences and filter properties are reported in Table 5 on the Online Materials. The filter properties, such as the central wavelength and zero point, are sourced from the respective telescope/instrument's papers and websites. For the cases in which only the name/system of the filter was reported, an average of the wavelengths (in \AA) and the zero points (in $Jy$) is made with the known values of the same filter from all the other telescopes and instruments in the catalogue. This approximation does not affect the wavelengths of the filters with a relevant systematic effect, and given that in the current analysis we do not employ the zero points, but we will use the magnitudes and wavelengths for the spectral analysis, we do not incur into major systematic effects.

\vspace{-0.8cm}
\section{GRB\lowercase{s} with peculiar \lowercase{k}-correction}\label{sec:kcorrectcases}
Following the procedure described in Section 3.2 of the paper, we applied the k-correction and host extinction correction to a subset of 138 GRBs out of the total 535. However, for five cases not included in the 138, a different investigation is required. Here, we briefly discuss them, considering the enlargement of the compatibility criterion of 2 $\sigma$ described in Section \ref{sec:spectral} up to 5 $\sigma$.

\textbf{GRB 021004A}. In their study, \citet{2005AA...443..841D} computed the $\beta_\mathrm{opt}$ value at 9 different epochs. We agree in 1 $\sigma$ with their results only at epochs 0.6380 days and 3.6520 days after the trigger. Despite the authors claiming no spectral evolution, our analysis using the 5 $\sigma$ criterion revealed significant deviations across six different epochs. Therefore, we computed the average of the $\beta_\mathrm{opt}$ values from each epoch and applied this value to the k-correction of the entire LC. Contrary to the literature, the colour evolution analysis showed no significant change for the $a=0$ fitting. At the same time, the variable slope fitting yielded the same number of filters with and without colour evolution.

\textbf{GRB 110205A}. \citet{2013MNRAS.432.1231C} provide the values $\beta_\mathrm{opt}=0.55^{+0.10}_{-0.12}$ and $\beta_\mathrm{opt,X}=0.55^{+0.14}_{-0.01}$. Our averaged $\beta_\mathrm{opt}$ is compatible with these values from $t = 5000\,sec$ to the late times. The authors do not specify the exact epoch where the spectral index is computed. Also in this case the 5 $\sigma$ criterion is not sufficient to apply the $k$-correction, thus we average the values of $\beta_\mathrm{opt}$ among the different epochs. The $a=0$ and variable slope fittings give no colour evolution as a result, agreeing with the literature.

\textbf{GRB 161219B}. \citet{2022ApJ...931...90L} find $\beta_\mathrm{opt}=0.32\pm 0.01$. In our analysis, we have $\beta_\mathrm{opt}$ values ($0.696\pm 0.619,1.326\pm 1.135$ respectively) at the epochs $t=25.66\,days,42.34\,days$. Our values are compatible with the literature in 1 $\sigma$. In this case, enlarging to 5 $\sigma$ is enough to solve the $k$-correction issue, and we can apply the average $\beta_\mathrm{opt}$ computed along all the LC. Since this is one of the four GRBs for which the rescaling factor fitting does not constrain the $P$ value, we will not discuss the colour evolution of this GRB.

\textbf{GRB 171010A}. \citet{2019MNRAS.490.5366M} estimate the $\beta_\mathrm{opt}=1.33^{+0.10}_{-0.28}$ at $t=1.45\,days$, and we find a $\beta_\mathrm{opt}$ compatible with it in 1 $\sigma$ up to $t=13.29\,days$. Furthermore, the authors prove the presence of a deviation of the GRB slope in the $g,r, i,z$ bands at around $10$ days due to the emerging SN, thanks to the spectral analysis. Indeed, at around $t\sim12\,days$ and later times, a deviation from the previous values of $\beta_\mathrm{opt}$ is visible in our analysis as well. Also, in this case, the 5 $\sigma$ criterion leads to the solution of the particular $k$-correction. 
The fitting with $a=0$ provides the same number of filters with and without colour evolution, while the variable slope fitting suggests that this GRB has colour evolution. Nevertheless, its chromatic or achromatic behaviour is not discussed in the literature.

\textbf{GRB 171205A}. \citet{2019Natur.565..324I} fit the optical and X-ray data together. Their value at $t\sim 0.03$ days is compatible with our early-time value in 1 $\sigma$, while our late-time value is incompatible with theirs. In this case, we average all the $\beta_\mathrm{opt}$ values estimated in the LC. The outcome of the $a=0$ fitting is no colour evolution, which agrees with the literature. Conversely, in the variable slope approach, this GRB is undetermined. 
In the Online Materials, we report the plots for these GRBs. See Figures 1 and 2 for the plot of the $\beta_\mathrm{opt}$ values as a function of the epoch and the colour evolution, respectively.

\vspace{-0.9cm}
\section{The spectral analysis formulation}\label{sec:ZaninoniKann}
We now show that the Spectral Energy Distribution formulations of \citet{kann2006signatures} and \citet{2013PhDT.......144Z} are equivalent in the host galaxy frame (or rest frame, where the factor $1+z$ is considered). To do so, we consider the Equations (1) and (2) from \citet{kann2006signatures}:

\begin{equation}
    F_{\nu} = F_{0} \nu^{-\beta_\mathrm{opt}} e^{-\tau(\nu_{host})},\,with\, \tau_{\nu_{host}}=(A_V/1.086)*[A_{\lambda,host}/A_V],  
    \label{eq:step1SED}
\end{equation}

together with the Equations (3.5) and (3.7) from \citet{2013PhDT.......144Z}, rewritten in one equation as

\begin{equation}
    F_{\nu} = F_{0} \nu^{-\beta_\mathrm{opt}} 10^{k \epsilon^{B}_{\nu} - k \epsilon^{B}_{\nu(1+z)}},
    \label{eq:step2SED}
\end{equation}

where $\epsilon^{B}_{\nu}=[A_{\nu}/A_{B}]$, $\epsilon^{V}_{\nu}=[A_{\nu}/A_{V}]$ and $k$ is the factor to convert the $\epsilon^{B}_{\nu}$ into the $\epsilon^{V}_{\nu}$. The following equivalence can be written:

\begin{equation}
    \log_{10} 10^{-k \epsilon^{B}_{\nu(1+z)}}=\log_{10}e^{-\epsilon^{V}_{\nu(1+z)/1.086}},
    \label{eq:step3SED}
\end{equation}

where the left-hand side is taken from Equation \ref{eq:step2SED}, the right-hand side from Equation \ref{eq:step1SED}. Transforming into linear the expression in Equation \ref{eq:step3SED} we obtain:

\begin{equation}
    -k \epsilon^{B}_{\nu(1+z)} = \frac{-\log_{10}(e)}{1.086} \epsilon^{V}_{\nu(1+z)}.
    \label{eq:step4SED}
\end{equation}

Thus, the two formulations are proven equivalent given the arbitrariness of the conversion factor $k$.

\vspace{-0.7cm}
\section{Examples of disagreement cases}\label{sec:disagreementexamples}
\begin{figure*}
  \centering
  \includegraphics[scale=0.26]{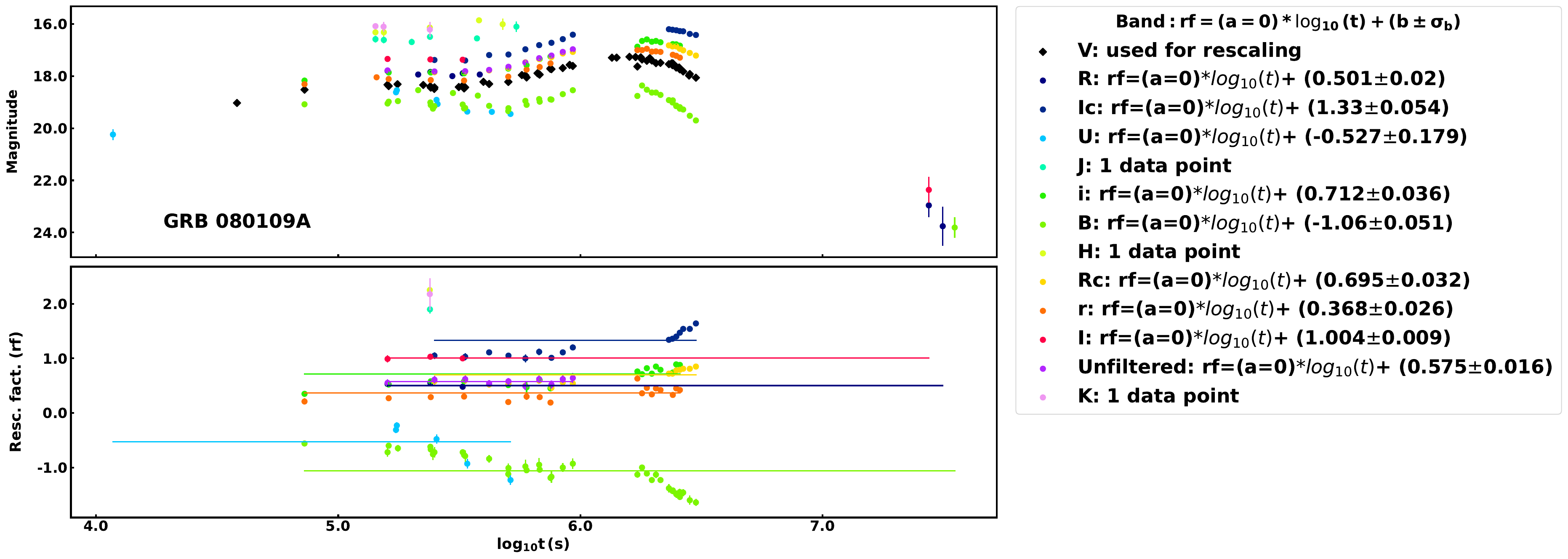} 
  \includegraphics[scale=0.26]{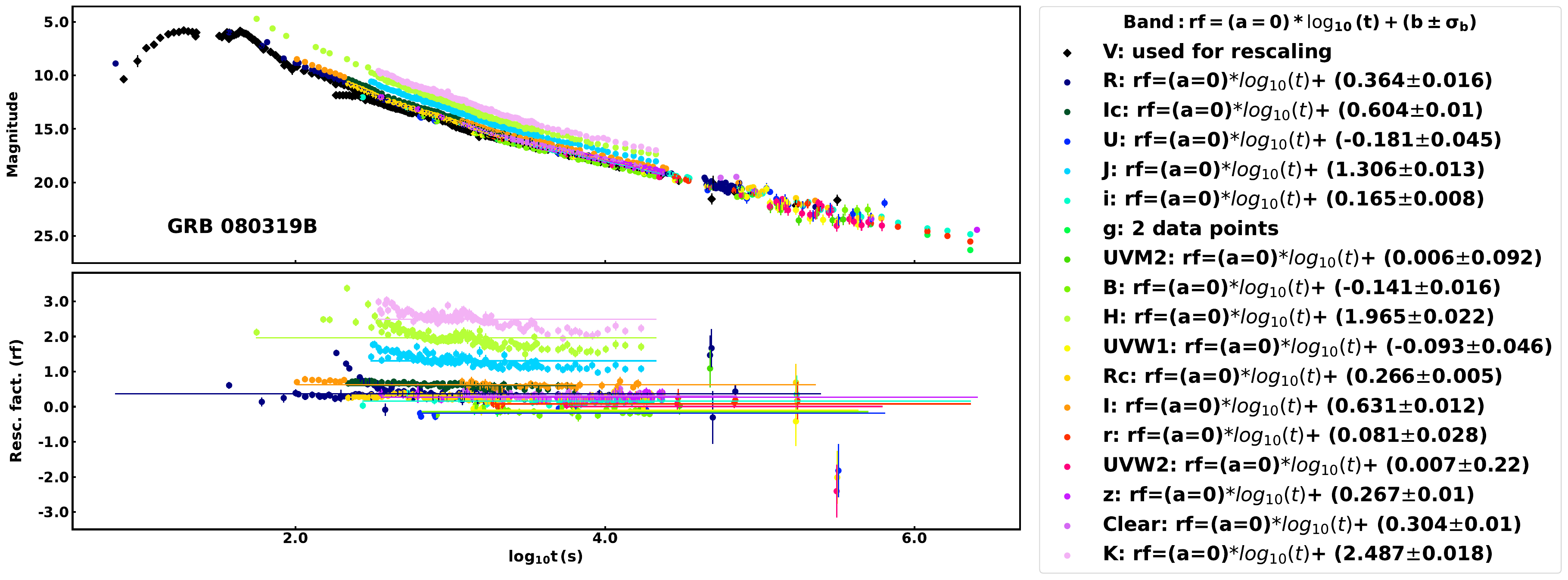}
  \includegraphics[scale=0.26]{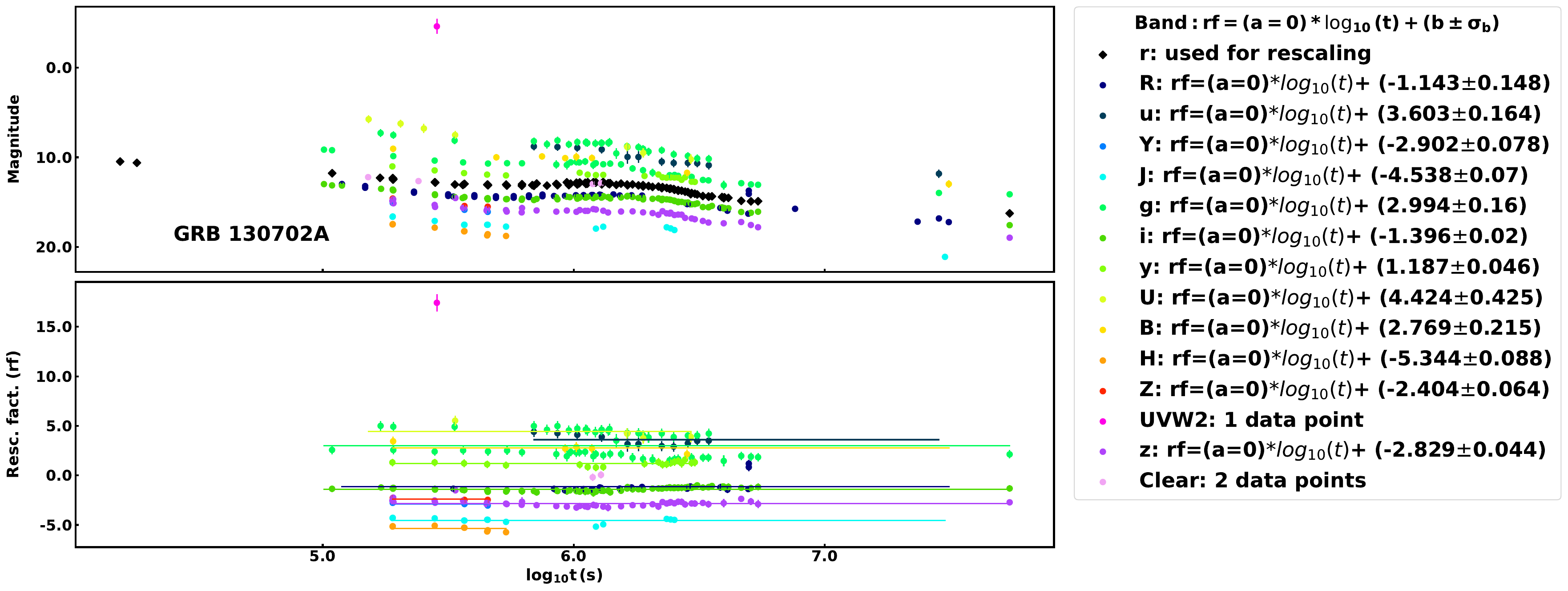} 
  \caption{\textbf{Upper panel.} The magnitude, in AB system corrected for Galactic extinction, as a function of $\log_{10}$ of time (in seconds) is reported for GRB 080109A, 080319B, 130702A showing an apparent disagreement with literature. \textbf{Lower panel.} The rescaling factors to the most numerous filter and the other filters as a function of $\log_{10}$ of time (in seconds). The most numerous filter, together with the fitting slope and its 1 $\sigma$ uncertainty are listed in the legend.}
  \label{fig:appdisagreea0}
\end{figure*}

We here report some examples of disagreement cases between our analysis and the literature. We discuss the 3 apparent disagreement cases for $a=0$ reported in Table \ref{tab:disagreement} (080109A, 080319B, and 130702A reported in Figure \ref{fig:appdisagreea0}). 

{\bf GRB 080109A.} \citet{Li2018a} find no colour evolution while we find colour evolution.

{\bf GRB 080319B.}  \citet{Wozniak2009} highlight a mild colour evolution of $V-I$ in the range of $100\,sec<t<10000\,sec$, from $(V-I)\sim0.8$ at $t=100$ down to $(V-I)\sim0.55$ at $t\sim2000$. In our case, by fitting all the rescaling factors versus time and the general trend suggests no colour evolution.

{\bf GRB 120119A.} In \citet{Morgan_2014}, the authors claim a colour change in the first $200\,sec$ after the trigger due to the dust destruction. In our analysis, the bands $I$ and $unfiltered$ show indeed small fluctuations in the same time range $t=200\,sec$, but we fit all the data in the LC and thus in average we find no colour evolution.

{\bf GRB 130702A}. For this GRB, \citet{Li2018a} highlight colour evolution but in our catalogue we do not have the $V$ band data that they use to find their results. Thus, we find no colour evolution.

{\bf GRB 230812B.} In this case, we have the same number of filters with colour evolution and without colour evolution. \citet{2024ApJ...960L..18S} highlight the colour evolution focusing only on the rescaling factors $g-i$ and $g-r$ and not all the filters like we find.

 

\bsp	
\label{lastpage}
\end{document}